\RequirePackage{rotating} 
\documentclass[fleqn,usenatbib]{mnras}

\usepackage{newtxtext,newtxmath}
 
\usepackage{mdframed}
\usepackage{graphicx}
\usepackage{caption}
\usepackage{cleveref}
\usepackage[T1]{fontenc}
\usepackage{ae,aecompl}
\usepackage{afterpage}

\usepackage{eso-pic}

\AddToShipoutPictureBG*{%
  \AtPageUpperLeft{%
    \hspace{0.75\paperwidth}%
    \raisebox{-3.5\baselineskip}{%
      \makebox[0pt][l]{\textnormal{DES-2020-0530}}
}}}%

\AddToShipoutPictureBG*{%
  \AtPageUpperLeft{%
    \hspace{0.75\paperwidth}%
    \raisebox{-4.5\baselineskip}{%
      \makebox[0pt][l]{\textnormal{FERMILAB-PUB-20-670-AE}}
}}}%

\usepackage{graphicx}	
\usepackage{amsmath}	
\usepackage{amssymb}	
\usepackage{xcolor}     
\usepackage{rotating}   

\DeclareCaptionFormat{cont}{#1 (cont.)#2#3\par}

\usepackage{longtable}
\usepackage{lscape}
\usepackage{multirow}
\usepackage{tabularx}
    \newcolumntype{L}{>{\raggedright\arraybackslash}X}
\usepackage{comment}
\usepackage[flushleft]{threeparttable} 

\usepackage{lineno}

\usepackage{xspace}

\newcommand{\code}[1]{\texttt{#1}\xspace}

\newcommand{\mof}{MOF\xspace}

\newcommand{\GalSim}{\code{GALSIM}}
\newcommand{\SExtractor}{\code{SExtractor}}

\newcommand{\PSFEx}{\code{PSFEx}}

\newcommand{\SWARP}{\code{SWarp}}
\newcommand{\swarp}{\SWARP}
\newcommand{\SCAMP}{\code{SCAMP}}
\newcommand{\scamp}{\SCAMP}
\newcommand{\ngmix}{\code{ngmix}}

\newcommand{\healsparse}{\code{healsparse}}

\newcommand{\gold}{\code{Y3 GOLD}}

\def\Sref#1{Sec.~\ref{#1}\xspace}
\def\Fref#1{Fig.~\ref{#1}\xspace}
\def\Tref#1{Table~\ref{#1}\xspace}
\def\Eref#1{Eq.~(\ref{#1})\xspace}

\title[DES Y3: Deep Fields]{Dark Energy Survey Year 3 Results: Deep Field Optical + Near-Infrared Images and Catalogue}

\author[W.~G.~Hartley, A.~Choi, A.~Amon et al.]{
\parbox{\textwidth}{
\Large{
W.~G.~Hartley,$^{1,*~}$
A.~Choi,$^{2,\dag~}$
A.~Amon,$^{3,\ddag~}$
R.~A.~Gruendl,$^{4,5}$
E.~Sheldon,$^{6}$
I.~Harrison,$^{7,8}$
G.~M.~Bernstein,$^{9}$
I.~Sevilla-Noarbe,$^{10}$
B.~Yanny,$^{11}$
K.~Eckert,$^{9}$
% ...
\begin{center} (DES Collaboration) \end{center}
}
\vspace{0.1cm}
\parbox{\textwidth}{ \small
\textit{Authors' affiliations are shown at the end of this paper.}\\
$^{*}$Corresponding author: william.hartley@unige.ch\\
$^{\dag}$Corresponding author: choi.1442@osu.edu\\
$^{\ddag}$Corresponding author: amon2018@stanford.edu\\
}}}

\date{Accepted XXX. Received YYY; in original form ZZZ}

\pubyear{2020}

\begin{document}
\label{firstpage}
\pagerange{\pageref{firstpage}--\pageref{lastpage}}

\maketitle

\definecolor{pink}{rgb}{0.848, 0.188, 0.478}
\begin{abstract}
We describe the Dark Energy Survey (DES) Deep Fields, a set of images and associated multi-wavelength catalogue ($ugrizJHKs$) built from Dark Energy Camera (DECam) and Visible and Infrared Survey Telescope for Astronomy (VISTA) data. The DES Deep Fields comprise 11 fields (10 DES supernova fields plus COSMOS), with a total area of $\sim30~$ square degrees in $ugriz$ bands and reaching a maximum $i$-band depth of 26.75 (AB, $10\sigma$, $2\arcsec$). We present a catalogue for the DES 3-year cosmology analysis of those four fields with full 8-band coverage, totalling $5.88~$ sq. deg. after masking. 
Numbering $2.8~$million objects ($1.6~$million post masking), our catalogue is drawn from images coadded to consistent depths of $r=25.7, i=25, z=24.3$ mag. We use a new model-fitting code, built upon established methods, to deblend sources and ensure consistent colours across the $u$-band to $Ks$-band wavelength range. We further detail the tight control we maintain over the point-spread function modelling required for the model fitting, astrometry and consistency of photometry between the four fields. The catalogue allows us to perform a careful star-galaxy separation and produces excellent photometric redshift performance (${\rm NMAD} = 0.023$ at $i<23$). The Deep-Fields catalogue will be made available as part of the cosmology data products release, following the completion of the DES 3-year weak lensing and galaxy clustering cosmology work.
\end{abstract}




\section{Introduction}

Deep survey fields have long been a mainstay of optical extragalactic astronomy \citep{kron1980, williams1996, dls2002,kashikawa2004, scoville07, furusawa08}. Given a limited amount of observing time, there is a necessary trade-off between survey depth and area. Studies of the most distant galaxies and faintest objects require images of extremely high sensitivity, which are expensive to build up and, as a result, the deepest fields typically cover just a few square degrees or smaller. Nevertheless, a square degree of area is sufficient to reduce sample variance uncertainties to acceptable levels in most studies of galaxy evolution at moderate and high redshifts \citep{somerville04}\footnote{Where sample variance is a concern, it is usually more efficiently overcome by similar-sized independent, widely separated fields, rather than through expanding area in one or two existing regions of the sky.}. Among the great many uses of these deep fields are statistical studies such as galaxy luminosity and stellar mass functions \citep[e.g.,][]{cohen2002, mortlock2015}, galaxy biasing \citep[e.g.,][]{coil2004, hartley2013}, morphological properties of intermediate and high-redshift galaxies \citep[e.g.,][]{lotz2006, elmegreen2007} and searches for peculiar or extreme sub-populations \citep[e.g.,][]{daddi2004, bouwens2007}. 

Over the decades of work using deep extragalactic fields, our understanding of distant galaxy populations has been accelerated by the addition of longer wavelength data \citep{lonsdale2003, lawrence07, foucaud2007, koekemoer11, mccracken2012, jarvis2013} and spectroscopy \citep{lefevre05, lilly07, coil11, bradshaw13, mclure13, lefevre15, mclure18}, which provide complementary information on the properties of the surveyed galaxies. Near-infrared data in particular are crucial for detecting galaxies at $z>1$, where even massive galaxies can drop out of optical images if they lack significant star formation. In combination, surveys at differing wavelength ranges are of far greater value than the sum of their parts, and as such the fields where these data already exist become the natural targets of further observations. We now have around half a dozen of these deep survey fields scattered across the extragalactic sky, containing panchromatic imaging data and abundant spectroscopy. Key amongst them are the Cosmological Evolution Survey (COSMOS, \citealt{scoville07}) / UltraVISTA \citep{mccracken2012} and Subaru/XMM-Newton Deep Survey (SXDS, \citealt{furusawa08}) / UKIDSS Ultra-Deep Survey (UDS, Almaini et al., in prep.) equatorial fields, and the Great Observatories Origins Deep Survey South (GOODS-S, \citealt{giavalisco2004}) field at declination, $\delta\sim$-28.

Among the images that have been built up over the last ten years in the key extragalactic fields are those collected using dedicated survey instruments or facilities, e.g., Wide-Field Camera (WFCam) on the UK InfraRed Telescope (UKIRT), the Visible and Infrared Survey Telescope for Astronomy (VISTA, \citealt{emerson2004}) and Hyper SuprimeCam (HSC, \citealt{aihara2018}). In such cases, the varied science goals envisioned for these instruments leads to a ``wedding-cake'' strategy, of multiple sub-surveys making different area-depth trade-offs \citep{lawrence07, aihara2018}. A further benefit of this type of survey strategy is that the deeper and richer survey fields can be used to better understand the shallower but wider-area survey, for instance by providing redshift information for colour-selected samples \citep[e.g.,][]{Kim2011}, or a true source list when estimating detection completeness \citep{y3-gold}. In this paper we describe the homogeneous set of images that we have constructed across eleven deep fields, and the catalogue derived from four of them in order to understand and overcome sources of systematic uncertainty in the Dark Energy Survey \citep[DES;][]{DES:2005,DES:2016ktf,desbook} three-year (Y3) cosmology analysis. Although developed with this specific use in mind, we also detail our careful star-galaxy separation and photometric redshifts that make the resulting catalogue a cutting edge data product for a broad range of extragalactic astronomy.

The DES has as its primary goal the characterisation of dark energy through measuring the expansion history of the Universe and growth of structure at late times. The principal probes used by the survey team are weak gravitational lensing, galaxy clustering (including the baryon acoustic oscillation scale length), abundance of galaxy clusters and light curves of distant type-1a supernovae (SN). The survey includes a main wide-field survey of the southern Galactic cap \citep{dr1} and a cadenced sub-survey optimised for SN detection \citep{snsurvey}. Observations were made using the Dark Energy Camera (DECam, \citealt{flaugher2015}), mounted on the Victor M. Blanco 4m telescope at the Cerro Tololo Inter-American Observatory near La Serena, Chile. DES observations began in August 2013 (Y1) and continued for five subsequent southern hemisphere spring and summer seasons until completion in January 2019. The wide survey (WS) spans 5,100 deg$^2$ and comprises 10 sets of exposures per photometric band ($g,r,i,z$ and $Y$) that each tile the entire footprint.  Tilings are offset by up to half of the 1-degree field diameter to fill focal-plane gaps and strongly tie the exposures' photometric calibration, such that a typical source is imaged 8 times per filter.

Concurrently, the SN sub-survey of ten fields was carried out with roughly weekly cadence, and sometimes chosen when observing conditions yielded a point-spread function (PSF) unacceptable for the WS\footnote{If the exposures had FWHM worse than 1.6\arcsec, the sequence was declared bad and was redone.} \citep{NeilsenScheduler2019}. As the goal of this sub-survey was to detect and characterise SN light curves, observations were minimally dithered. During the five years that this sub-survey was active, roughly $30\%$ of the total observing time was taken in these ten supernova pointings. The locations of the SN fields lie within the main survey footprint and were chosen to coincide with a number of pre-existing deep extragalactic survey fields, particularly the VISTA Deep Extragalactic Observations survey (VIDEO; \citealt{jarvis2013}) and spectroscopic surveys. Combined with DECam observations of the COSMOS / UltraVISTA field, the SN fields have already provided high-quality redshift information for photometric redshift (photo-z) calibration in DES (\citealt{sanchez14, bonnett16}, \citealt*{hoyle18}).

In the DES Y3 weak lensing and clustering cosmology analysis, the DES Deep Fields (SN fields plus COSMOS) are used to understand sources of systematic uncertainties in the main survey area and how they propagate through the cosmology analysis. Sources measured in the deep stacked coadd images act as an effective truth table of the galaxy population that is present in the main survey, and thus provide a highly accurate input for the ``Balrog'' \citep{suchyta16, y3-balrog} and weak-lensing image simulation \citep{y3-imagesims} pipelines and will, in future, be used in the ``Bayesian Fourier Domain'' \citep[BFD;][]{bernsteinarmstrong2014,bernstein2016} shear measurement approach. In the Balrog process, copies of Deep-Fields galaxies are injected into the main survey images and their response to the extraction pipeline measured. Similarly, the BFD method of measuring weak gravitational lensing (WL) magnification and shear requires a prior on the distribution of galaxy moments in the measurement bands from higher $S/N$ images ($S/N \ge\sqrt{10}\times$ the summed WS images, see \citealt{bernstein2016}).

Furthermore, the inference of galaxy redshift distributions in the DES Y3 analysis relies heavily on the multi-layered ``SOMPZ'' methodology (\citealt*{buchs19}; \citealt{Sanchez19}; \citealt*{y3-sompz}). 
The SOMPZ method is a photometric redshift approach based upon the self-organising map algorithm (SOM, \citealt{kohonen1982, masters2015}), and takes as input a set of galaxy photometry.
The more photometric bands used as input, the narrower the intrinsic dispersion of galaxy types and redshifts within a single cell of the derived map, and thus the more accurate it is expected to be in terms of mean redshift. To further facilitate the use of the Deep Fields for this purpose, DES and community DECam exposures in complementary bands ($u$ and $Y$) in the SN fields and COSMOS field were also obtained. The final cosmology catalogue contains photometry in eight bands, $ugrizJHKs$, for 1.6 million sources across four fields with a total area of $5.88$ square degrees (after masks are applied).

Similar area and, in some cases, deeper optical data already exist in our Deep Fields. The great value in the present work is the ability to combine the WS and Deep Fields data on a consistent footing. Having deep and wide data taken using the same filter bands is essential for the aforementioned Balrog, SOMPZ and BFD to reach sub-percent accuracy. With these goals in mind, we take great care in PSF modelling, photometric calibration with respect to the WS and consistency of tools used, to ensure the highest level of homogeneity between the two data sets that we can achieve.

Though designed with WL cosmology as the primary use, the Deep Fields are anticipated to add to the long tradition of deep extragalactic science performed in these regions of the sky \citep[e.g.,][]{coppin2006, williams2009, kovac2010, ilbert2013, lee2013, hartley2015, galametz2018, maltby2019, 2020MNRAS.493.2059B, momose2020}, and to enhance non-dark energy science in the WS. The $ugrizY$ DECam observations represent an update to the optical data that have previously been available to users of the VIDEO data, while a highly consistent dataset covering both VIDEO and UltraVISTA / COSMOS will open up possibilities to more easily estimate and overcome the sample variance present in studies that have previously relied only on one field. Though often not a major limitation, the sample variance caused by large-scale over-densities in degree-scale fields impacts our understanding of the even basic quantities, such as the stellar mass function at intermediate redshift \citep{2017A&A...605A..70D}. 

The remainder of the paper is structured as follows: in \Sref{sec:data} we describe the input images, their processing, creation of coadded images, masking and basic tests of image quality. Catalogue extraction from the detection images is presented in \Sref{sec:catalogue} and is followed by details of the PSF modelling in \Sref{sec:psfmodel}. \Sref{sec:photometry} provides details of the multi-band photometry measurements, based on model fits across the $r$, $i$ and $z$-bands. We then perform a fine-tuning of the photometric zero-points in the four Deep Fields in \Sref{sec:calibration} and go on to describe our star-galaxy separation and photometric redshifts of the extracted sources in \Sref{sec:stargal} and \Sref{sec:photoz}. We conclude in \Sref{sec:conclusions}. In the Appendices we provide additional information about the images and software configurations, an alternative astrometric solution approach and the final catalogue columns. We expect to make this final catalogue available to the community alongside other data products released from the DES Y3 cosmology analysis.

\section{Input data and image construction}
\label{sec:data}

The scientific utility of deep survey fields is enhanced by the combination of bands covering different wavelengths, both for the DES cosmology analysis and ancillary science. In this section we describe the construction of the optical DES Deep-Fields images (\Sref{sec:optobs}) and coincident near-infrared images (\Sref{sec:irdata}) that will later be combined in our cosmology catalogue. 

\subsection{Optical (DECam) observations}\label{sec:optobs}

\begin{figure*}
    \centering
    \includegraphics[height=0.245\textheight]{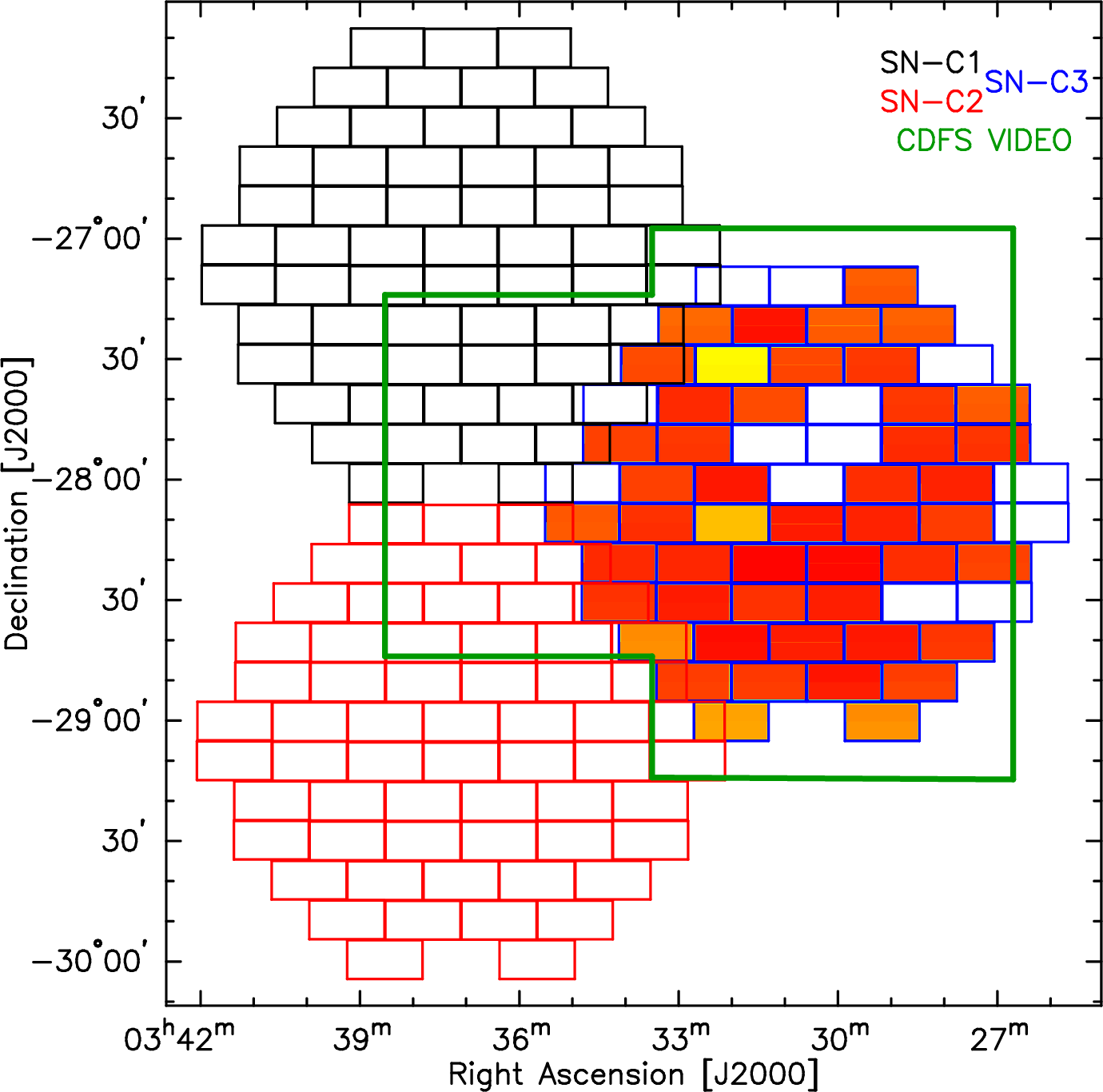}
    \includegraphics[height=0.245\textheight]{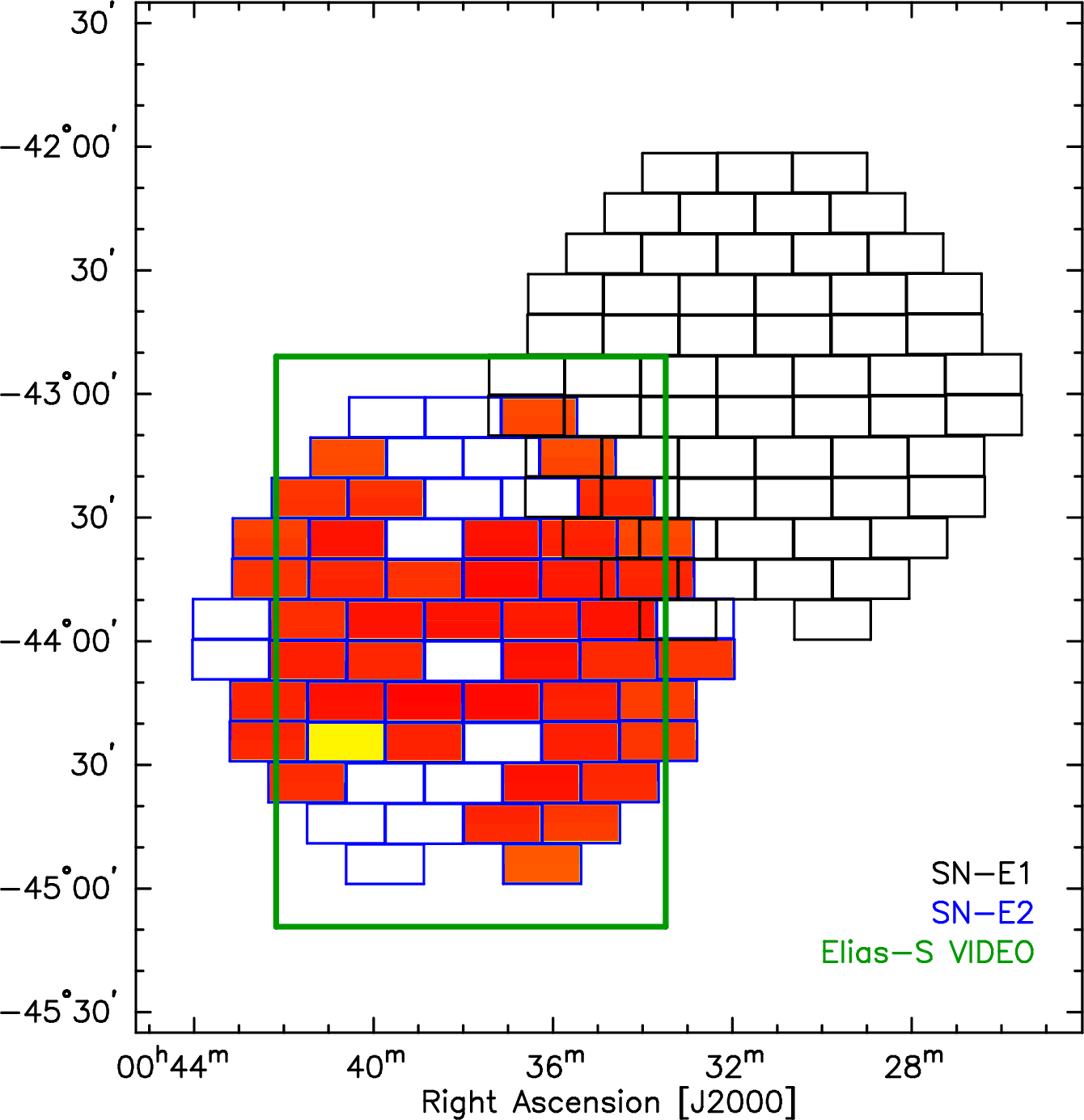}
    \includegraphics[height=0.245\textheight]{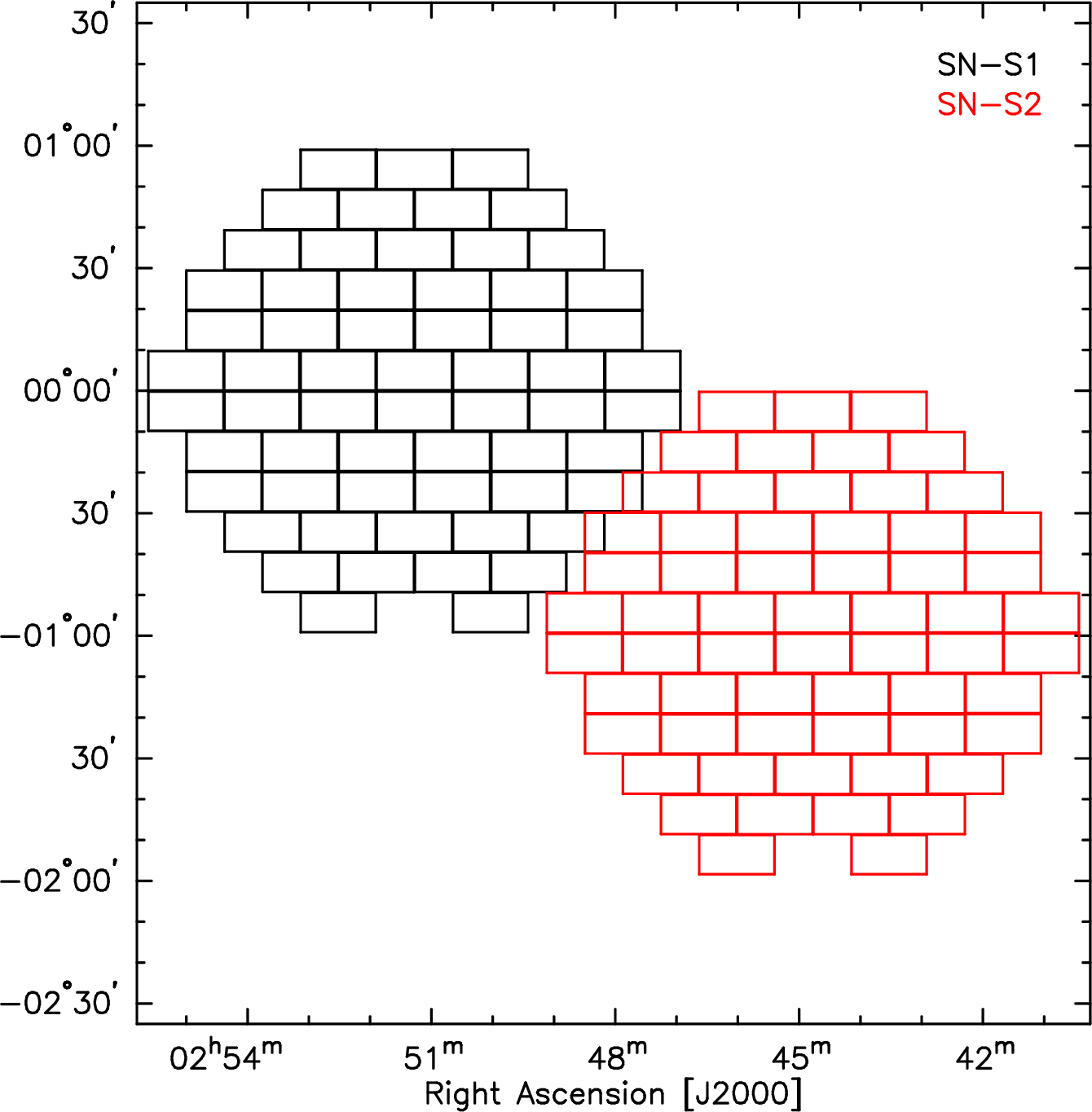}
    \includegraphics[height=0.245\textheight]{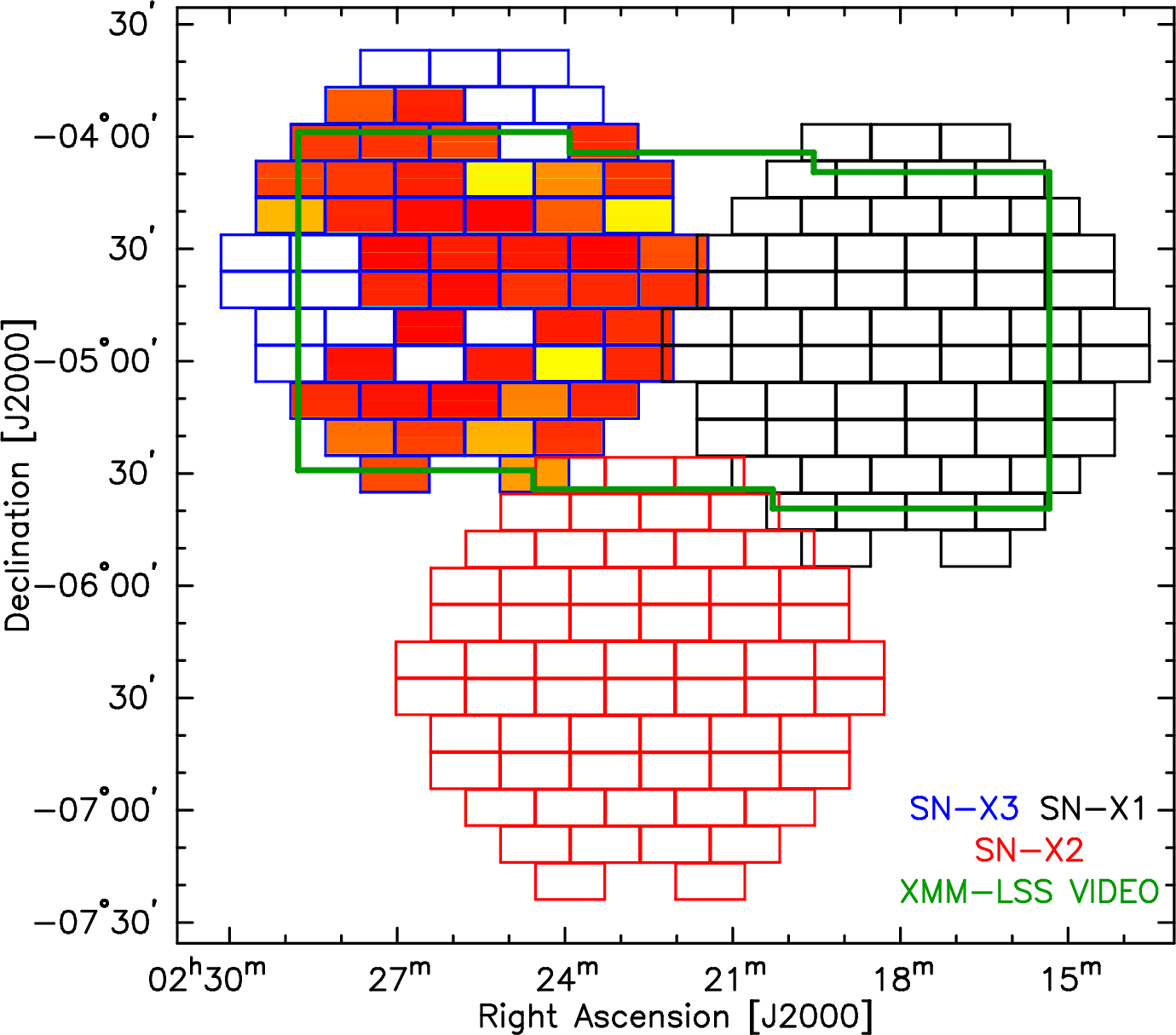}
    \includegraphics[height=0.245\textheight]{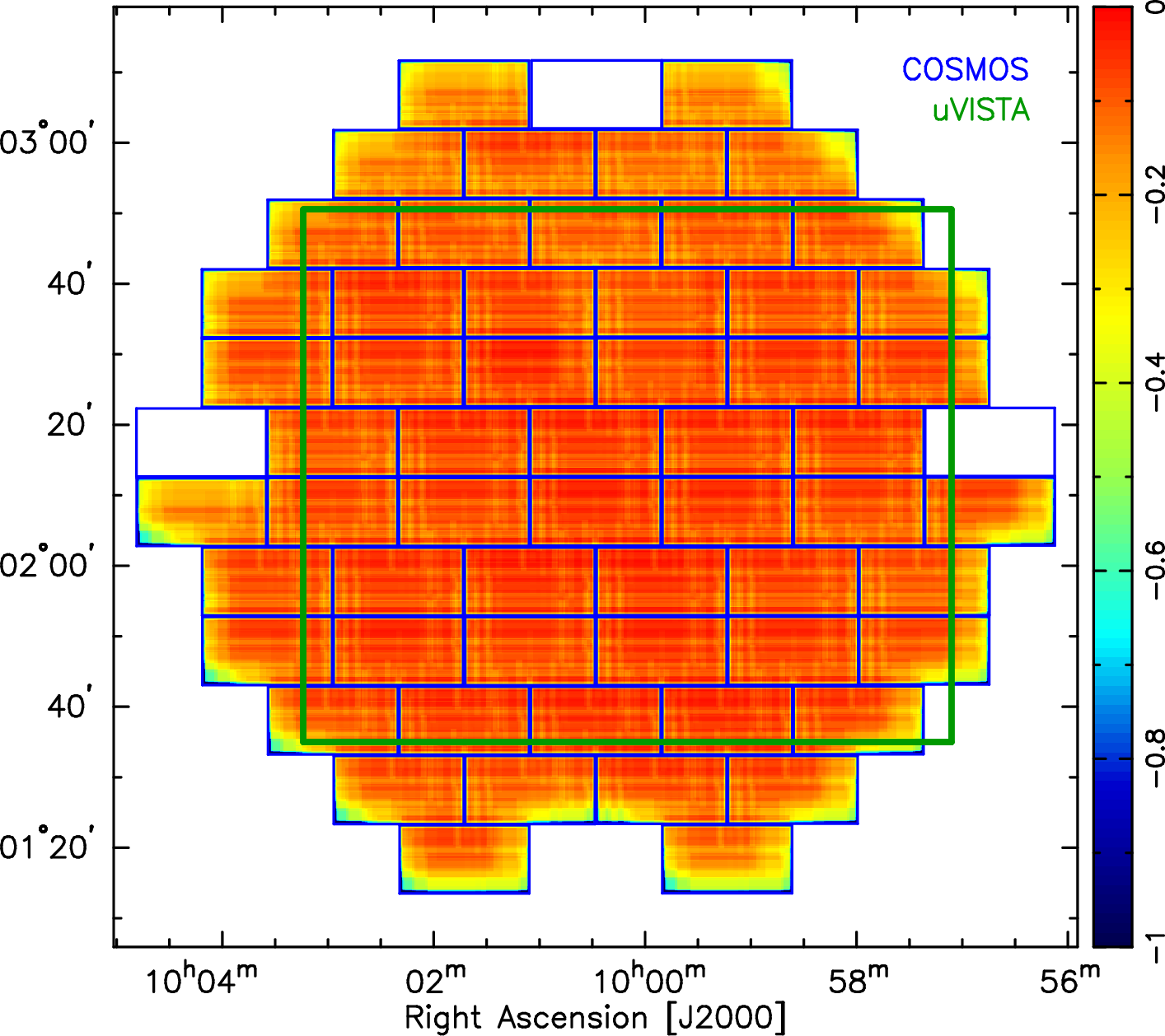}
    \caption{Locations and layout of the DES Deep Fields, comprising ten supernova survey pointings (SN-C, X, E and S) and one pointing covering the COSMOS field. Filled rectangles represent the subset of the Deep Fields' DECam chips that were used in building our Deep-Fields catalogue for the DES Y3 cosmology analysis (see \Sref{sec:field_selection}), colour coded by difference in $i$-band $10\sigma$ limiting magnitude relative to the maximum depth in that field. Green outlines show VISTA near-infrared coverage from VIDEO and  UltraVISTA. Empty chips in SN-C3, SN-X3, SN-E2 and COSMOS indicated those that were excluded entirely in our catalogue due to scattered light or other large defects (see \Sref{sec:masks}).}
    \label{fig:deepfield_schematic}
\end{figure*}

Construction of the optical-band deep-field images begins with a query to the DECam exposure database for all images of duration $>30$~s that are within $\approx1\arcmin$ of the nominal field centers of the 10 DES supernova fields (see \Fref{fig:deepfield_schematic}) in the $ugrizY$ bands.  For the $griz$ bands, a very large number of exposures are available from the DES SN search programme, which observes these fields with roughly weekly cadence through their entire periods of visibility. 
The DES SN program limits dithers to $\lesssim20\arcsec$ from the nominal pointing, to avoid inhomogeneity of depth or PSF characteristics in the stacked images. Homogeneity is also desirable for the Deep Fields, hence the restriction to well-aligned pointings.  DES also conducted single-night observational campaigns in the $u$-band (for all SN fields) and $Y$-band (for a subset of the fields). 
The DES SN data have been supplemented with all publicly available community observations satisfying the pointing constraints.  Community exposures substantially improve the total depth of the $u$-band observations of the SN-E2, SN-C3, and SN-X3 supernova fields.  We construct $ugrizY$ images of the COSMOS field from DES and community DECam observations, again constrained to be near the nominal pointing so that stacks of each DECam CCD's exposures yield a homogeneous image \citep[for full details, see][]{NeilsenScheduler2019}.  
A summary for the fields targeted in this paper is given in the Appendix, a subset of which is provided in \Tref{tab:fields} for the four fields that comprise our cosmology catalogue.

The selected exposures were all processed through the DES single-epoch pipelines \citep{Morganson2018DESDM} with the same configurations as those used for the DES Wide Survey (WS), and the \emph{griz} DES SN observations were included in the global photometric calibration \citep[see][]{burke2018}. 
The photometric calibration for the remaining images were obtained by bootstrapping through a set of tertiary standards. The DES pipelines also assess the data quality of each observation which was used to select images for coaddition.  The quality assessment yields for each exposure $i$ of duration $T_i$ an estimate of the PSF size $FWHM_i$, the sky noise variance level $s_i$, and the atmospheric transmission $\eta_i$ relative to a clear night.  From these we create an "effective exposure time" \citep{Neilsen:2016},
\begin{equation}
    T_{\rm eff,i} \propto T_i \frac{\eta_i^2 }{s_i \times FHWM_i^{2}}.
\end{equation}
$T_{\rm eff,i}$ is proportional to the $(S/N)^2$ of a faint point source of fixed magnitude for that exposure.  We normalize $T_{\rm eff}$ for each band such that a value of unity is obtained for a typical 90~s DES WS exposure taken in clear, dark conditions.

\begin{table*}
\begin{center}
\caption{\label{tab:fields}  DES Y3 Deep Fields position, area, $i$-band FWHM, exposure time and depth, number of sources detected in the COADD\_TRUTH detection images (after applying masks) and number of chips excluded due to scattered light or other large scale image quality defects. Exposure time, FWHM and depth information for all pointings, depth levels and DECam filters is given in \Tref{tab:full_table}.}
\begin{tabular}{|c|c|c|c|c|c|c|c|c|}
\hline \hline
Field   & RA    & dec   & Mask-free Area & FWHM$_i$ & Exp. Time$_i$ & Depth$_i$ & N sources & N excl. chips\\ 
& (J2000) & (J2000) & (sq. deg.) & ($^{\prime\prime}$) & (s) & ($10\sigma$, $2\arcsec$) & \\
\hline
 SN-C3     & 52.6484 (03:30:35.6) & -28.1000 (-28:06:00.0) & 1.70  & 0.72 & 5,862 & 25.06 & 462,739 & 13\\ 
 SN-X3     & 36.4500 (02:25:48.0) & -4.6000 (-04:36:00.0)  & 1.52 & 0.77 & 4,122 & 25.04 & 430,555 & 16\\ 
 SN-E2     & 9.5000 (00:38:00.0)  & -43.9980 (-43:59:52.8) & 1.42 & 0.83 & 5,620 & 25.06 & 386,946 & 17\\ 
 COSMOS & 150.1166 (10:00:28.0) & +2.2058 (+02:12:21.0)  & 1.24 & 0.94 & 8,885 & 25.54 & 403,147 & 3\\ \hline 
 Total & - & -  & 5.88 & - & 24,489 & - & 1,683,387 & 49\\ \hline
\end{tabular}    
\end{center}
\end{table*}

\begin{table}
\begin{center}
\caption{DECam community programmes used in conjunction with DES imaging to construct the Deep-Fields images.}
\label{tab:extdata}
\begin{tabular}{|l|l|l|l|} 
\hline \hline
Program   & PI & Field & Bands \\ 
\hline
 2014B-0613 & Cooke        & SN-C3     & $u$      \\
 2015B-0603 & Infante      & SN-C3     & $Y$      \\
 2016A-0620 & Curtin       & SN-C3     & $u$      \\
 \hline
 2013A-0351 & Dey          & COSMOS & $ugrizY$ \\
 2013A-0529 & Rich         & COSMOS & $u$      \\
 2014A-0073 & Kilic        & COSMOS & $g$      \\
 2014B-0146 & Sullivan     & COSMOS & $riz$    \\
 2014B-0404 & Schlegel     & COSMOS & $gz$     \\
 2015A-0107 & Belardi      & COSMOS & $g$      \\
 2015A-0608 & Forster      & COSMOS & $gri$    \\
 2016A-0104 & Sullivan     & COSMOS & $griz$   \\
 2016A-0386 & Malhotra     & COSMOS & $zY$     \\
 2016A-0610 & Infante      & COSMOS & $z$      \\
 \hline
 2014A-0239 & Sullivan     & SN-E1     & $u$      \\
 2014B-0146 & Sullivan     & SN-E1     & $u$      \\
 2016A-0104 & Sullivan     & SN-E1     & $u$      \\
 \hline
 2014A-0191 & Hildebrandt  & SN-E2     & $u$      \\
 2016A-0104 & Sullivan     & SN-E2     & $u$      \\
 \hline
 2016B-0260 & Curtin       & SN-X3     & $u$      \\
 \hline
\end{tabular}    
\end{center}
\end{table}


\subsubsection{Selection of input images}\label{sec:decam_input_selection}
Each of the science goals for the DES Deep Fields is optimized by building the lowest-possible-noise image at or below a given target effective resolution (\texttt{FWHM}), or the smallest possible \texttt{FWHM} at a given effective noise level.  To optimize the depth/resolution trade-off in a given band, we order the candidate exposures for a given field and band by increasing \texttt{FWHM}.  An optimally weighted sum of the first $N$ images in this ordered list will then achieve the minimal noise for its output \texttt{FWHM}, and the minimal \texttt{FWHM} for its noise level.  We quantify the expected depth and resolution of such a coadd by
\begin{align}
    T_{\rm eff}(N) & = \sum_{i=1}^{N} T_{{\rm eff},i} \\
    FWHM(N) & = \frac{\sum_{i=1}^{N}T_{{\rm eff},i}FWHM_i}{T_{\rm eff}(N)}
\end{align}
Varying $N$ thus describes the best available trade-off between noise and resolution for each filter and band. \Fref{fig:depthvsres} plots the available trade-off in $i$-band for all the SN fields.

\begin{figure*}
    \centering
    \includegraphics[width=\linewidth]{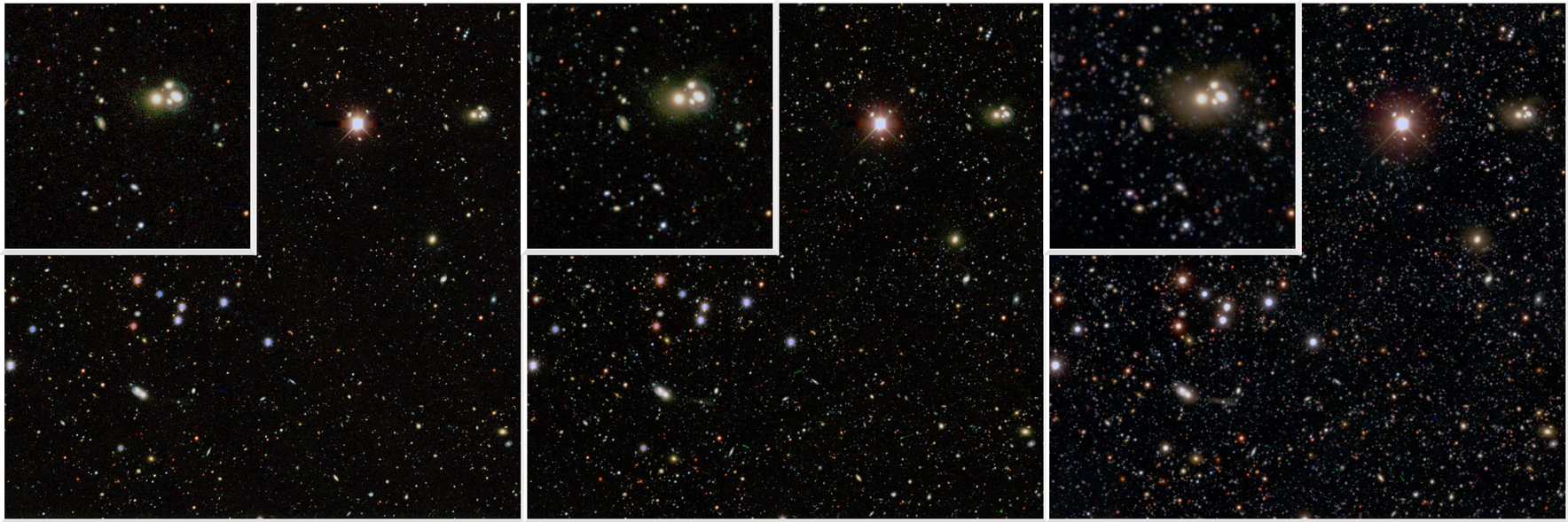}
    \caption{$grz$ image of approximately one half of a DECam chip from the SN-X3 field, showing the three depth level coadd images we produce: \texttt{SE\_TRUTH}, \texttt{COADD\_TRUTH} and \texttt{DEEPEST}, left to right respectively. Inset figures are a factor 2 zoom-in of the interacting group located on the far right in the main image.}
    \label{fig:colourimg}
\end{figure*}

\begin{figure}
    \centering
    \includegraphics[width=\linewidth]{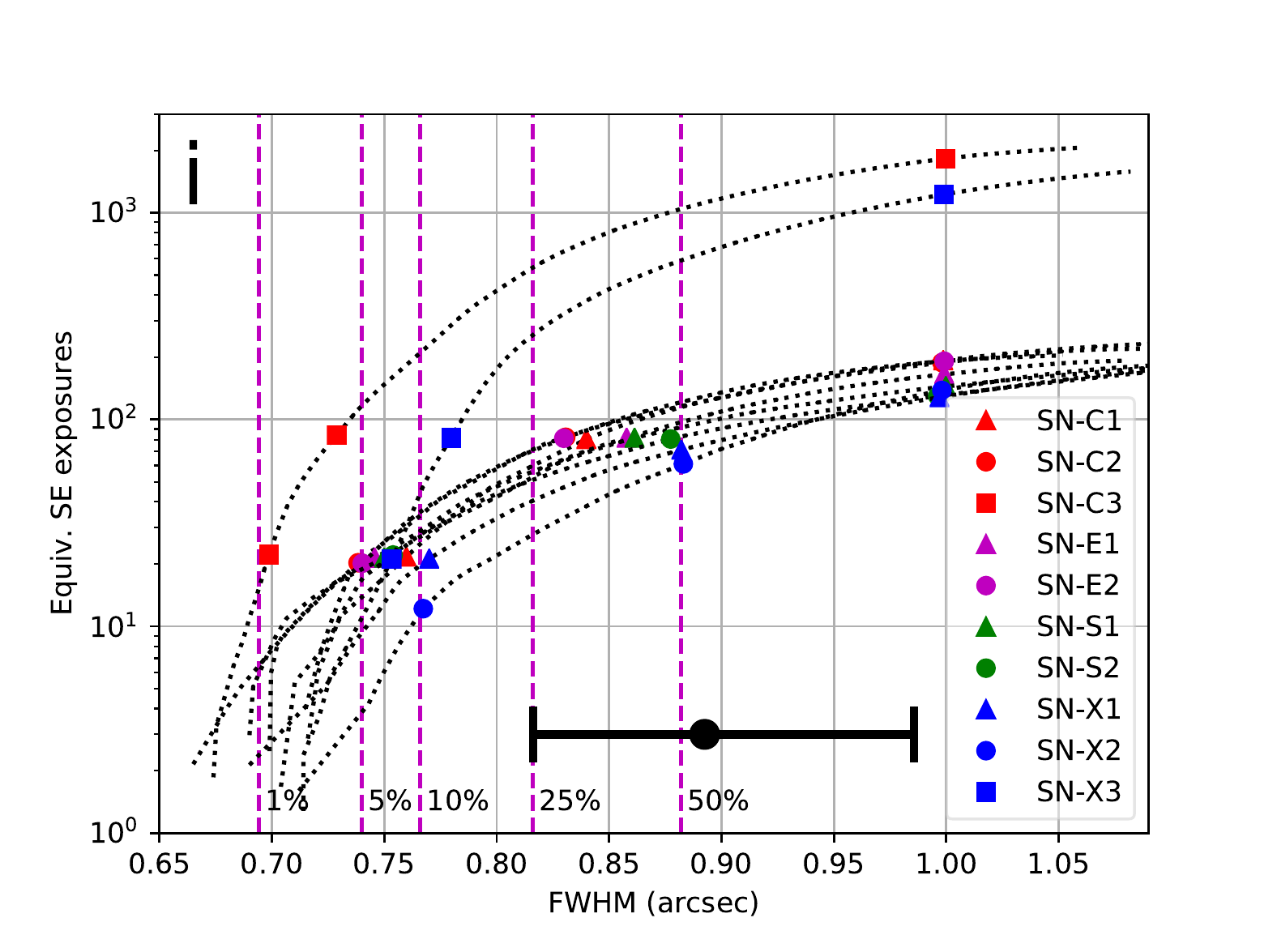}
    \caption{The dotted lines describe the depth-vs-resolution tradeoff available for the $i$-band in each of the 10 DES SN fields. The vertical axis gives depth in terms of effective exposure time, in units of a typical DES WS exposure. Dashed vertical lines mark the percentiles of \texttt{FWHM} for all WS single exposures in this band used in the Y3 data release. The black dot and error bar mark the median and 10--90 percentile range of seeing \texttt{FWHM} obtained in Y3 coadd images for this band. The colored symbols mark the choices made in each field for depth/resolution of the three varieties of Deep-Fields coadd images. The three groups (from lower left to upper right) are SE\_TRUTH, COADD\_TRUTH, and DEEPEST.}
    \label{fig:depthvsres}
\end{figure}

For each of the ten SN fields we create 3 images in the $griz$ bands that represent different choices in this depth / FWHM trade-off:
\begin{itemize}
\item The {\bf DEEPEST} coadd is targeted at maximizing the detectability of faint sources.  It includes all exposures until relaxing the \texttt{FWHM} cutoff no longer substantially increases depth.  The resultant targeted deep-image $FWHM$'s are $1.2\arcsec$, $1.0\arcsec$, 1.0\arcsec, and 0.9\arcsec in $g,r,i,$ and $z$ bands, respectively.  The DEEPEST coadds for the SN-C3 and SN-X3 fields are exceptionally deep in the red: in $z$ band, they are $\approx3000\times$ deeper than a typical single WS exposure ("SE depth"), summing $\approx250,000$~s of exposure time, and reaching 3.2~mag deeper than the DES WS coadds.
\item The {\bf COADD\_TRUTH} Deep Fields are intended to provide galaxy images for Balrog injection, and priors on the "noiseless" distributions of properties of galaxies detectable in DES WS coadds.  We target a depth $10\times$ the expected WS coadd depth, i.e., $T_{\rm eff}=80$ in units of typical SE depth, with typically $38$ input images per coadd.  But we place a higher priority that the deep-field $FWHM$ be no worse than the median SE $FWHM$ of 1.10\arcsec, 0.96\arcsec, 0.88\arcsec, and 0.84\arcsec in $griz$ bands, so some of the shallow SN fields do not attain COADD\_TRUTH depth of $T_{\rm eff}=80$ in $gri$ bands.
\item The {\bf SE\_TRUTH}: coadds are intended to produce galaxy images for injection into the single exposures with very good seeing, i.e., those for which the COADD\_TRUTH images have significantly higher $FWHM$ than the target SE's.  The input images to the SE\_TRUTH Deep Fields are chosen to meet the most restrictive of two criteria: $T_{\rm eff}=20,$ and $FWHM$ better than 90\% of all WS exposures.  This places an upper limit of 0.93\arcsec, 0.82\arcsec, 0.77\arcsec, and 0.72\arcsec on the $griz$ $FWHM$'s of SE\_TRUTH coadds.  All of the SE\_TRUTH Deep Fields can attain sufficient resolution with $T_{\rm eff}\ge10$ except for SN-X2 $r$-band.
\end{itemize}

Example $grz$ colour images for these three depth levels covering half of one of the chips in the SN-X3 field are shown in \Fref{fig:colourimg}. The available data for the SN-field $u$ and $Y$ bands, and for the COSMOS field, come from a handful of nights and do not represent as wide a range of seeing conditions as is available for the SN-field $griz$ data.  As a consequence, for each of these we create just a single deep coadd, using nearly all of the available data.  For the COSMOS field, this results in $FWHM$ values of 0.94\arcsec--1.20\arcsec for the $ugrizY$ bands.

\subsubsection{Pipeline changes with respect to DES Y3 GOLD}
The Deep Fields required some modest configuration changes to the DES multi-epoch (coaddition) pipeline used for the Y3A2 reductions on which the DES Y3 GOLD sample is built (see \citealt{y3-gold}).  The standard DES COADD tiles are defined as a fixed set across the sky with extent of $\sim 0.75\degr \times 0.75\degr$. For the Deep Fields, we defined new coaddition "tiles" that are each slightly larger than a single CCD $18.41^{\prime} \times 9.64^{\prime}$ and centered at the median location of each CCD within the DECam focal plane for a given field.  A summary of the changes in the processing follows.

First, we added constraints to more strictly enforce which images contributed to the resulting coadd image.  In the nominal WS pipeline, all calibrated survey images that meet the survey quality cuts are included.  For the SN Deep Fields the input images at \emph{griz}-bands were restricted to come from the same CCD. At \emph{u}- and \emph{Y}-bands, this restriction had to be lifted because the images were drawn from a much smaller population that included large offsets between the constituent exposures.  For the COSMOS Deep Field, un-dithered data do not exist, however, many exposures exist that have telescope pointings with bore-sights within $\frac{1}{2}$ of a DECam CCD width/height of one-another. Thus we loosened the restriction compared to that applied to the SN fields, and allowed adjacent CCDs to also contribute to a given COADD image. This procedure results in a more uniform depth across each image in the COSMOS field than we would have otherwise achieved. 

Second, astrometric solutions are refit using all images in all bands simultaneously (using the AstrOmatic utility \texttt{SCAMP}, \citealt{Bertin:2006}).  For the WS this function uses the objects from each exposure that overlap a given COADD tile. For the Deep Fields, only catalogues from the individual CCDs that overlap the tile are used.  We used GAIA-DR1 \citep{gaiadr1}, which was the best astrometric reference catalog available at the time when this processing occurred, and ensure that our astrometric accuracy is at least as good as the WS ($\sim150~$mas).

In the Y3 DES COADD pipeline detection, images were formed by constructing a combined \emph{r,i,z}-band image using \texttt{SWarp} with \texttt{COMBINE\_TYPE CHI\_MEAN}. For the Deep Fields this was found to produce less robust detection of faint objects in the presence of diffuse emission, therefore the configuration was altered to  \texttt{COMBINE\_TYPE AVERAGE} (i.e., a simple average of the images). Finally, alterations in object detection and PSF measurement are detailed in \Sref{sec:catalogue} and \Sref{sec:psfmodel} respectively.

\subsubsection{Background subtraction tests}

\begin{figure*}
    \centering
    \includegraphics[scale=0.37]{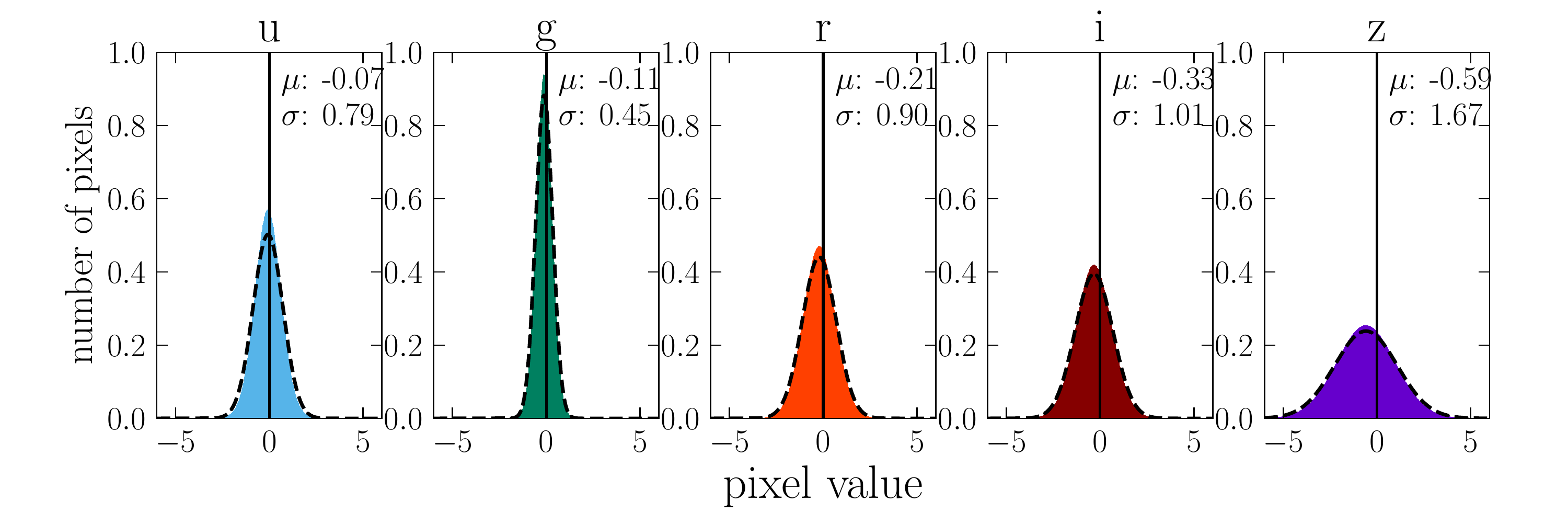}\includegraphics[scale=0.37]{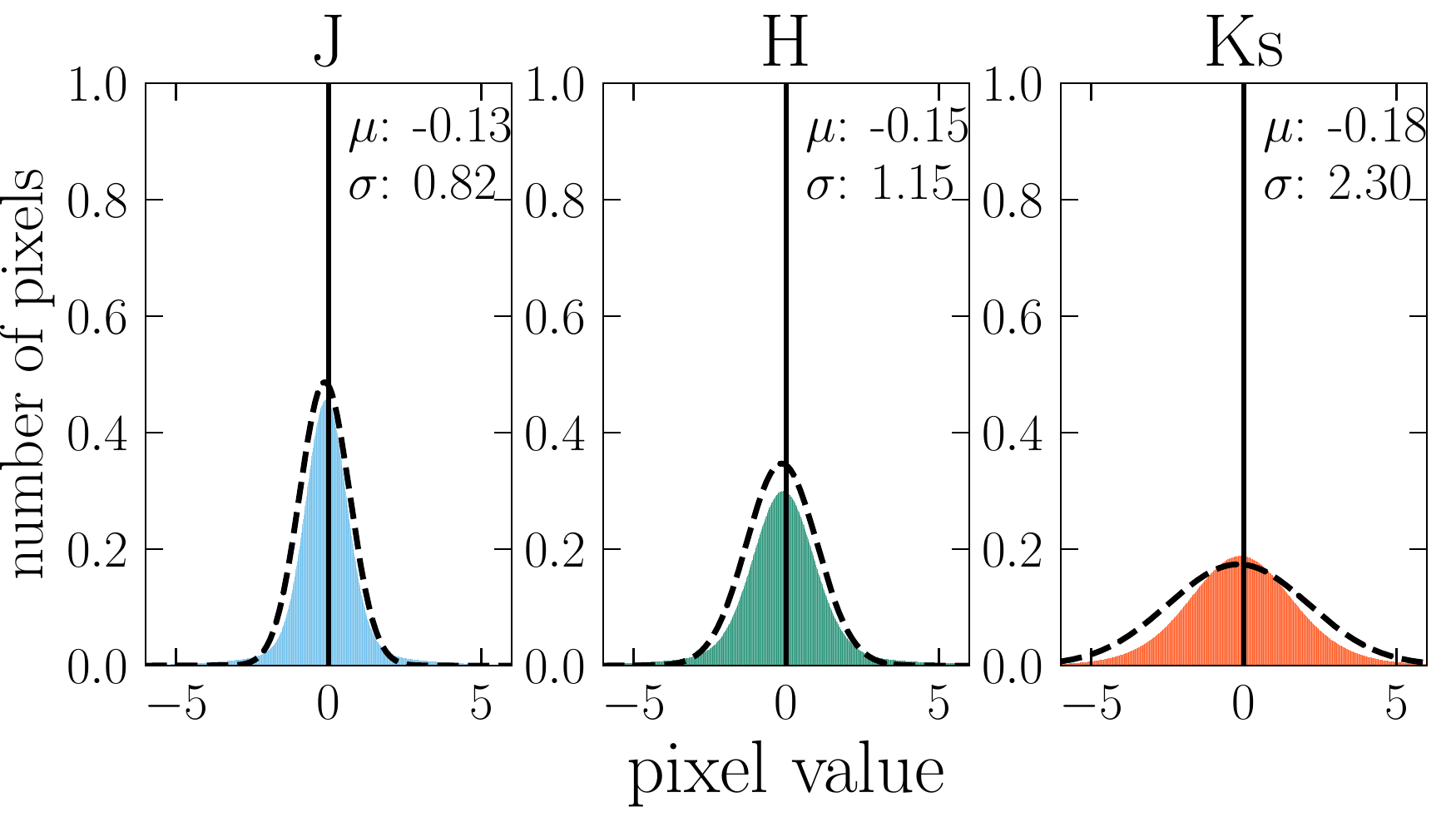}
    \caption{Histograms of pixel values for the optical DECam filters used in this work, after removal of pixels coincident with sources and masks. Here we show a stack over all chips for each band, together with Gaussian function fits to the distribution. The background is very slightly over-subtracted: mean values correspond to surface brightnesses of $30.0, 29.3, 28.7, 28.2, 27.6, 29.3, 29.2$ and $29.0~{\rm mag/sq. arcsec}$ for $u,g,r,i,z,J,H$ and $Ks$ respectively.}
    \label{fig:decam_bg_test}
\end{figure*}

Residual sky background estimation for the optical DECam images is performed using \texttt{SExtractor}, which can slightly over-estimate the background due to the faint outskirts of some objects not being identified as source pixels. Once that background is subtracted, such an over-estimation would lead to a small negative background when photometric measurement is performed. This is particularly important for our model-based photometry (see \Sref{sec:photometry}), as one of the components that is fit to the galaxy images follows a de Vaucouleurs profile, with $I\propto(1/R)^4$. A poorly estimated background could therefore have a significant impact on the measured flux.

We check the accuracy of the sky subtraction by analysing the values of the pixels that remain after removing masked regions (see \Sref{sec:masks}) and source pixels identified in \texttt{SExtractor}'s segmentation check image. The distribution of the remaining pixel values should be approximately Gaussian and centred on zero, with the possible addition of a tail of high pixel values due to those source pixels that have not been included in the segmentation map and a low level of residual artifacts, which we can remove via clipping.

The left-most five panels of \Fref{fig:decam_bg_test} show normalised histograms of pixel values (scaled to a common zeropoint, ${\rm ZP}=30~$AB) of all chips in the COSMOS area for each DECam band. Overlaid are best-fit Gaussians, with mean $\mu$ and sigma $\sigma$ reported in the top right corners. In each case the pixel distribution is well represented by the Gaussian fit, and the mean value well within the standard deviation with respect to the zero (black solid line). Across these DECam chips, the background seems to be very slightly over subtracted, with mean values equivalent to surface brightnesses of $30.0, 29.3, 28.7, 28.2$ and $27.6~{\rm mag/sq. arcsec}$ for $u,g,r,i$ and $z$ respectively. Our test does not probe any positional dependence of background over or under subtraction, and for simplicity we leave the minor offsets uncorrected in the following.

\subsection{Near-IR images}\label{sec:irdata}

The VISTA infrared camera (VIRCam; \citealt{dalton2006}) on the VISTA telescope is a sparse array of 16 detectors within its field of view, arranged to allow a contiguous mosaic to be constructed from six pointings, covering $1.5 \times 1.23$ square degrees. The VIDEO survey contains a total of eight such mosaics across three extragalactic fields: XMM, ELAIS and CDFS. UltraVISTA comprises a single mosaic, but with half of the pointings taking almost seven eighths of the total survey exposure time, resulting in stripes that are $0.5-1$ magnitudes deeper than the remaining three pointings. Further details of the VIDEO and UltraVISTA data are given in \citet{jarvis2013} and \citet{mccracken2012} respectively, and point-source depths and mean seeing values are provided in \Tref{tab:nearir}.

To build our near-infrared images we begin with all non-deprecated frame stacks in all broad-band filters\footnote{Our final catalogue contains just the set of $J$, $H$ and $Ks$ bands that cover all of our selected fields.} taken as part of the VIDEO \citep{jarvis2013} and UltraVISTA \citep{mccracken2012} public surveys that were available as part of the VIDEO and UltraVISTA DR4 release. Images are served via the Vista Science Archive\footnote{http://horus.roe.ac.uk/vsa/} following data reduction at the Cambridge Astronomical Survey Unit (CASU). The CASU reduction pipeline stages include corrections for detector reset, linearity, darks, flat field and sky background. Images are further destriped and jitter stacked to account for bad pixels. These steps are described in detail in the CASU online documentation\footnote{http://casu.ast.cam.ac.uk/surveys-projects/vista/technical/data-processing}. The two surveys were carried out over 15 semesters between October 2009 and February 2018 (DR4 includes data up to and including semester 2014B), and contain 8,987 (VIDEO) and 1,485 (UltraVISTA) such frame stacks across all filters.

\subsubsection{Image processing and coaddition}\label{sec:ircoadds}

VISTA frame stacks come in the form of multi-extension Flexible Image Transport System (FITS) files, with one extension for each of the 16 chips in a single pointing. These files were unpacked into individual chip images, fluxes scaled to a common zero-point of 30 mag. using the image header zero-point and the noisy edges of each image were trimmed by setting the corresponding weight map to zero in the 100 pixel region around the border. In selecting the frames to go into our final coadded images, we chose to follow the quality assurance scheme used in the VIDEO survey and excluded only those frames that have reported seeing greater than $1~$arcsec FWHM. Note that the release mosaics in the UltraVISTA survey were built from a lower-level data product with stricter quality control and an improved treatment of the background subtraction. One of our chief concerns is producing a consistent set of photometry across our different fields, and so we treated the UltraVISTA frames in the exact same way as we did the VIDEO ones, rather than attempting to replicate their process.

Coadded images were produced for each combination of pointing and chip number. We chose to build distinct coadds for each pointing in order to keep the variation in PSF across each image as simple as possible, without sharp discontinuities at the joins of different chips and pointings. To build the coadded images we used \texttt{Swarp} with \texttt{sigma\_clipping}$=3.5\sigma$, i.e., input pixel rejection for values further than $3.5\sigma$ from the mean value in that pixel. 

The resulting images contain a number of cosmetic defects, including an electronic effect called detector column pull-down which has previously been observed in Spitzer data and the UKIDSS UDS. We followed Almaini et al. (in prep.) in correcting for this effect by median filtering the background, excluding source pixels identified using \texttt{SExtractor}, with a thin rectangular tophat kernel of dimensions $200 \times 3~$pix$^{2}$. The filter is applied twice, with a transposed filter during the second pass. This process does a good job in correcting the visible stripes of background decrements and also corrects small amounts of over-subtraction that can occur around bright objects or moderately clustered regions of the images. However, it comes at the cost of further correlating the image noise, which is already correlated due to the \code{SWarp} resampling during coaddition. Any cosmetic issues that remain at this point were masked out (see \Sref{sec:masks}).

\subsubsection{Background subtraction tests}

Although the background filtering algorithm that we employ has been tested in detail on the UKIDSS UDS (Almaini et al., in prep.), we nevertheless perform the same check that we applied to the optical data to measure the sky subtraction accuracy. 
We re-run \texttt{SExtractor} on the VIDEO and UltraVISTA single-chip coadd images, this time with background estimation fixed to a constant value of zero. The results are shown in the right-most three panels of \Fref{fig:decam_bg_test} for the three near-IR bands. As expected, the median background level is extremely low in all three cases ($29.3$, $29.2$ and $29.0~{\rm mag/sq. arcsec}$ for $J$, $H$ and $Ks$ respectively). This is by construction through the background correction process. The pixel distributions are clearly non-Gaussian, likely owing to depth variation from the dithered observations.

\subsubsection{Astrometry}\label{sec:astrometry}

A shared astrometric alignment among the images in each band for each object is central to the multi-band fitting algorithms used for galaxy photometry models.  In the optical bands the DES pipelines accomplish this by simultaneously fitting astrometric solutions with \texttt{SCAMP} for all the images (in all bands) that contribute to a coadd tile. In order to achieve the same for the near-IR images, we used the DES catalogs as an astrometric reference and simultaneously fit all the constituent near-IR band images/catalog to produce a bootstrapped astrometric solution for each near-IR image. We derived an alternative solution for the COSMOS/UltraVISTA field via a Gaussian Process regression of object positions using stars from the UltraVISTA DR3 catalogue (see Appendix~\ref{sec:gp_astrom}). This solution worked extremely well and was fully consistent with our \texttt{SCAMP} astrometry, providing further confidence.

\subsection{Image masking}\label{sec:masks}

Because the deep coadds combine far more exposures than the WS coadds, there are many more opportunities for the coadd to become contaminated by unmasked single-exposure anomalies, particularly streaks from asteroids, meteors, and satellites.  As a consequence, we implemented several improvements to the streak-detection algorithms to make the single-exposure masks more complete.

Masking was conducted on parallel but different tracks for the optical DECam and near-IR VIRCam images. In both optical and near-IR cases, we started with automated masks and supplemented them with an extensive manual masking campaign.  For the DECam images, \healsparse\footnote{https://github.com/lsstdesc/healsparse} masks combine automated masks from the DESDM processing and manual masks into a single file, which are applied at the catalog-level. 

For the DECam images, the manual masking was performed by viewing color images, so that artifacts were more easily identifiable.  Typical artifacts and transients included cosmic rays, artificial satellites, meteors, and asteroids. DES 3-color images of the Deep Fields were viewed on CCD-sized tiles by a team consisting of a mix of undergraduates and more experienced astronomers. Transient objects were flagged using a tool that recorded the center of the transient, the size of the circular patch that spanned it, and a comment about the shape of the transient, e.g., “rainbow streak”, “single band point”, “weird thing”, etc. A long streak could be covered by a sequence of small circular patches. Each tile was scanned several times. Images that were heavily contaminated by scattered light or any other large or extensive defects were identified at this point and later excluded during catalogue construction.

\begin{figure}
    \centering
    \includegraphics[width=\linewidth]{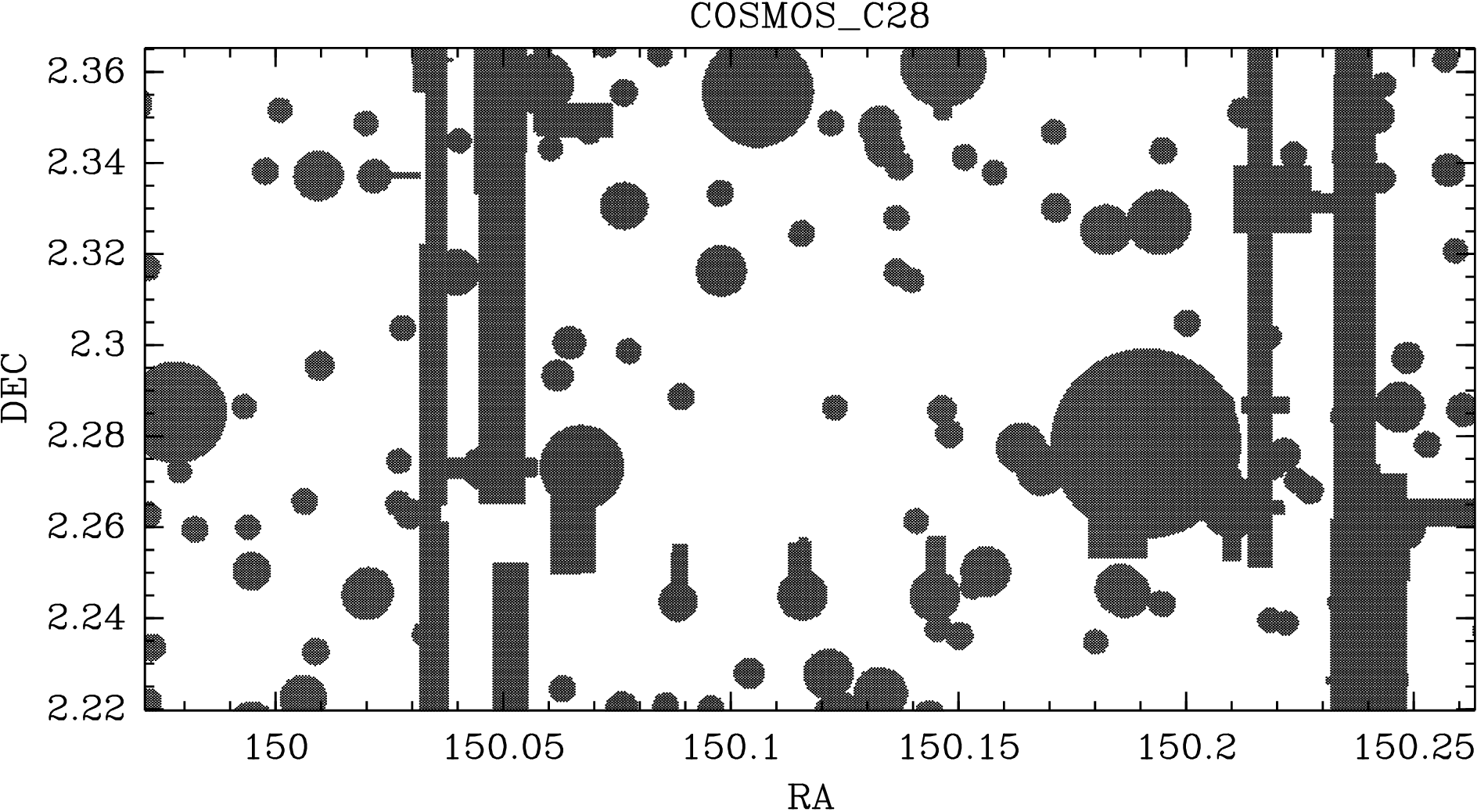}
    \caption{An example of masked regions in COSMOS, specifically on the C28 chip. This is an intersection of \healsparse masks constructed for the DECam coadd and the UltraVISTA manual masks.}
    \label{fig:masking}
\end{figure}

For the VIRCam images, we begin by constructing an automated mask around all infrared-bright stars, as follows:
\begin{itemize}
    \item Using the Two Micron All Sky Survey \citep[2MASS;][]{2006AJ....131.1163S} Point Source Catalog (PSC), we identify RA and Dec positions of point sources in $J$, $H$, and $Ks$.
    \item We then inspect a few randomly-selected coadd images and manually apply circular regions to the point source positions using SAOImageDS9 \citep{2003ASPC..295..489J}, masking out unusable areas of the images.
    \item With these regions, we fit a second-order polynomial to obtain a function that determines the radius of a circular region given the magnitude of the point source.
    \item Finally, we create masks for all near-IR images using this function.
\end{itemize}
Following the automated process, we then manually inspected all of the single-chip coadds, adjusting the base masks and applying additional polygons when necessary to areas affected by scattered light, stellar halos, and other artifacts. \Fref{fig:masking} shows an example of the intersection of DECam and VIRCAM masked regions on a single chip in the COSMOS/UltraVISTA region, C28.

\subsection{Image depths and seeing}
\label{sec:decam_depth}

\begin{table}
\begin{center}
\caption{\label{tab:nearir} Median $10\sigma$ depth in $2\arcsec$ apertures and PSF FWHM for the VISTA data used in this work.}
\begin{threeparttable}
\begin{tabular}{|c|c|c|c|c|c|}  \hline \hline
Field   & Programme & J & H & K \\ \hline
 SN-C3 & VIDEO &23.5, $0.8\arcsec$&22.7, $0.8\arcsec$&22.8, $0.8\arcsec$\\ \hline
 SN-X3  & VIDEO &23.8, $0.8\arcsec$&23.3, $0.8\arcsec$&22.9, $0.8\arcsec$\\ \hline
 SN-E2 & VIDEO &23.0, $0.8\arcsec$&23.2, $0.8\arcsec$&22.9, $0.8\arcsec$\\ \hline
 COSMOS &UVISTA$^{*}$&24.0, $0.77\arcsec$&23.6, $0.76\arcsec$&24.0, $0.78\arcsec$\\ \hline 
\end{tabular}
\begin{tablenotes}
    \item[$^{*}$] Numbers refer to the deep, rather than ultra-deep, stripes.
\end{tablenotes}
\end{threeparttable}
\end{center}
\end{table}

Seeing FWHM, exposure time and $10\sigma$ limiting depth ($2\arcsec$) for each DECam band, field and coadd depth level are provided in the Appendix, \Tref{tab:full_table}. In \Fref{fig:deepfield_schematic} we show the depth variation in the $i$-band for each field. Variations between chips of up to $0.3$ magnitudes can be seen in the SN fields, which are the result of a fraction of the input images being excluded from the final coadds due to cosmetic defects that were either not identified or were not able to be corrected for earlier in our pipeline. Meanwhile the effect of dithered pointings is clear in the COSMOS field. Here, the small gaps between chips and the field edges result in variations of up to $0.2$ magnitudes within individual chips, and a fall-off of depth at the field edges.

Depths and seeing FWHM for VIDEO and UltraVISTA are described at length in the relevant papers \citep{mccracken2012,jarvis2013}. Our images are single-chip images built from the same input frames, and as such, match closely in depth and seeing. Average values for VIDEO and UltraVISTA are reported in \Tref{tab:nearir}, where the values given for UltraVISTA represent the shallower stripes (see \Sref{sec:irdata} and \citealt{mccracken2012}).

\section{Catalogue extraction and photometry}
\label{sec:catalogue}

We generate catalogues from the detection images of all depth level COADD images using \SExtractor \citep{Bertin:1996}. We use the same configuration as used for the DES Y3 processing (see \citealt{Morganson2018DESDM}), but with lower thresholds for detection and more aggressive deblending to accommodate the relative increase in objects in the Deep Fields compared to the DES main survey. The specific parameter changes used were: \texttt{DETECT\_THRESH} 0.8, \texttt{ANALYSIS\_THRESH} 0.8, \texttt{DEBLEND\_MINCONT} 0.00001, and \texttt{DEBLEND\_NTHRESH} 64. The values of these parameters for the WS processing were arrived at by inspecting fields that include galaxy clusters or bright stars, and balancing the deblending and detection of true objects in the clusters against spurious sources near bright stars. In the present work, we adjust them in order to better identify faint sources, as bright stars are later masked out. In the remainder of this section we detail the process of constructing the DES Deep-Fields catalogue for our Y3 cosmology analysis, beginning with the sub-selection of the fields we use.
 
\subsection{Field selection for the Y3 cosmology catalogue}
\label{sec:field_selection}

\begin{figure}
    \centering
    \includegraphics[width=\linewidth]{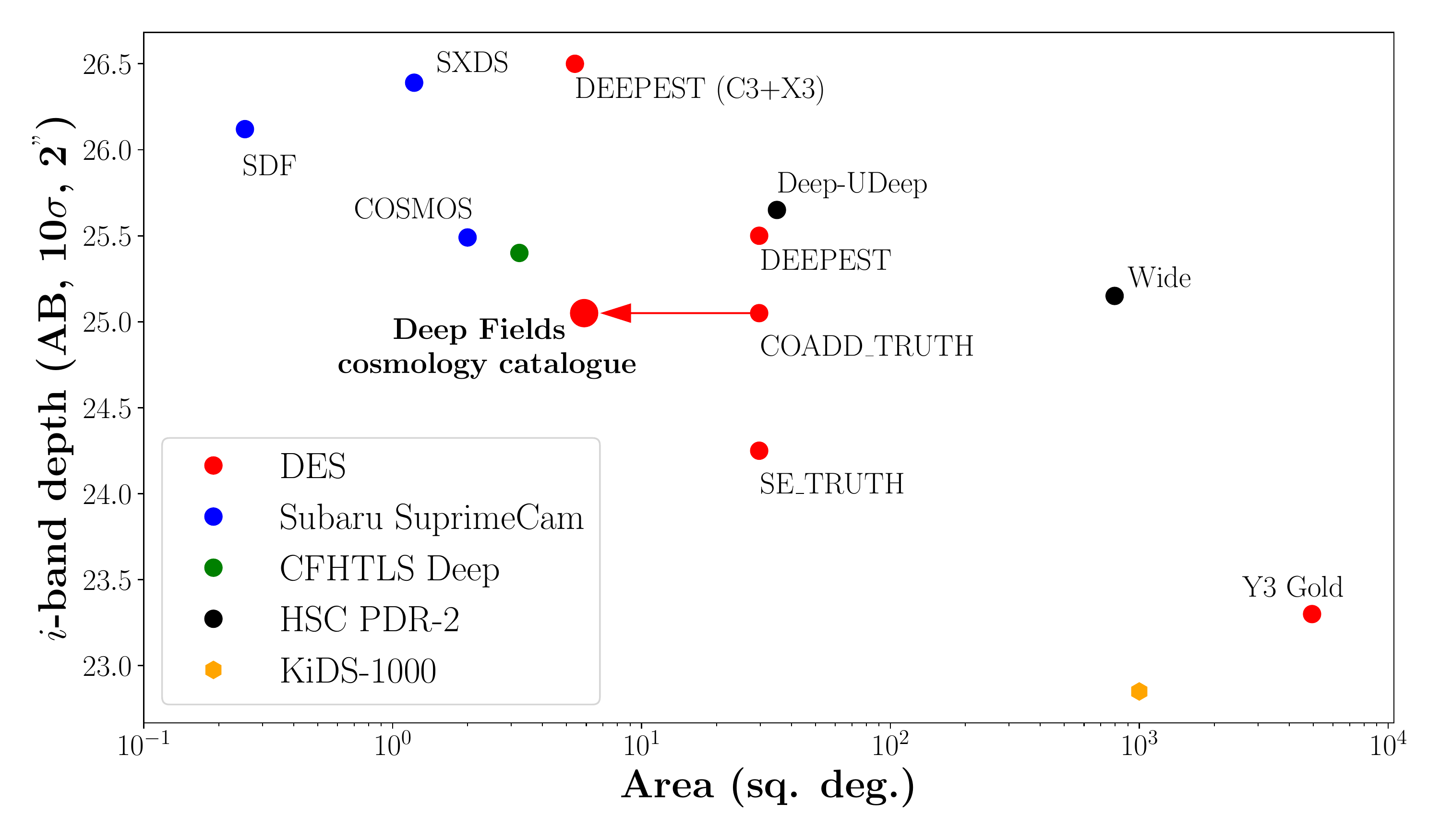}
    \caption{$i$-band depth (AB, $10\sigma$, $2\arcsec$) versus area for the DES Deep Fields and sub-selections relevant to this work, alongside a few other contemporary optical surveys.}
    \label{fig:surveys}
\end{figure}

The scientific goals of the Deep Fields in part require the existence of longer-wavelength near-infrared data to complement the optical DECam images, but also sufficient area to minimise sample variance uncertainties. We therefore choose a sub-set of the Deep Fields, with the aim of balancing the number of available photometric bands against the area of coverage. We select four pointings of the eleven we have at our disposal from which to build our main Deep-Fields catalogue: SN-C3, SN-X3, SN-E2 and the COSMOS field. 

The SN-C3 and SN-X3 fields are the two deep DES SN fields which were also chosen to overlap with pre-existing spectroscopic data sets, such as the VVDS Deep 02hr field. All four fields substantially overlap with VISTA data, either from VIDEO or UltraVISTA, as can be seen in \Fref{fig:deepfield_schematic}. The SN-X1 field is also largely covered by the VIDEO programme, and can be used to take advantage of spectroscopic redshifts from the UDSz \citep{bradshaw13,mclure13}. However, at the time of data collection $J$-band images were not available for this pointing (VIDEO DR4). The field is also covered by the UKIDSS UDS, but the WFCam filters used for the UDS differ slightly from their VIRCam equivalents. For simplicity we do not use the SN-X1 field in building our catalogue, but intend to do so in future iterations. 

As previously mentioned, one of the goals of the catalogue is to provide a list of objects that can be injected into the main DES survey area. This is in order to measure the transfer function relating noiseless true photometry to the photometry recovered by the detection and photometric measurement pipeline. The catalogue used for this purpose must be highly consistent in terms of filter responses and measurement pipelines, sufficiently deep as to be effectively noiseless, but also have high enough resolution in seeing FWHM to be able to reflect the characteristics of the main survey. These competing needs lead us to produce the COADD\_TRUTH level of coadded images, and it is these that we use to construct our catalogue. The remainder of this paper concerns only these chosen images and the multi-wavelength catalogue we build from them. A shortened summary of the characteristics of our selected fields is provided in \Tref{tab:fields}, with fuller information provided in \Tref{tab:full_table}. The $i$-band depth and area of the images from which we draw our catalogue is shown in the context of other optical surveys in \Fref{fig:surveys}. These include the previously mentioned SXDS \citep{furusawa08}, COSMOS \citep{scoville07}, HSC \citep{2019PASJ...71..114A}, in addition to the Subaru Deep Field \citep[SDF;][]{kashikawa2004}, the Canada-France-Hawaii Telescope Legacy Survey Deep\footnote{\url{https://www.cfht.hawaii.edu/Science/CFHTLS/T0007/CFHTLS_T0007-TechnicalDocumentation.pdf}}, and the Kilo-Degree Survey \citep[KiDS;][]{Kuijken_K1000}. Note that KiDS has a similar wavelength coverage as described here, but includes a $Y$-band \citep[for details, see][]{Wright_KV450}.

\subsection{Point Spread Function Models}
\label{sec:psfmodel}

We model the point spread function using a combination of \SExtractor and \PSFEx for both the optical and near-IR data, with some differences in the settings used for star selection that are detailed in the following.

\subsubsection{Optical PSFs}\label{sec:optpsf}

The procedure is derived from the one used for the standard DESDM pipeline described in Sec. 4.5 of \citet{Morganson2018DESDM}.  As an initial step, we run \SExtractor on the coadds described in \Sref{sec:optobs} with settings geared toward detection of point sources (\texttt{DETECT\_THRESH} 5.0;  \texttt{DETECT\_MINAREA} 3) \citep[c.f. Table 12 in ][]{Morganson2018DESDM}.  \PSFEx is then run on the resulting \SExtractor catalogue, which contains sub-images, or ``vignettes'', corresponding to each detection. A key difference compared to the settings provided in Table 13 of \citet{Morganson2018DESDM} is that when running \PSFEx to automatically select stars, \texttt{SAMPLE\_MINSN}\footnote{More information on \PSFEx configurable parameters can be found at \url{https://psfex.readthedocs.io/_/downloads/en/latest/pdf/}.} is increased from 20 to 70.  \texttt{SAMPLE\_MINSN} controls the minimum S/N of the vignette allowed into the star sample. Values of 30, 50, 70, and 100 were all tested, and 70 was the value determined to best balance the goal of getting more than 100 PSF stars per DECam chip whilst not including galaxies, with the latter assessed by inspecting plots of \texttt{FWHM} (pixels) vs S/N. The default \texttt{PSFVAR\_DEGREES} is the same as for the standard pipeline and is set to 2, so that the mapping of the PSF variations over pixel coordinates is done with a quadratic polynomial.

\begin{figure*}
    \centering
    \includegraphics[width=\linewidth]{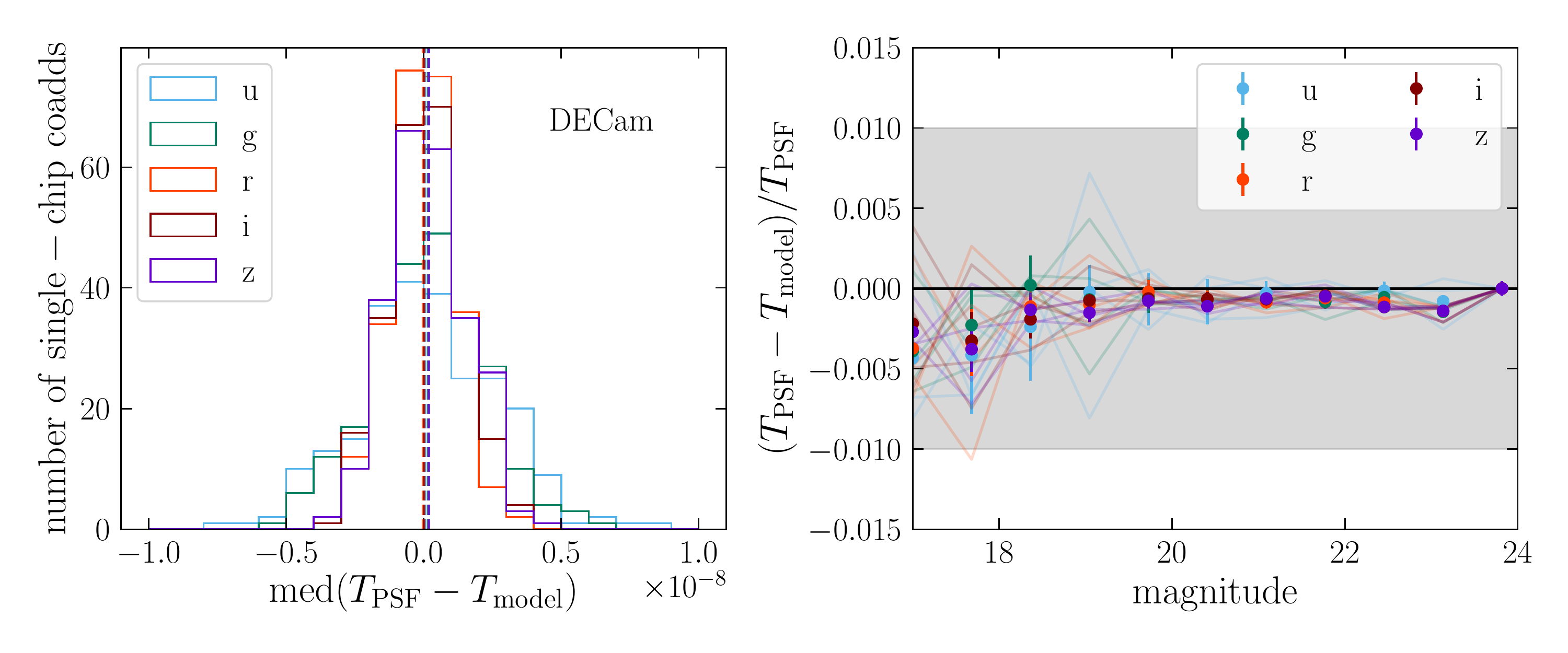}
    \caption{COSMOS and SN field PSF residuals. Left panel: a histogram of the median PSF residuals for each single-chip coadd in the $u, g, r, i, z$ filters.  The median values of each histogram are plotted as dashed lines. Right panel: Fractional residual PSFs binned by magnitude, which is taken here as \texttt{MAG\_AUTO} in the given filter.  Median values over all magnitude bins are plotted as solid lines (in this case not visible as they are so close to zero).  The shaded grey band indicates a 1\% region.  Solid lines are overlaid that indicate the spread of the four different fields, while the filled data points indicate the averages of all four fields.}
    \label{fig:optpsf}
\end{figure*}

To validate the performance of the PSF models produced by \PSFEx we investigated size residuals, where the size $T$ is given by $$T=I_{xx}+I_{yy},$$ and $I_{xx}$ and $I_{yy}$ are second moments of the light distribution \citep{1997A&A...318..687S}. For each of \PSFEx's selected stars, we made cutouts of the image and weight files at the given location and fit the object using \ngmix\footnote{\url{https://github.com/esheldon/ngmix}}, a software package to fit Gaussian models to the light distribution, to obtain $T_{\rm PSF}$. We ingested the PSF model at the given location, drew it into a cutout using \GalSim \citep{2015A&C....10..121R}, and fit the object using \ngmix to obtain $T_{\rm model}$.  A histogram of the median residual for each single-chip coadd $T_{\rm PSF}-T_{\rm model}$ is shown in the left panels of \Fref{fig:optpsf} for COSMOS COADD\_TRUTH, and fractional residuals $(T_{\rm PSF}-T_{\rm model})/T_{\rm PSF}$ are shown in the right panel binned as a function of magnitude (\texttt{MAG\_AUTO}) for the given filter.  The data points indicate the fractional residuals averaged over all four fields fields shown, while the solid lines show the fractional residuals for each field for a sense of variation. These figures show that, overall, the PSF models meet the main stipulation of fractional residuals $<1\%$ that is required to obtain accurate photometry measurements, shown as the shaded gray region.

\subsubsection{Near-IR PSFs}
\label{sec:nirpsfs}

For the near-IR data, we ran \SExtractor with \texttt{DETECT\_THRESH} 5.0 
and  \texttt{DETECT\_MINAREA} 3 as before, although in this case background subtraction is turned off as it has already been subtracted during coadd image creation and processing. We then ran \PSFEx to automatically select stars with \texttt{SAMPLE\_MINSN} set to values ranging from 60 to 500 for UltraVISTA single-chip coadds and 70-100 for VIDEO single-chip coadds.  
These values were chosen after visual inspection of the stellar locus in plots of \texttt{FWHM} (pixels) vs S/N that are automatically produced by \PSFEx, and we revised the choice of \texttt{SAMPLE\_MINSN} to account for different noise levels in some of the single-chip coadds. We imposed an additional criterion of $16<$ \texttt{MAG\_AUTO} $<20$ and re-ran \PSFEx on this selection of objects with \texttt{PSFVAR\_DEGREES} set to 3.

\begin{figure*}
    \centering
    \includegraphics[width=\linewidth]{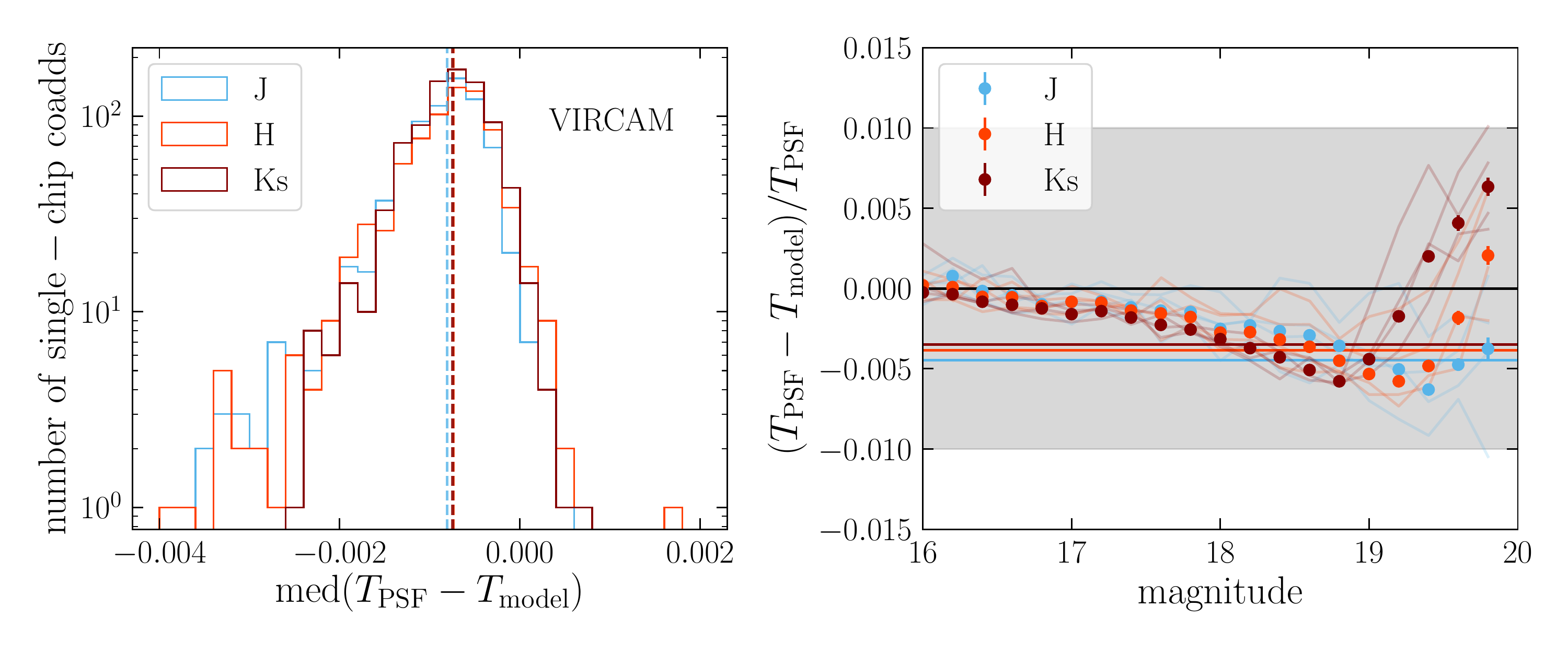}
    \caption{UltraVISTA and VIDEO PSF residuals, similar to Fig~\ref{fig:optpsf} but for UltraVISTA and VIDEO $J, H$, and $Ks$ filters.}
    \label{fig:nirpsf}
\end{figure*}

Histograms of the median residual for each single-chip coadd $T_{\rm PSF}-T_{\rm model}$ are shown in the left panel of Figure~\ref{fig:nirpsf} (UltraVISTA, SN-C3, SN-X3, and SN-E2) with the three different colors representing the different filters $J$, $H$, or $Ks$.  Fractional residuals $(T_{\rm PSF}-T_{\rm model})/T_{\rm PSF}$ are shown in the right panel of Figure~\ref{fig:nirpsf} binned as a function of magnitude (\texttt{MAG\_AUTO}) for the given filter. Comparison of the PSF residuals in the DECam optical bands (Figure~\ref{fig:optpsf}) and the VIRCAM NIR bands (Figure~\ref{fig:nirpsf}) indicates that the \PSFEx models are much more accurate for DECam, possibly owing to imperfections in non-linearity correction during the near-infrared data reduction.

\subsection{Photometric measurement}
\label{sec:photometry}

Our photometric measurements are based on a pipeline similar in spirit and approach to the Multi-Object Fitting \citep[\mof,][]{y1gold} used to create \gold but with some differences, necessary to handle the greater source density and addition of near-infrared data. The DES photometric pipeline is a development upon the \texttt{ngmix} shape measurement software
for weak lensing measurements and was adopted chiefly to enable consistent galaxy model fits across a set of single-epoch images, rather than a single coadded image where the PSF may vary discontinuously across the image. In the case of the DES SN fields the single-epoch images are minimally dithered with respect to one another, and discontinuities in the PSF across a coadded image are therefore not a concern.
Nevertheless, our key science goals require that the Deep Fields photometry is extracted in the same manner as the main survey (for the $griz$ bands at least), and thus we use a pipeline built upon the same software principles, adjusting where necessary. The main steps to produce photometry measurements are:

\begin{itemize}
\item Object detection (see \Sref{sec:decam_input_selection}).
\item PSF measurement (see \Sref{sec:psfmodel}).
\item Reformatting input images into postage stamp image cutouts for fitting and collation of required meta-data.
\item Formation of neighbouring object groups and deblending. 
\item PSF and galaxy model fitting, with simultaneous photometry measurement for the bands being fit.
\item Forced photometry measurements on the remaining bands.
\end{itemize}

\noindent We now describe the last four of those steps.

The DES photometry pipeline is built around Multi Epoch Data Structures \citep[MEDS;][]{2016MNRAS.460.2245J,2018MNRAS.481.1149Z}\footnote{\url{https://github.com/esheldon/meds}}, a \texttt{fits} data format developed to assist object-by-object operations, such as photometry and shape measurements, across large survey data. MEDS comprise the collated information for each object's single-epoch image cutouts, weight images, bit masks, segmentation footprints and associated meta-data. One such file is produced per band and contains the information for all objects detected in a coadd $riz$ detection image. The MEDS files are produced with a minimum cutout size of 32 pixels per side, up 256 pixels depending on the object's \texttt{FLUX\_RADIUS} measurement from \SExtractor (see \citealt{2016MNRAS.460.2245J}, Appendix A for details). The \SExtractor-estimated background is also subtracted at this point, meaning that we are free to fix the background to zero during object fitting. MEDS files are produced automatically as part of the pipeline processing for our DECam data, but were built separately for the VISTA near-IR data.

We construct near-IR MEDS files based on the $riz$ detection catalogue, using the optically-determined RA and Dec for each of the objects to make cutouts of the corresponding locations in the near-IR images and weights. Rather than use each individual frame for the near-IR, we use the coadded images as a single epoch. The individual exposures in the near-IR are shallow, due to the bright sky background, and many such exposures are required in order to build up image depth. We consider the additional cost of processing so many epochs to far outweigh the possible gain in information retention. Note that the near-IR weight files have already had the masks described in \Sref{sec:masks} applied and thus contain zeros in any pixel that coincides with a mask. Furthermore, we save the corresponding PSF models described in \Sref{sec:psfmodel} and segmentation maps that we generate with \SExtractor. To create these segmentation maps, we run \SExtractor over all the near-IR images and weights with \texttt{-DETECT\_THRESH 1.1}, \texttt{-ANALYSIS\_THRESH 1.1}, and \texttt{-DEBLEND\_MINCONT 0.001}. Finally, astrometric solutions from \scamp, as described in \Sref{sec:astrometry} are converted into JSON format and saved in the MEDS files. We make no attempt to re-centre the near-IR image cutouts during photometric measurement, relying on the SCAMP astrometric solutions to be sufficiently accurate.

In a parallel operation to MEDS file construction, we form associations of neighbouring sources that may contaminate one another's measurements, and deblend them. For speed, this operation is performed on the coadded images with \texttt{Shredder}\footnote{\url{https://github.com/esheldon/shredder}}. The aim of the deblending at this stage is to form model representations of all the objects in a group simultaneously, using a set of Gaussian Mixture Models (GMMs). Each GMM representing a single galaxy consists of ten components, and the PSF is similarly represented by a five-component model. Models are initialized as a de Vaucouleurs profile, but are free to evolve during fitting --- only the centre is fixed.

Groups of objects are defined by forming friends-of-friends (FoF) groups of objects using the \SExtractor segmentation image constructed from the DECam detection image, where a ``friend'' link is defined as two objects having touching segmentation footprints. The vast majority ($\sim~95\%$) of FoF groups in unmasked areas are singletons, and large groups are very few in number. Overall, $\sim~10\%$ of unmasked galaxies are in groups of two or more members. This group forming stage, togther with GMM initialisation based on \SExtractor measurements, is performed with \texttt{Shredx}\footnote{\url{https://github.com/esheldon/shredx}}. All groups with two or more members are processed with the deblender and their resulting model fits are catalogued to be used during photometric measurement. Deblending objects with \texttt{Shredder} in this way has been found to work well even for object groups of up to 48 members (DES et al., in prep.).

\subsubsection{Model-fitting photometry}

For each object we return three sets of photometric measurements: PSF fluxes, bulge+disk model fluxes and Gaussian aperture fluxes, across all eight bands. Measurements are performed using the single-epoch images, with the exception of the near-IR images where the coadded images are used. When model fitting, the neighbours identified and deblended by \texttt{Shredder} are subtracted from the images using their catalogued GMMs.

PSF fluxes are a simple freely-varying amplitude fit of the individual-epoch PSF models (described in \Sref{sec:optpsf} and \Sref{sec:nirpsfs}) at the position of the source in question, which is then integrated over the extent of the PSF model. In the case of the optical bands the final measurement is therefore a signal-to-noise weighted mean flux, while for the near-IR we have just a single effective epoch constructed from the coadded data. Our main photometric measurement is a bulge+disk galaxy model fit, described below. Finally, the Gaussian aperture fluxes that we produce are an analytic estimate, rather than a direct measurement, that applies a Gaussian-weighted aperture to an object's pre-seeing bulge+disk model, previously fit. These fluxes are robust against the possible noise or biases caused by the de Vaucouleurs component in bulge+disk measurements if the background is not perfectly estimated. They therefore enable important checks and comparisons during the Balrog process \citep{y3-balrog}. The FWHM of the Gaussian aperture is fixed to $2.5^{\arcsec}$.

\subsubsection{Bulge+Disk fluxes}

\begin{table*}
\begin{center}
\caption{Prior ranges for the parameters used in our Bulge+Disk model fits.}
\label{tab:mof_priors}
\begin{tabular}{|l|l|l|} 
\hline \hline
Parameter   & Prior type & Prior range \\ 
\hline
$\delta$ RA & Gaussian & $\sigma=0.263\arcsec$ \\
$\delta$ Dec & Gaussian & $\sigma=0.263\arcsec$ \\
Shape parameters, {\bf g} & \citet{Bernstein2014} & $\sigma=0.2$ \\
Effective area, T & Flat & $-1, 1\times10^{5}$ \\
Fracdev & Truncated Gaussian & $\mu=0.5$, $\sigma=0.1$, min=0, max=1 \\
Flux & Flat & $-1000, 1\times10^{9}$ \\
\hline
\end{tabular}    
\end{center}
\end{table*}

Our main flux measurements are derived from a bulge+disk model fit to the $griz$ optical multiband, multi-epoch data, where the bulge and disk components are Sersic profiles with $n=4$ and $n=1$ respectively. The model is parameterized by the object centre (ra,dec), effective area (T), bulge-to-total ratio (``fracdev''), overall model flux for each band ($griz$) and the parameters governing the elliptical deformation matrix ({\bf g}, which encode the position angle and axis ratio). The prior ranges for these parameters are given in \Tref{tab:mof_priors}. All of the individual epochs across the four $griz$ optical bands are fit with the same model parameters for any given object. In the cases where an object is a member of a group, its neighbours are subtracted off before fitting, according to the model derived previously with \texttt{Shredder}. This model fitting step is performed by a new package, \texttt{fitvd}\footnote{\url{https://github.com/esheldon/fitvd}}. However, it should be noted that, similar to \mof, \texttt{fitvd} is built on top of the core functionality of \ngmix and thus they naturally produce highly consistent output measurements. One crucial difference between the two packages is the ability of \texttt{fitvd} to operate in a forced-photometry mode.

To enhance the stability of solutions and to reduce degeneracies in the parameter space, we restrict the freedom of the model space by fixing the relative effective radii of the bulge and disk components to be unity. We refer to the fluxes measured in this way as bulge + disk with fixed size ratio (BDF). While there are certainly galaxies for which this is a poor approximation, our main focus is on obtaining consistent and robust colours for the relatively faint galaxies that enter into the weak-lensing cosmology analysis, and where there is insufficient information to constrain a fully free bulge+disk model. Fitting was initialized based on the previously measured PSF fluxes and the second moments of the light distribution computed by \SExtractor. Overall, only $1.26\%$ of unmasked objects result in a failed fit, with a bias towards objects in FoF groups with many members.

With final model fits in the main optical bands obtained, the final step is to perform the forced-photometry measurements on the remaining bands. \texttt{fitvd} takes the morphological parameters of the model that were determined from the $griz$ bands, convolves the model with the appropriate PSF, renders it and fits the amplitude to the data. This forced-photometry mode was applied to the u-band and near-IR photometry. All members of a FoF group are fit at the same time. 

\subsubsection{Similarities with alternative software solutions}

The \texttt{ngmix} software is not the only package that performs object fitting simultaneously across many input images for a large sample of objects. The most similar alternative software available is \texttt{The Tractor} \citep{Lang2016}. Both packages represent Sersic profiles and PSFs with combinations of Gaussians to allow fast convolution and share the same parameterization of object shapes (though with differing notation). \texttt{The Tractor} requires a driver script to operate on samples of more than a few to several objects. This script should make image cutouts around the objects of interest, handle the position-dependent PSF and initialize the fitting process. In our pipeline, building the MEDS format files performs an equivalent role, and so it is possible to adapt those files for use in \texttt{The Tractor}. We ran a small test area of one DECam chip to compare the photometry extraction via model fitting the four main DES bands, $griz$, and found excellent agreement.

Similarly, performing forced photometry on an image using a previously-determined model for a galaxy or set of objects is not a novel concept. Perhaps the most commonly-used code for this purpose in recent years is \texttt{T-PHOT} \citet{Merlin2015, Merlin2016}, notably in the deep survey areas of the CANDELS HST programme \citep{Galametz2013, Guo2013}. The use case for \texttt{T-PHOT} is often ensuring robust object colours from images with very different PSF sizes, e.g. combining HST images with ground-based imaging, or optical / near-IR images from 4 to 10-m class telescopes with Spitzer data. In such cases the object footprints from the higher-resolution image can be used directly, with a suitable convolution kernel to take account of the PSFs. In our case we do not have a band that is consistently of higher resolution (tighter PSF) than our other bands and so model fitting is the appropriate solution. \texttt{T-PHOT} can take the parameters of a model, such as the restricted bulge+disk model we use, and convolve it with the PSF to produce the object light profile to be fit against the data. Multiple objects can be fit simultaneously to deblend objects, and the intended functionality is therefore identical to our \texttt{fitvd} pipeline. Our decision not to employ one or both of \texttt{The Tractor}\footnote{For completeness, we note that \texttt{The Tractor} can also operate in a forced-photometry mode.} and \texttt{T-PHOT} is based on ease of pipeline implementation and consistency of code base with the main survey pipline.

\subsubsection{Variant for input to weak lensing image simulations}
We also constructed a special version of the photometry catalog for use in the Y3 weak lensing image simulations described in \citet{y3-imagesims}.  These simulations require realistic and well-constrained morphology as input to test the impact of blending on the shapes and photometric redshift distributions for the galaxies used for Y3 cosmology analysis.  The input catalog focuses on the COSMOS region and only contains photometry measured from the DECam $griz$ bands.  It is not simply a subset of the main catalog described above because detections and fits to morphological parameters (half-light-radius, bulge-to-disc-ratio, ellipticity) are based on the HST imaging, with bulge and disk components approximated by de Vaucouleurs and exponential profiles, respectively. These model fits are used to estimate fluxes in the DES $griz$ filters from the DECam imaging described in \Sref{sec:optobs}. Fig. 3 of \citet{y3-imagesims} validates this catalog by showing that the simulated distributions of quantities like magnitude, color, S/N, and size match the observed quantities of the WS galaxies.

\subsection{Galactic reddening correction}
Interstellar extinction corrections to the $ugriz$ photometry were applied following \citet{2011ApJ...737..103S} and described fully for the DES main survey GOLD sample in \citet{y3-gold}. The procedure applied in the Deep Fields is the same as used for the fiducial correction in the DES main survey and we refer to that paper for details and associated uncertainties. Briefly, we compute the total extinction in a specific band $A_b$ for each object using the associated $E(B - V)_{\rm SFD}$ reddening map value \citep[SFD;][]{1998ApJ...500..525S}. To compute the $A_b / E(B - V)_{SFD}$ coefficients in each band, we assumed a \citet{1999PASP..111...63F} extinction law with $R_V = 3.1$ and used the total transmission curves of each band for the calculation. For $J, H$ and $Ks$ we use the coefficients from \citet{gonzalez2018}. Table \ref{tab:photom_tune} provides the $A_b$ coefficients for each band $b$.

\subsection{Photometric calibration}
\label{sec:calibration}

The DES redshift methodology hinges on the consistency of the Deep-Field photometry across the different fields. This is especially true between the SN-VIDEO fields and the COSMOS-UltraVISTA field, as many of the key spectroscopic surveys are located in COSMOS but it is not covered by the main DES survey. The near-infrared data are calibrated against the 2MASS survey \citep{skrutskie2006} in all fields by the CASU reduction pipeline, though with relatively few stars per field available. Meanwhile, we could choose to tie the optical DECam data to the PanSTARRS survey \citep{chambers2016}, but due to differences in filter transmission curves it is not clear that doing so would be an improvement over the tertiary standards that we used when building our images. We nevertheless compare our PSF photometry with that from HSC \citep{aihara2018}, which is tied to PanSTARRS, in Appendix \ref{sec:append_data}.

Our approach for ensuring consistency between the four Deep Fields instead relies on matching stellar and galaxy loci between the fields, before tying to the DES WS using stellar sources. The remainder of this subsection details the following major steps:
\begin{itemize}
\item Compute expected differences in the colours of the stellar loci in our fields due to their sky position, using the Besan\c con Galaxy Model (BGM; \citealt{robin1986, czekaj2014}).
\item Compute median offsets in individual band and field combinations via the positions of stellar and red galaxy loci in a wide variety of colour-colour diagrams.
\item Perform an absolute calibration of the now relatively-calibrated fields to the DES WS using stellar PSF photometry.
\item Estimate residual calibration uncertainties through comparing galaxy photometry from the Deep Fields and WS, and the residual scatter from the stellar and galaxy loci diagrams. 
\end{itemize}
Final calibration adjustments and uncertainties are given in \Tref{tab:photom_tune}. 

\subsubsection{Stellar locus matching}
\label{sec:besancon}

Stars form a tight sequence in many colour-colour plots and in the absence of variations in population colours due to differing ages or metallicities, stellar sequences in different fields should lie on top of one another. Of particular use are combinations of colours where the stellar locus lies orthogonal to one of the axes, and extended along the other. In such cases we are able to perform a regression along the locus to check for offsets in colour between fields, and any colour dependence in such offsets as are found.

The observed population of Galactic stars in any given field depends upon the line of sight through the Milky Way, and in particular the relative contributions of bulge, disk and halo stars. Differences in metallicity (amongst other properties) at fixed stellar type result in small differences in the observed colours. If we assume that the stellar loci in our four fields are identical, and derive fine-tuned offsets to enforce this assumption, then we may actually introduce a systematic difference in the galaxy colours between one field and another.

The BGM is a four-component model of our galaxy, allowing an observer to compute the expected surface density, magnitudes and colours of stars along any line of sight. We extract simulated stellar populations for each of our fields, using the web interface\footnote{\url{https://model.obs-besancon.fr/}}, and compute offsets in the position of the stellar locus in colour-colour plots of various combinations of our $ugrizJHKs$ filters. We find no significant predicted offsets between the four fields ($<0.2\%$), indicating that the stellar population (including metallicity) is similar across our fields. We now proceed to match the stellar loci in our observed data. 

We perform a non-parametric regression by first estimating the density of stars in colour space through applying a kernel density estimate (KDE) to the population of stars in each field, using two colours at a time. We then identify the ridge-line of the sequence by searching for the maximum density and its associated value in one colour, conditioned on a value of the second colour in finely spaced intervals. That is, we find
${\rm max}({\rm Dens}(c_1,c_2) ~|~ c_{2,i})~{\rm for }~i \in \{1...k\}$, where $c_{2,i}$ are the finely sampled values of the second colour axis, numbering $k$ samples. We then compute residual differences for each field, relative to the mean of the four fields. Finally, a colour offset is determined though a median of the residuals over the useful range in $c_2$, determined by inspection of the fields' KDEs. The residuals are noisy due to intrinsic variation in the relative numbers of stars of different stellar types in each field, as well as shot noise and small photometric errors. 

\begin{figure*}
    \centering
    \includegraphics[width=\linewidth]{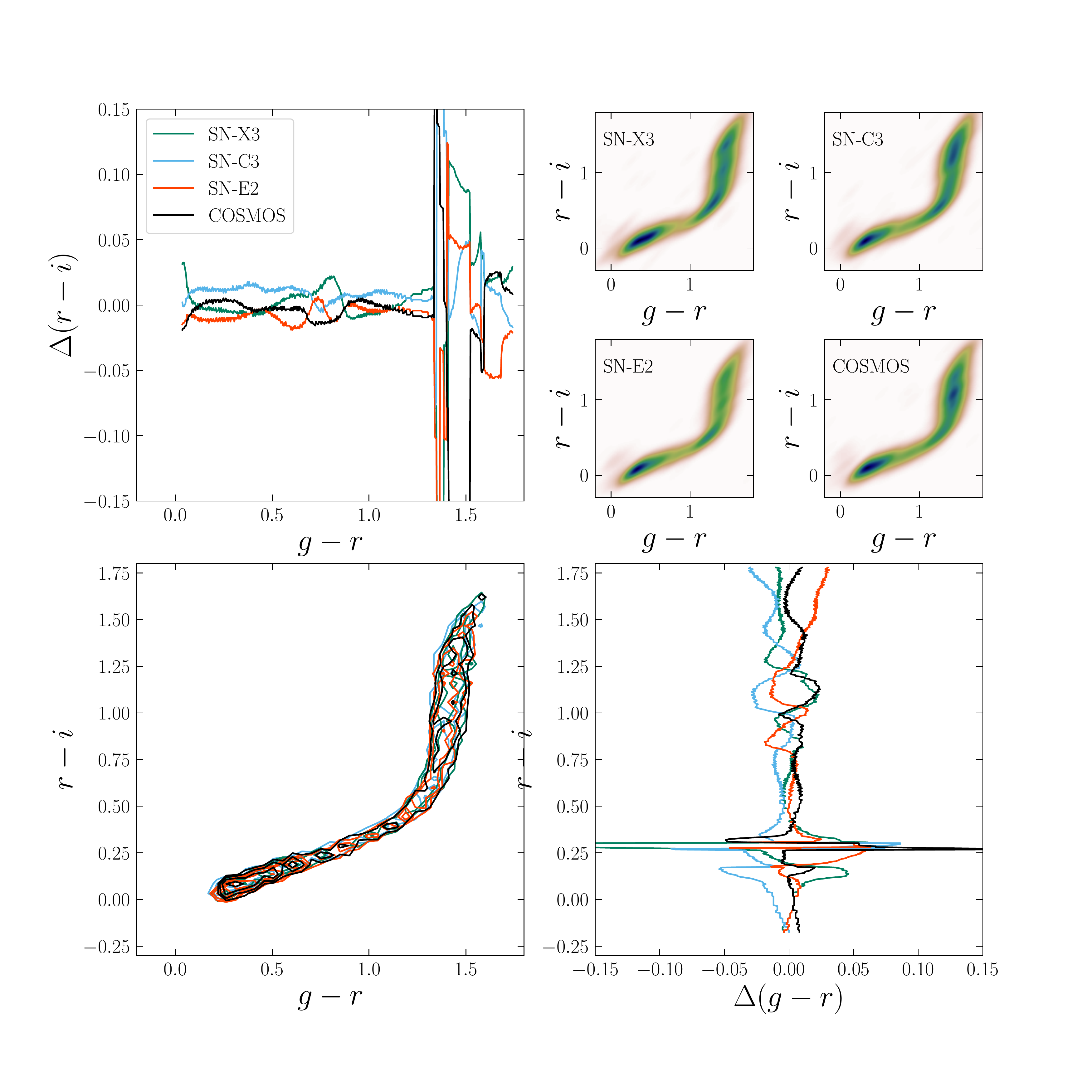}
    \caption{Example colour-colour residual diagnostic plot for stars in the $g-r$, $r-i$ colours space. In each set of four major panels we show two-dimensional histograms in colour-colour space by way of number density contours (lower left) and a kernel density estimate (upper right), for each of the four fields. The remaining two panels show offsets of the location of maximum density for one of the colours, as a function of the other colour, versus the mean of the four fields.}
    \label{fig:col_residuals}
\end{figure*}

An example of one such colour-colour plot is shown in the upper half of \Fref{fig:col_residuals}. The four smaller panels in the upper right show the individual fields' KDE of their stellar population in the $g-r$ and $r-i$ colours. The lower left panel shows that the populations do indeed lie on top of each other, as expected, while the remaining two panels show the residuals after subtracting the mean of the fields for the two colours. 

\begin{figure*}
    \centering
    \includegraphics[width=\linewidth]{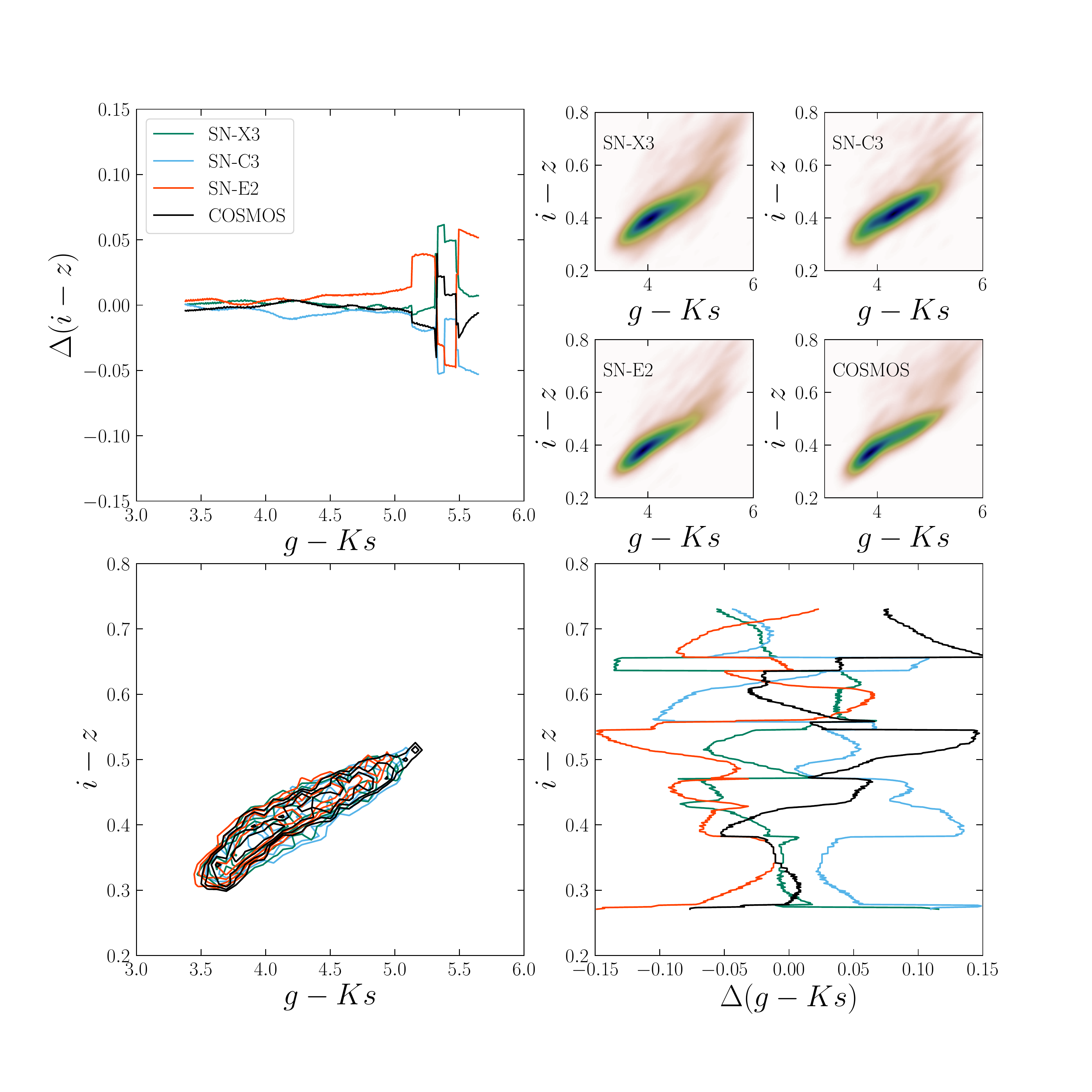}
    \caption{Same as \Fref{fig:col_residuals}, but for red sequence galaxies in the $g-Ks$, $i-z$ space}.
    \label{fig:gal_col_residuals}
\end{figure*}

\subsubsection{Red sequence galaxy locus matching}
\label{sec:rs_locus}
Red sequence galaxies have very similar spectral energy distributions (SEDs) at fixed redshift, and a strong $4000${\AA} break that allows them to be separated from the wider population, even in observed frame colours. For this reason, a red sequence galaxy sample drawn from a range in redshifts forms a tight sequence in some colour-colour combinations, and can be used in a similar way to the stars in a photometric regression. Over the area of each field, one or more square degrees, we expect general agreement in the location of the sequence in dereddened colour-colour space, with any sample variance entering as a source of noise.

\begin{figure}
    \centering
    \includegraphics[width=\linewidth]{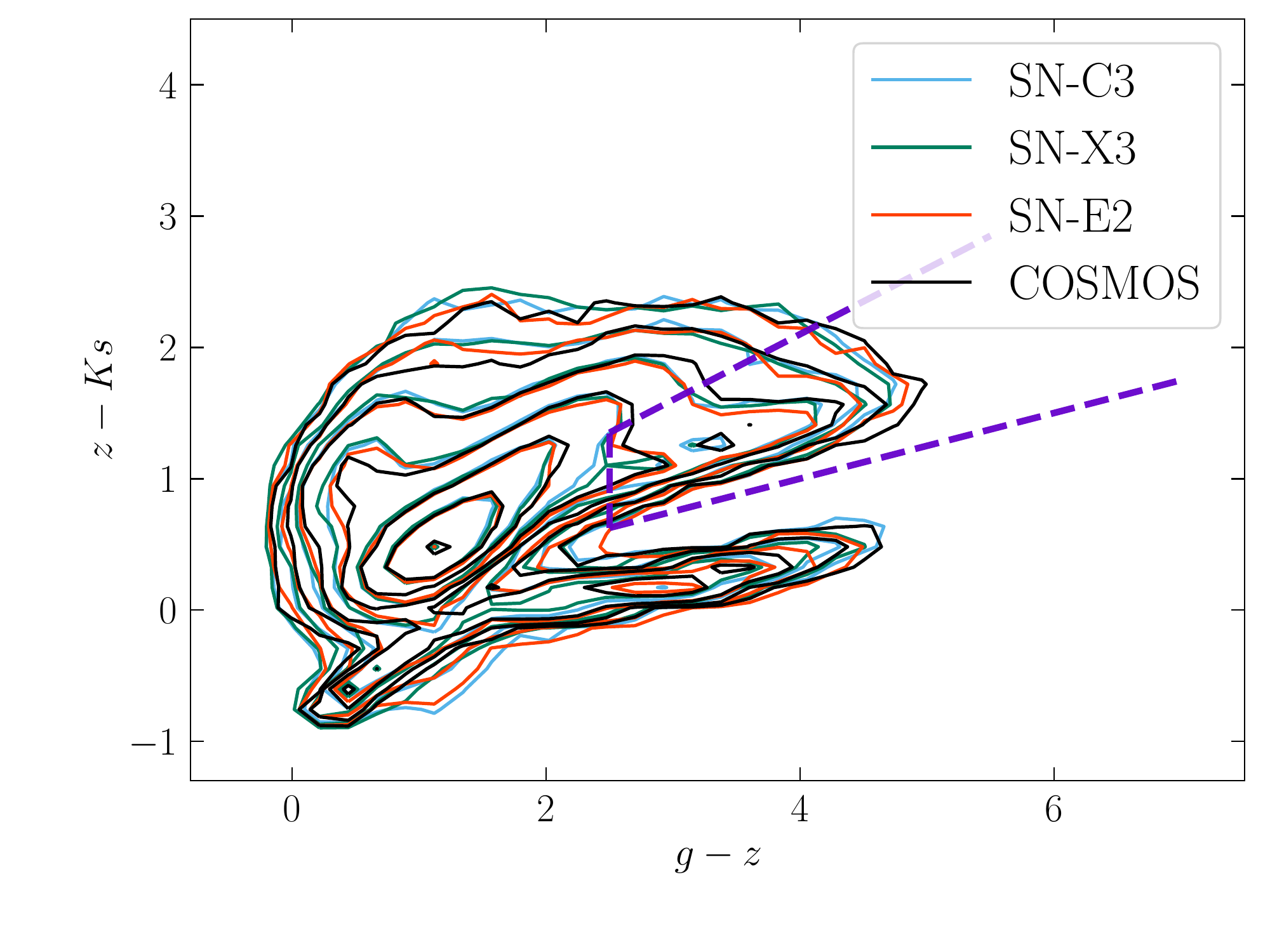}
    \caption{Selection diagram for the red-sequence galaxy population used in fine-tuning the inter-field calibration. Similar to a $BzK$ diagram \citep{daddi2004}, the $g-z$ and $z-K$ colours cleanly separate the stellar locus from the galaxy population and allow the isolation of the red sequence at intermediate redshifts \citep{lane2007}, shown by the purple dashed selection region. Galaxies in this sequence should have the same colours in different fields, and are used as part of our regression analysis.}
    \label{fig:gal_selection}
\end{figure}

We use the $g-z$ vs $z-Ks$ colour space to select red sequence galaxies in analogy with the $BzK$ diagram \citep{daddi2004,lane2007}. The selection is shown in \Fref{fig:gal_selection}. In \Fref{fig:gal_col_residuals} we show the $i-z$ vs $g-Ks$ colour space and residuals for our red sequence galaxy selection. Although far noisier than the equivalent figure for the stellar locus, it is clear that the $i-z$ colour is already very consistent across our four fields. Note that although there is a clear sample variance between the fields (particularly visible in the density plot for the SN-C3 field, top right of \Fref{fig:col_residuals}), it does not translate to a discrepancy in the colour of the locus.

We compute residual offsets with respect to the mean of the four fields for many combinations of colours for both the stellar locus and red galaxy sequence. Where we find a consistent need for a change in the zero-point across a few or more such diagrams, we compute it as the mean of the indicated offsets. Operationally, we begin by looking for consistent shifts in the $g, r, i$ and $z$ bands. Following that, we calibrate the near-IR bands, and finally the $u$-band. These small adjustments (typically $\le 1\%$) are stored in table of band-field combinations and applied when computing differences with respect to the WS, which we come on to now.

\subsubsection{Calibration to DES Y3 GOLD}
\label{sec:gold_comp}

The uniformity of the photometric calibration across the DES footprint is at a level $< 0.3\%$ \citep{y3-gold}, and roughly $1\%$ in absolute calibration (DES collaboration, in prep.).
The SN fields overlap the main survey and so for bright sources the main survey data provide an excellent anchor. Moreover, a key target use of the Deep Fields is to provide source injections to the main survey, and so consistency between the Y3 WS GOLD sample of the main survey and the Deep-Fields catalogue is a firm requirement. We have already performed a photometric matching of our fields to one another in terms of their colours. The remaining freedom in calibration is therefore in the form of either coherent shifts across all four fields in a single band, or shifts in all the bands of a single field. 

To perform the anchoring to the WS we position match the Deep-fields catalogue with stellar sources from the GOLD catalogue \citep{y3-gold} with a tolerance of $0.5\arcsec$ and then cut the sample to $18<i<20$ on the Deep-Fields PSF photometry. This cut is to reduce the impact of photometric errors and to ensure that any residual defects in background subtraction have minimal influence. Using PSF magnitudes for both the Deep Fields and WS, we then compute median offsets for each SN field against their WS counterparts and use the mean of these field-based offsets to perform the two calibration freedoms described above. Final photometric adjustments are reported in \Tref{tab:photom_tune}.

\begin{figure*}
    \centering
    \includegraphics[width=0.24\textwidth]{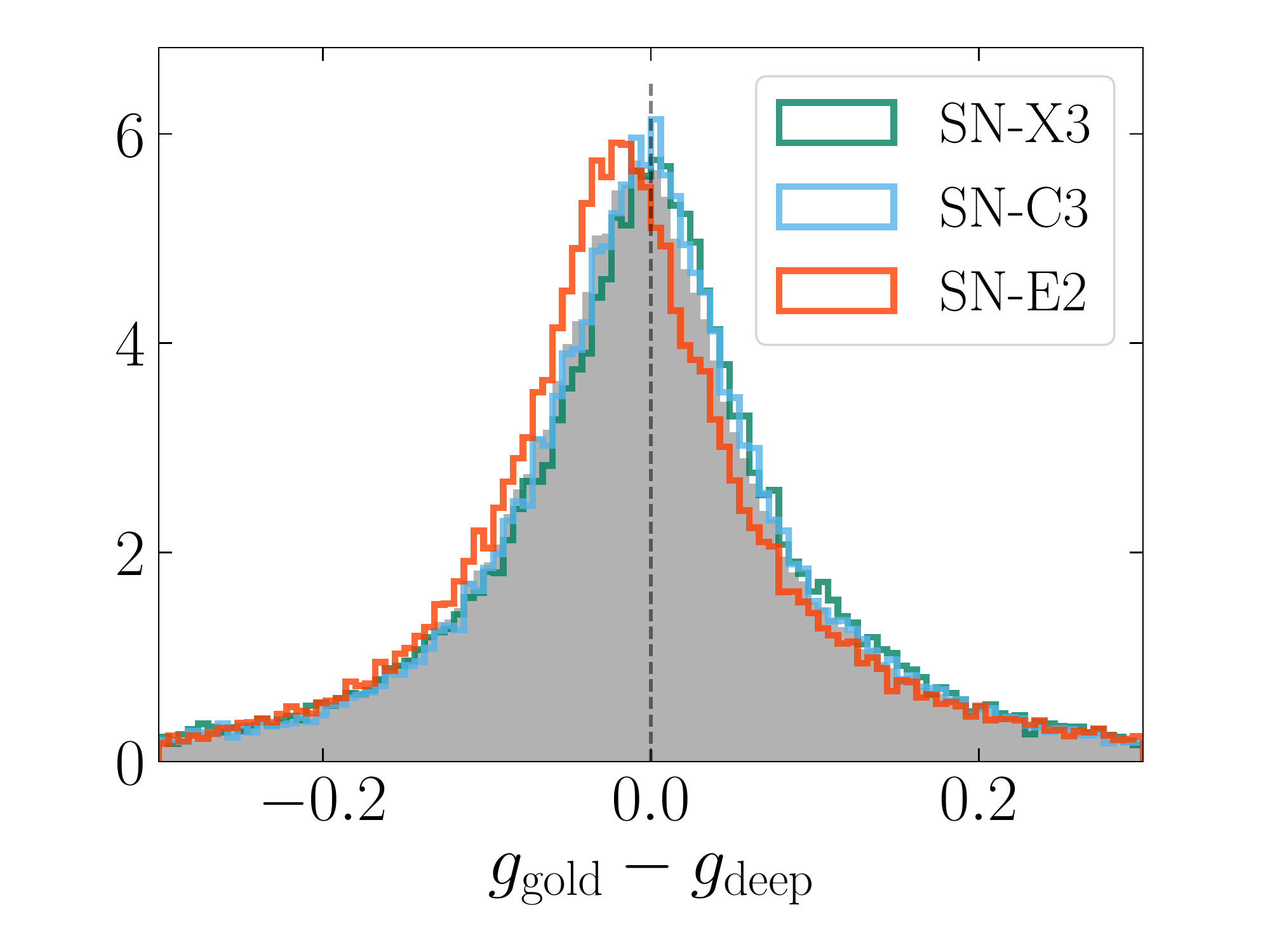}
    \includegraphics[width=0.24\textwidth]{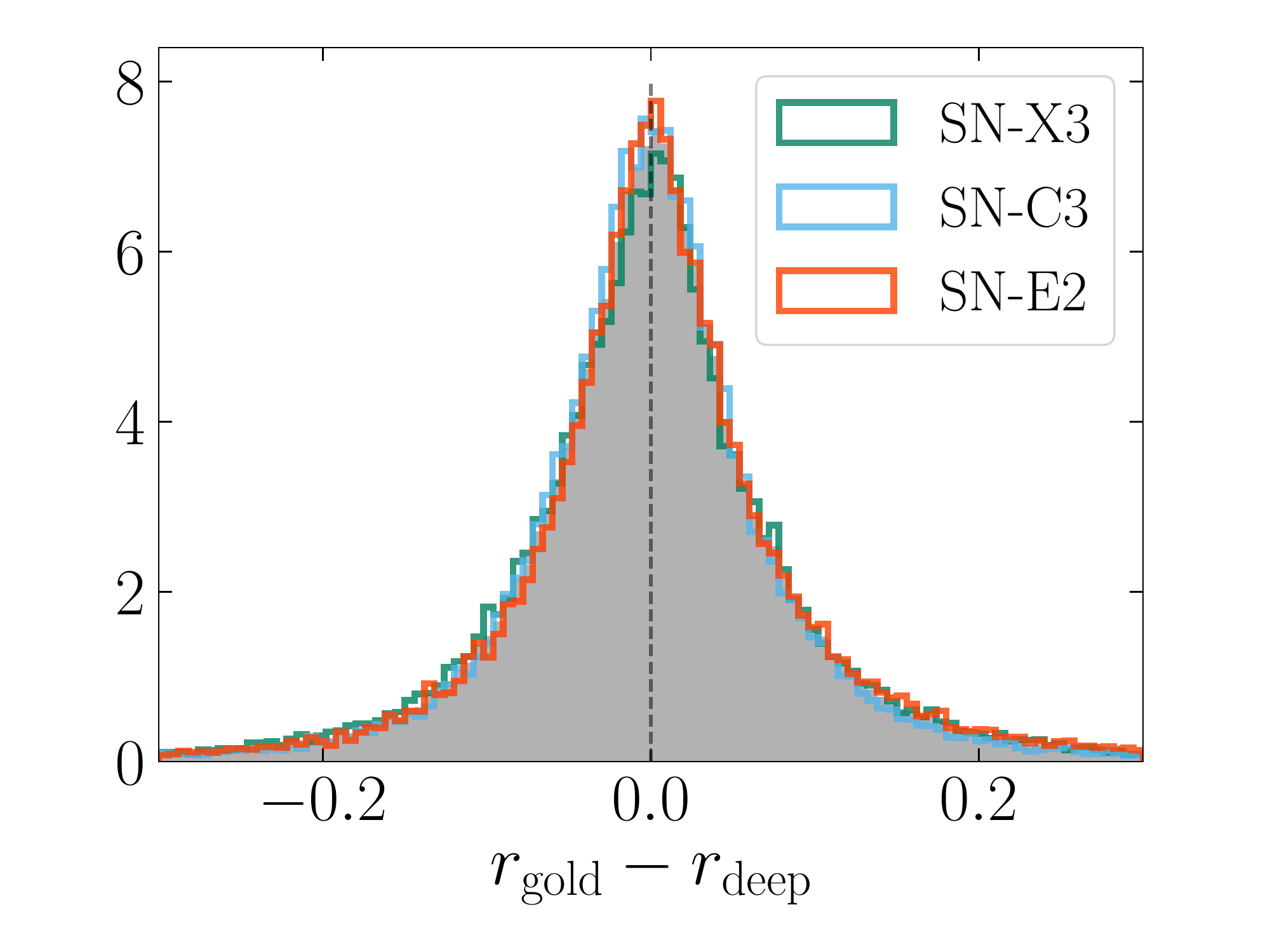}
    \includegraphics[width=0.24\textwidth]{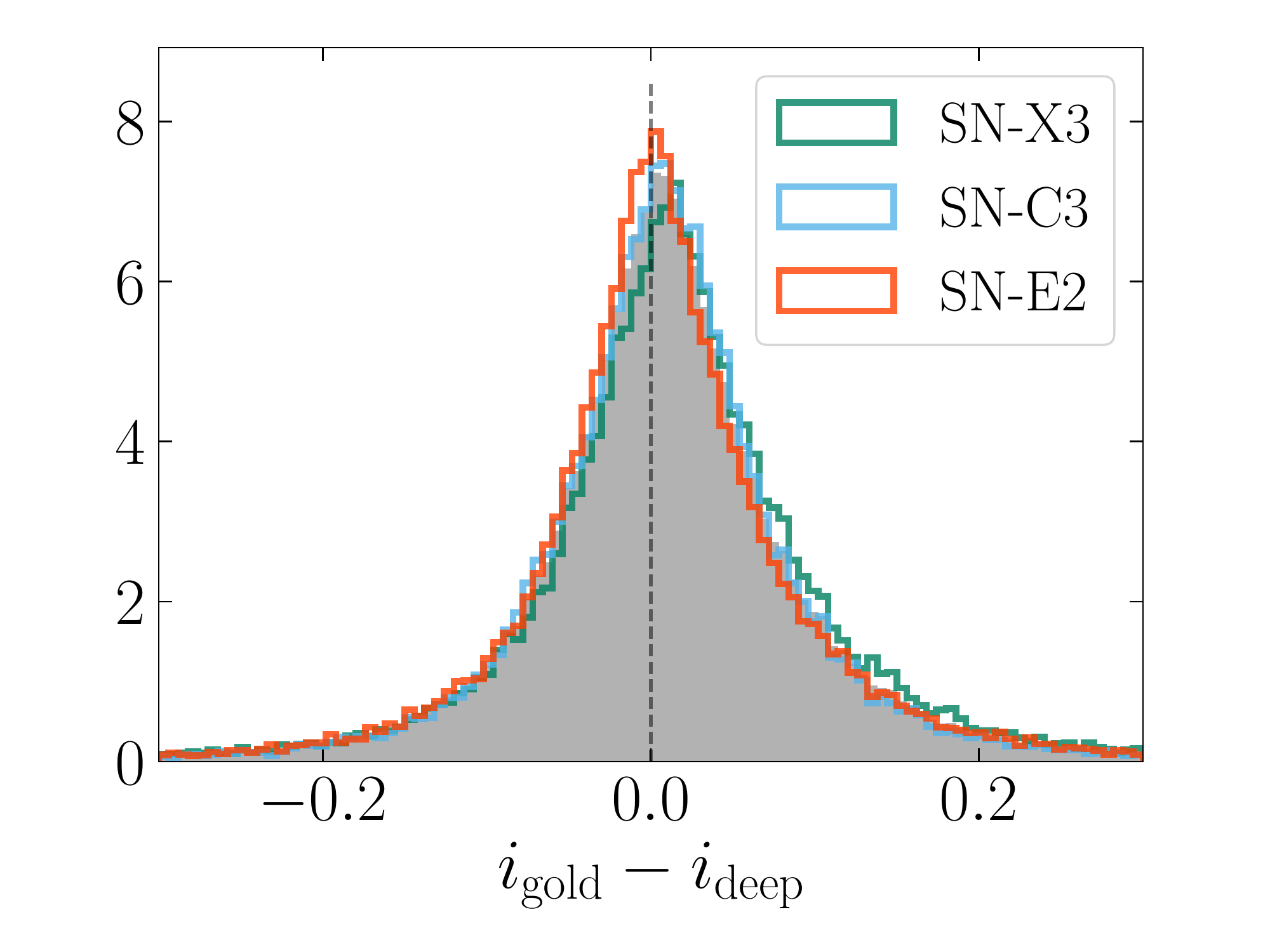}
    \includegraphics[width=0.24\textwidth]{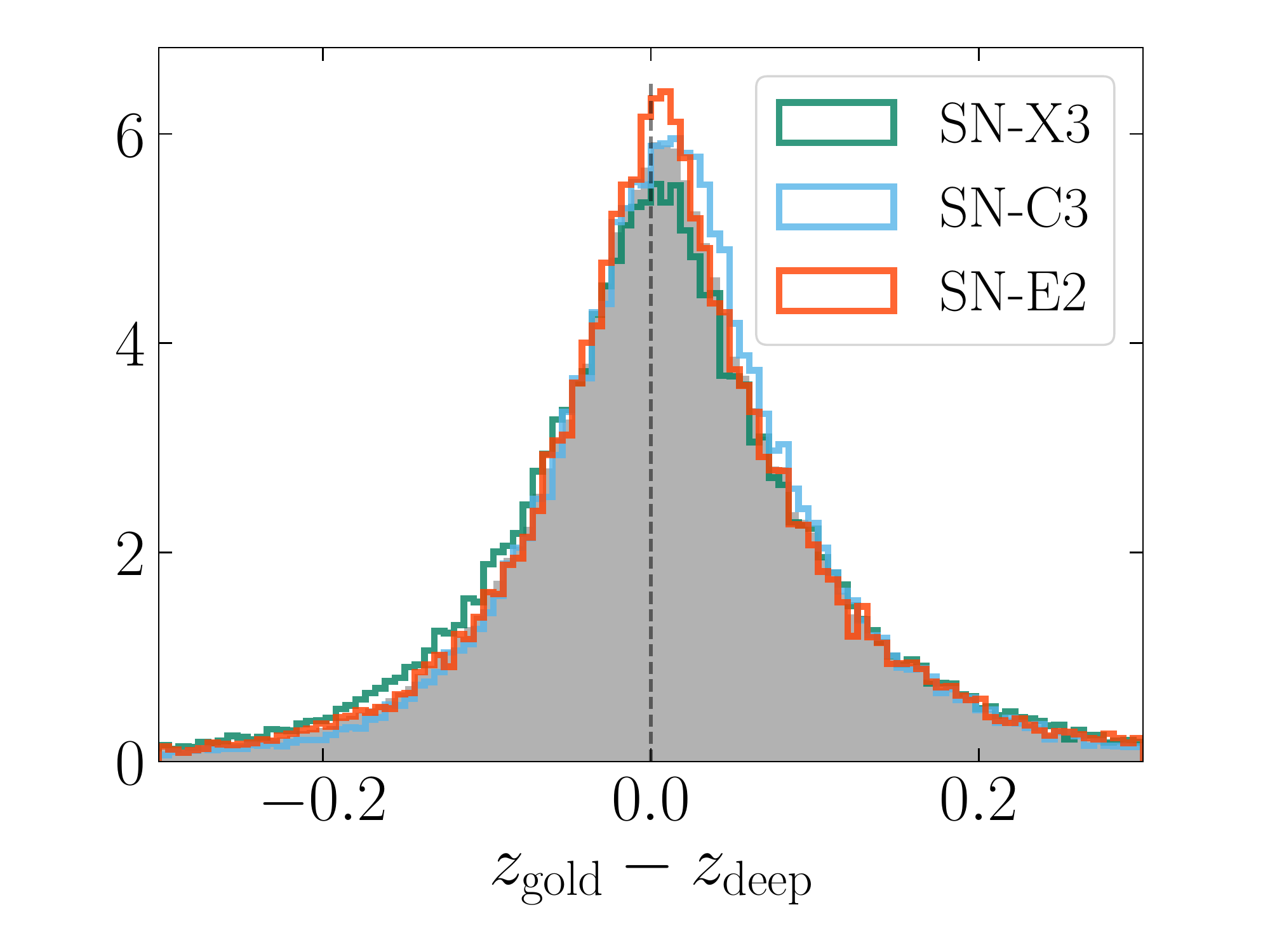}
    \includegraphics[width=0.24\textwidth]{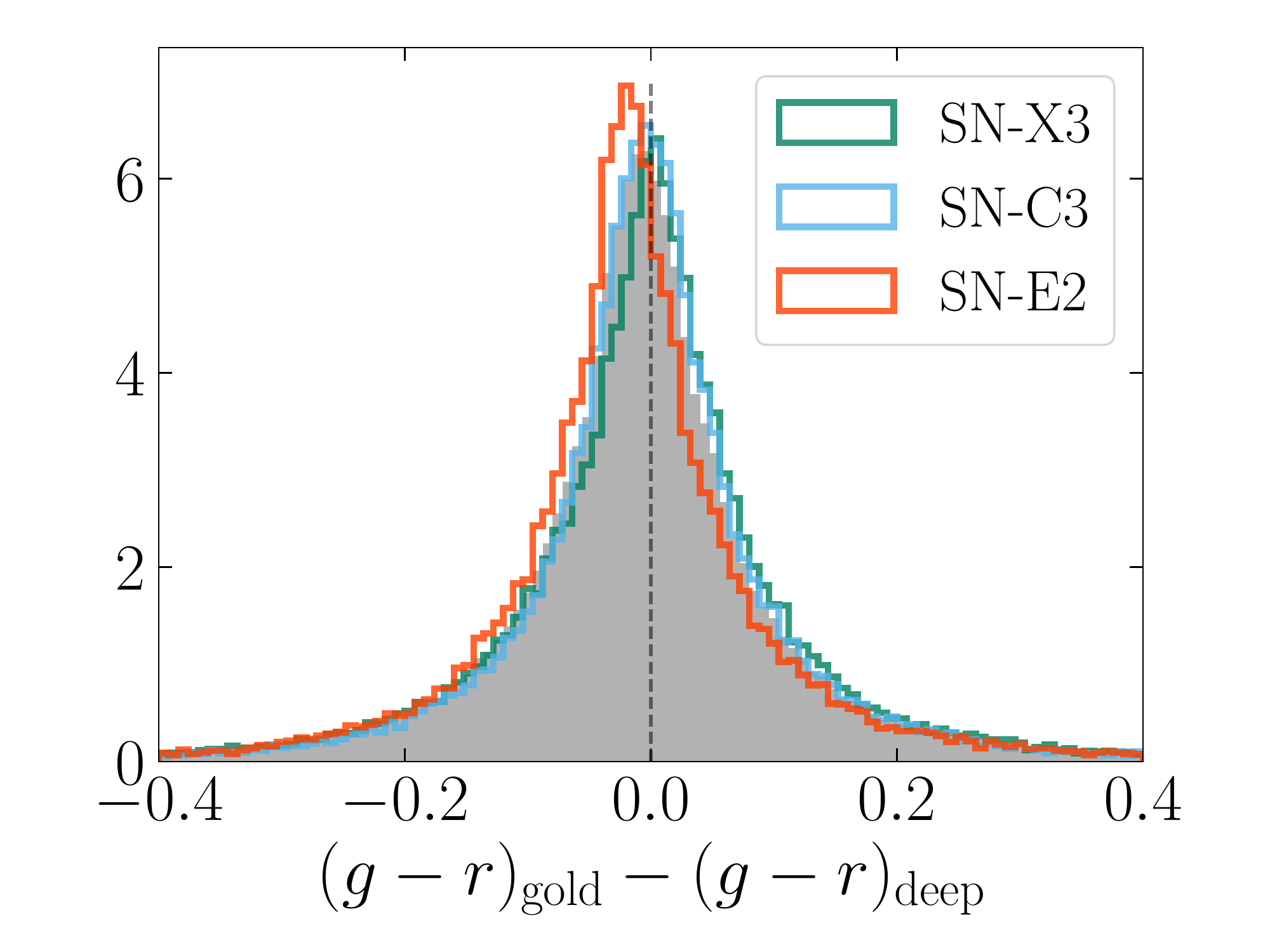}
    \includegraphics[width=0.24\textwidth]{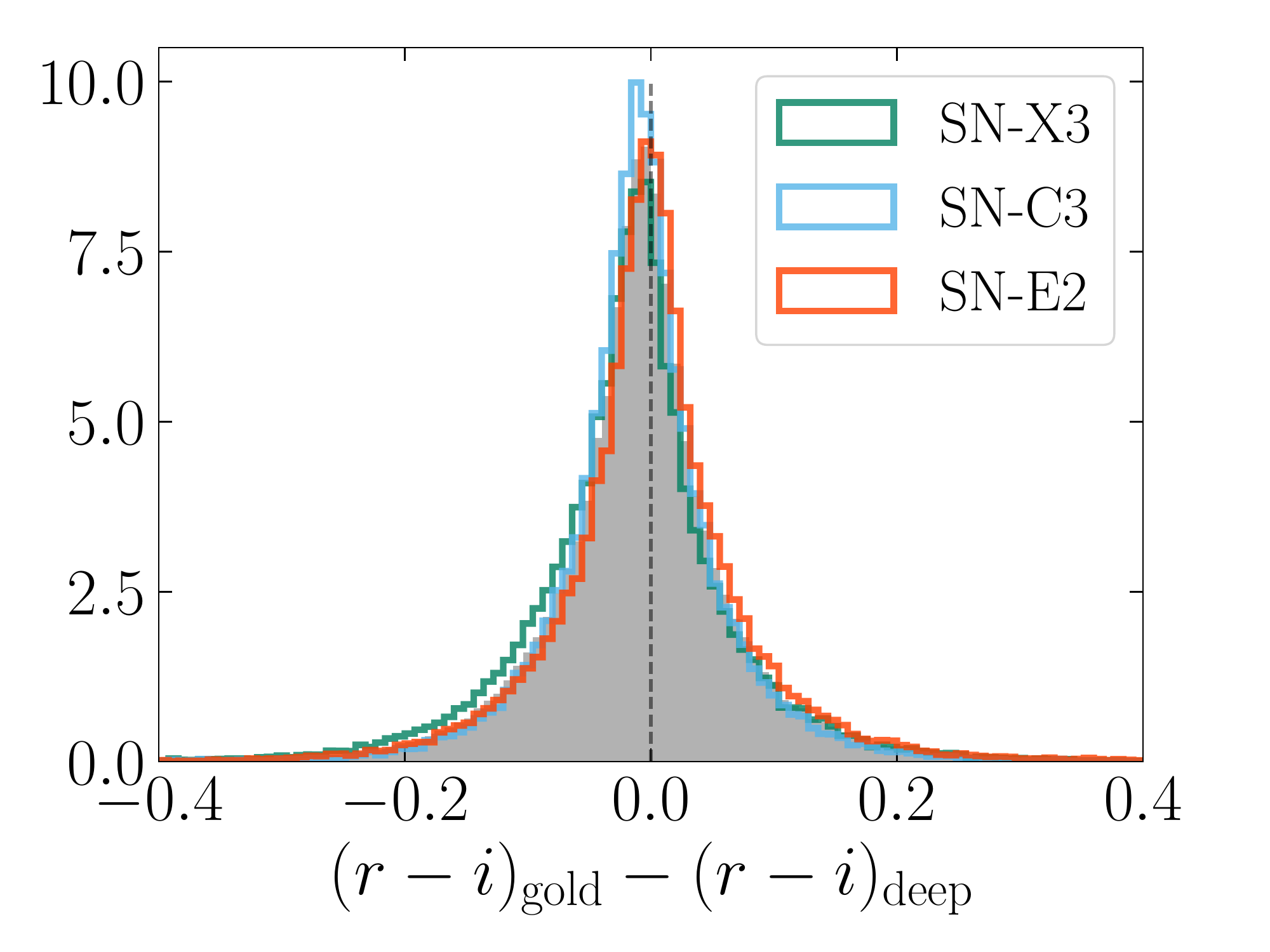}
    \includegraphics[width=0.24\textwidth]{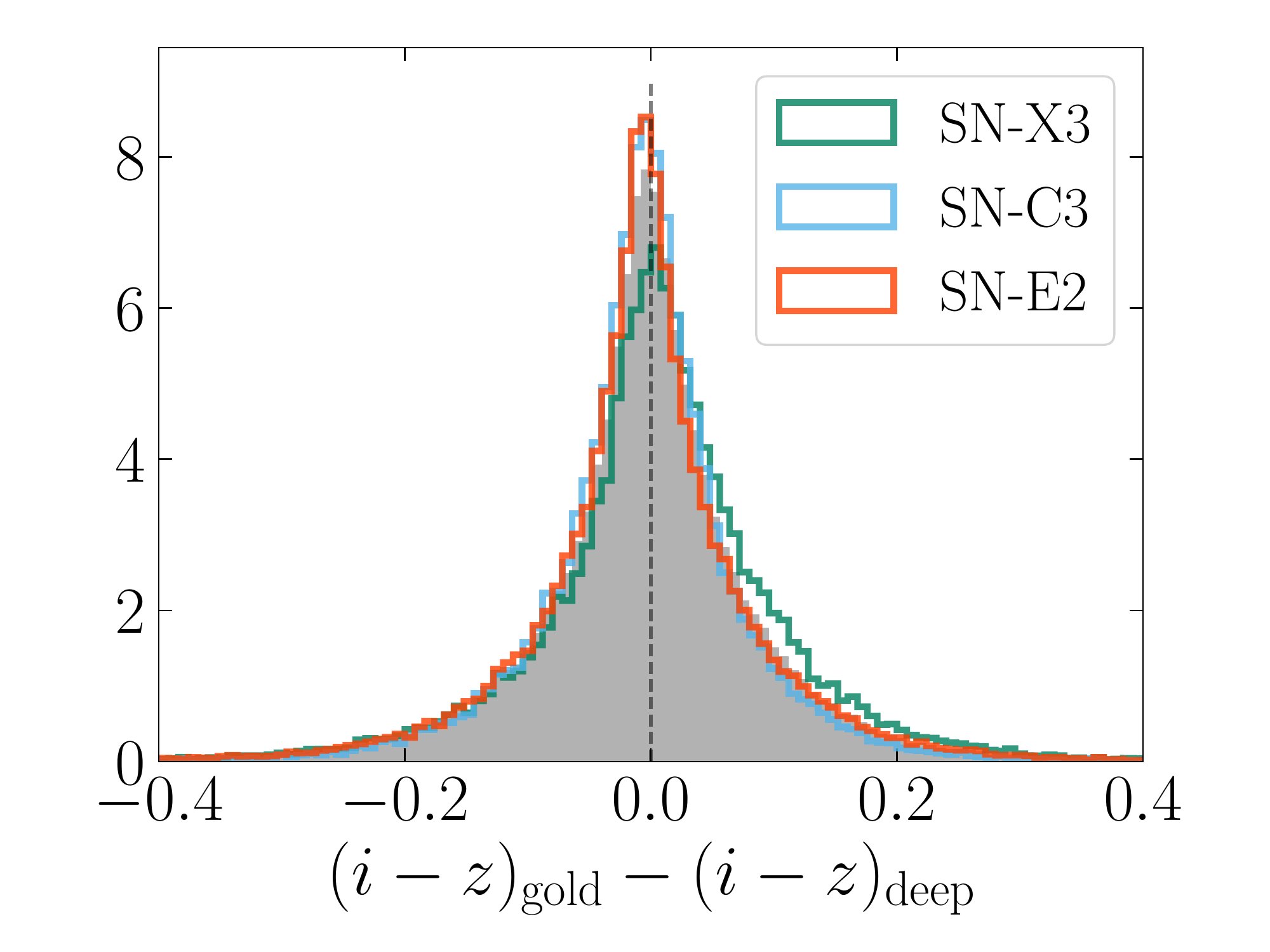}
    \caption{Comparison of the final Deep-Fields photometry with matched galaxies from the DES main survey Y3 GOLD catalogue, at $i<22.5$. Upper panels show single band magnitude differences, while the lower row of plots shows colours. Individual fields are as labelled, and the combination of all three SN fields is shown by the grey histogram. As the COSMOS field lies outside of the DES footprint it cannot be included in this figure. Histograms are normalised to emphasise differences in the median values.}
    \label{fig:gold_comparison}
\end{figure*}

\subsubsection{Final zero-point offsets and uncertainties}
\label{sec:photo_errors}

\begin{table*}
\begin{center}
\caption{\label{tab:photom_tune} Galactic reddening coefficients and magnitude zeropoint (ZP) adjustments following field-to-field calibration and anchoring to the GOLD main survey sample. Adjustments are in the sense, ${\rm mag}_{{\rm calib}} = {\rm mag}_{{\rm meas}} + {\rm adjustment}$. Flux correction factors used during photo-z estimation are also provided.}
\begin{threeparttable}
\begin{tabular}{|c|c|c|c|c|c|c|c|}  \hline \hline
 band & A$_{b}$ & SN-X3 & SN-C3 & SN-E2 & COSMOS & ZP uncertainty$^{a}$ & Photo-z factor$^{b}$ \\ \hline
 u & 3.963 & 0 & 0 & 0.034 & -0.03 & 0.055 & 1.105 \\
 g & 3.186 & 0.011 & 0.011 & 0.025 & 0.011 & 0.005 & 1.015 \\
 r & 2.140 & 0.006 & -0.004 & 0.02 & 0.006 & 0.005 & 0.977 \\
 i & 1.569 & 0 & 0.005 & 0.014 & 0 & 0.005 & 0.987 \\
 z & 1.196 & 0.004 & 0.004 & 0.019 & 0.004 & 0.005 & 0.993 \\
 J & 0.705 & 0 & 0.015 & 0.006 & -0.014 & 0.008 & 1.013 \\
 H & 0.441 & 0 & 0.015 & -0.002 & -0.004 & 0.008 & 1.025 \\
 Ks & 0.308 & 0 & 0.015 & -0.002 & -0.004 & 0.008 & 0.984 \\
\hline
\end{tabular}
\begin{tablenotes}
    \item[$^{a}$] Relative zero-point uncertainties.
    \item[$^{b}$] Multiplicative factors applied to catalogue fluxes during photometric redshift computation, derived from the fitting template spectra to PRIMUS galaxies (see \Sref{sec:photoz_zp_calib}).
\end{tablenotes}
\end{threeparttable}
\end{center}
\end{table*}

Using, for example, the mean of the $i$-band offsets computed for stellar sources in each field does not guarantee that they will all be perfectly aligned in terms of photometry to the GOLD sample. Any remaining differences can be used to quantify the accuracy of our calibration process, and to do so we repeat the above procedure but this time with galaxies. We allow a positional match within $1\arcsec$ to account for the greater centroid uncertainty in fainter, extended sources, and cut the resulting matched catalogue at $i<22.5$ in model magnitude. Histograms of bulge+disk model magnitude and colour differences for matched galaxies are shown in \Fref{fig:gold_comparison}. Coloured steps represent individual fields, while the filled grey histogram shows the combination of the three SN fields. Source numbers are normalised to ease comparison of the fields. 

The standard deviation of the median offsets between the Deep Fields and WS visible in \Fref{fig:gold_comparison} is the main contributor to our estimate of final zeropoint uncertainty for the $griz$ bands. In addition, we return to the stellar and galaxy loci diagrams shown in \Sref{sec:besancon} and \Sref{sec:rs_locus}, together with many different permutations of band combinations not shown in these figures. We used multiple diagrams to determine zeropoint adjustments for each band, taking the mean of all usable such diagrams. The standard deviation from these measurements is therefore a second way in which we estimate the calibration uncertainty. For $uJHKs$ bands, these diagrams are our only source of information to estimate the residual zeropoint uncertainty.

Individual band estimates of uncertainties are noisy, and so we group the bands into three sets for which the pipeline processing was homogeneous and combine their values: the main survey bands, VISTA bands and the $u$-band. Our estimated uncertainty in $griz$ is $0.5\%$, and if corrected for residual error in the wide field calibration, shows that the \texttt{fitvd} deep field version of the model fitting photometry works extremely well. For the VISTA bands we find a slightly larger error at $0.8\%$, largely due to the fact that we account for the global optical - NIR error in the VISTA bands. The $u$-band was found to be rather more difficult to calibrate consistently across fields because the associated colour-colour plots are very noisy, resulting in a $5.5\%$ uncertainty that reflects our inability to estimate the error as much as the calibration itself. The final uncertainties in our photometric calibration are reported in \Tref{tab:photom_tune}, and as an additional check we compare our photometry with HSC Deep-UltraDeep photometry in Appendix~\ref{sec:append_data}.

\subsection{Raw number counts}
\label{sec:counts}

\begin{figure}
    \centering
    \includegraphics[width=\linewidth]{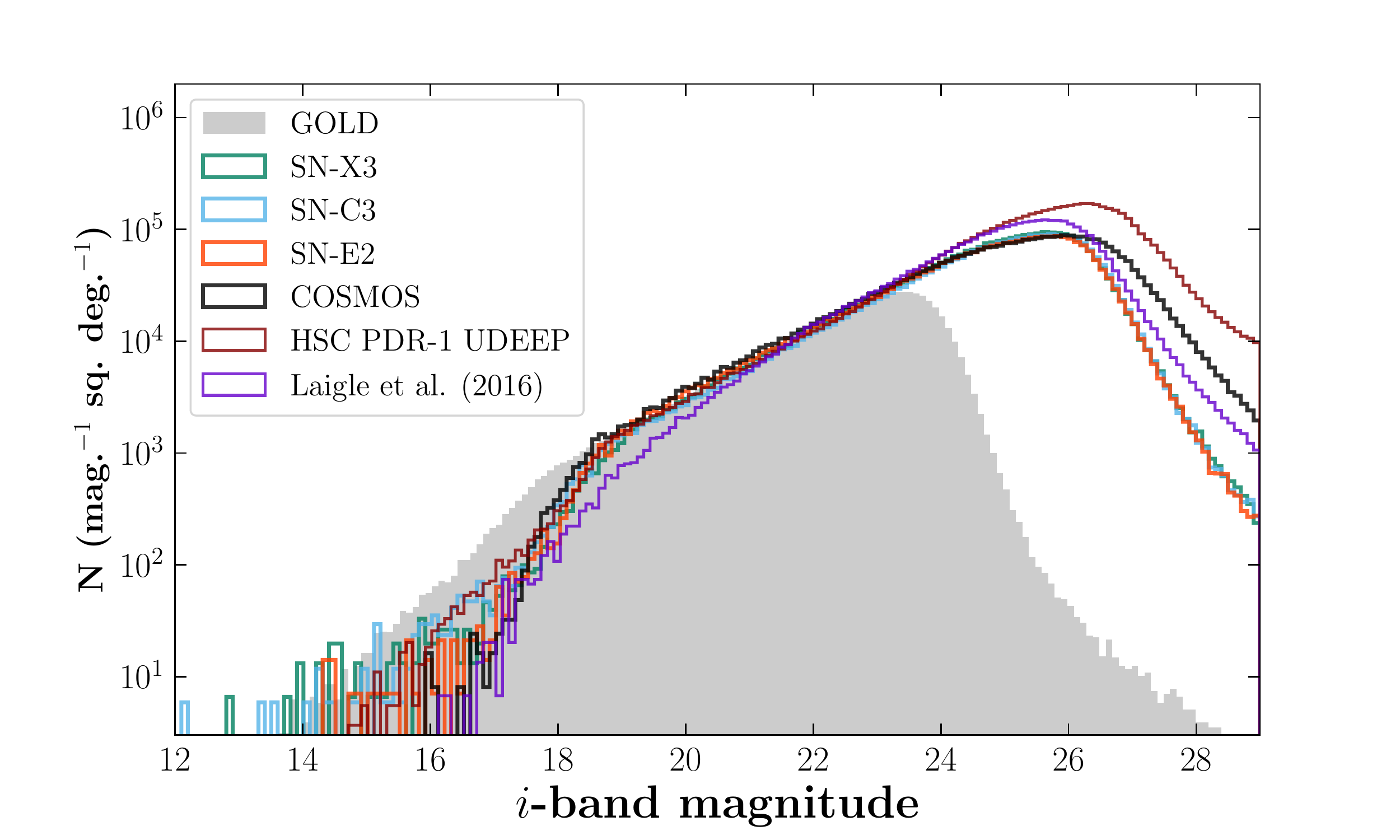}
    \includegraphics[width=\linewidth]{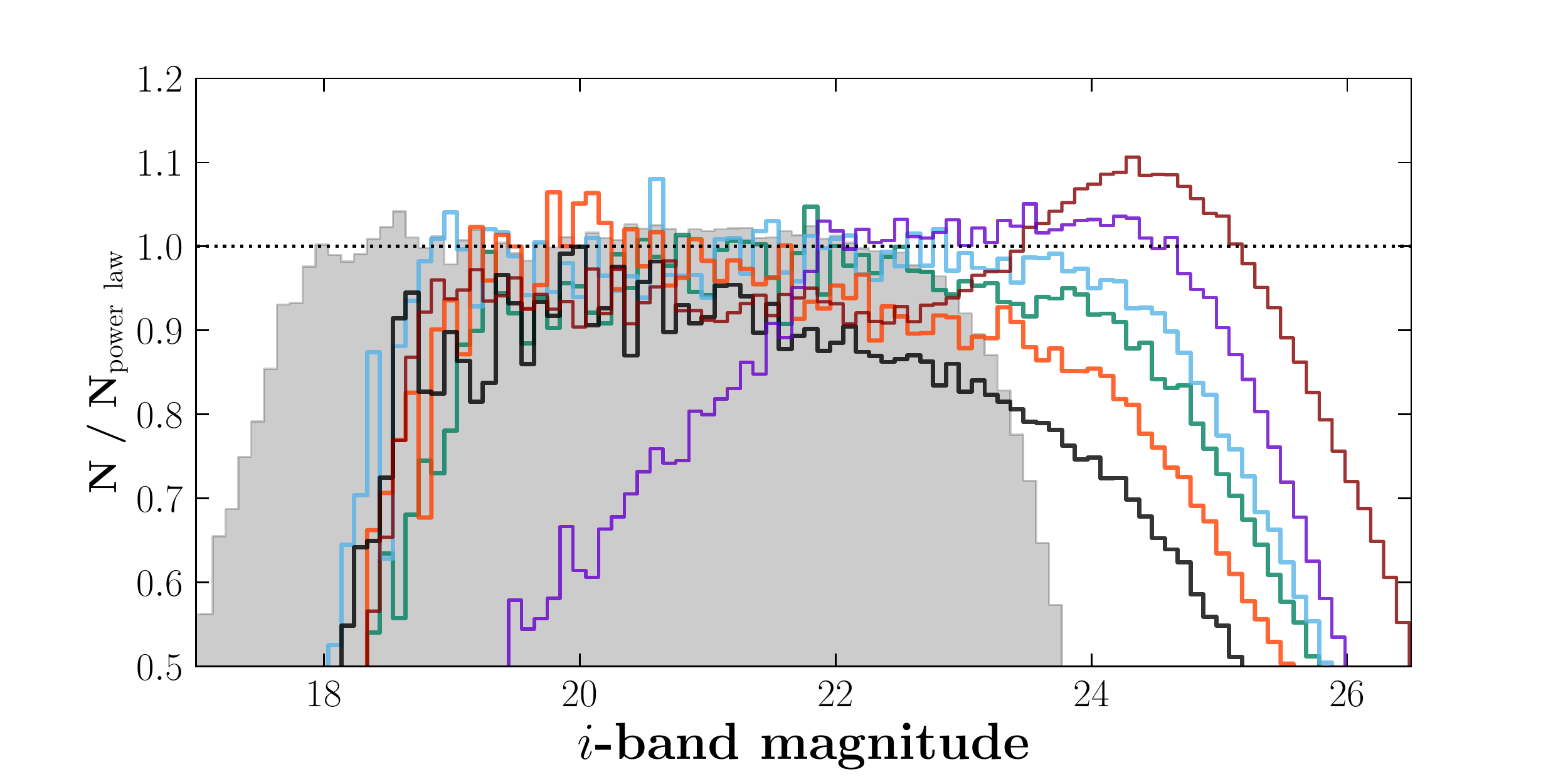}
    \caption{{\em Upper:} Raw number counts (uncorrected for incompleteness) in the $i$-band for our four Deep Fields (\texttt{COADD\_TRUTH} depth level), together with the main DES survey, HSC PDR-1 UltraDeep (COSMOS + SXDS) and the \citet{laigle2016} UltraVISTA / COSMOS catalogue. The fall off in number counts at the bright end with respect to the GOLD sample is largely due to masking of bright stars and saturation (for the \citealt{laigle2016} catalogue). {\em Lower:} Ratio of number counts to a fiducial power law distribution.}
    \label{fig:number_counts}
\end{figure}

\Fref{fig:number_counts} shows raw source number counts, uncorrected for incompleteness, as a function of $i$-band magnitude for our four Deep Fields (\texttt{COADD\_TRUTH} depth level, colour stepped histograms) and the main WS GOLD sample (filled grey histogram). The peak of the Deep Fields' number counts is $\sim1.25~$magnitudes fainter than for the GOLD samples, as expected for images with ten times the exposure time. The four Deep Fields are highly consistent with one another at bright magnitudes, but begin to diverge somewhat at fainter magnitudes.  In particular, the SN-E2 and COSMOS fields fall away from the reference power-law behaviour (lower panel of \Fref{fig:number_counts}) at brighter magnitudes than the SN-C3 and SN-X3 fields. Moreover, the number counts at the faintest magnitudes are highest in the COSMOS field.  This latter observation is again expected, due to depth variation across the COSMOS set of images. Overall, the exposure time in COSMOS is a little higher than the three SN fields, but the data are drawn from observations with different pointings, which leads to shallower regions caused by the gaps between the DECam chips and the outer regions of the field. The slight deficit in objects at $i\sim23$ in SN-E2 and COSMOS may well be a result of source blending, as the PSF is significantly broader in these two fields ($\sim0.9\arcsec$ vs $\sim0.7\arcsec$ \texttt{FWHM}).

\subsection{Completeness}
\begin{figure}
    \centering
    \includegraphics[width=\linewidth]{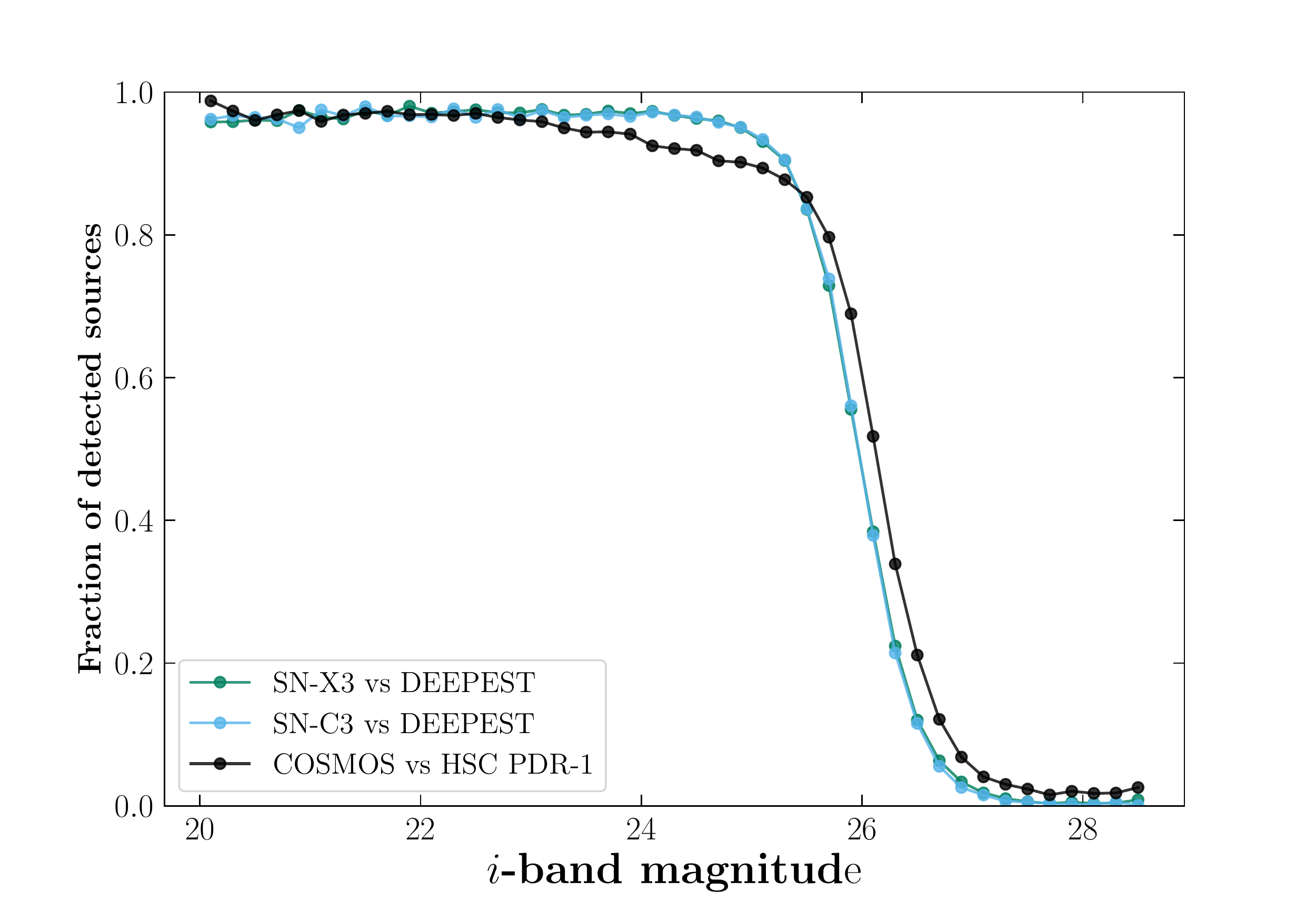}
    \caption{Detection completeness of sources in our Deep-Fields catalogue (\texttt{COADD\_TRUTH} depth level) versus the DEEPEST level of coadd images (SN-X3, SN-C3) or HSC-PDR1 (COSMOS) as a function of magnitude. Deep-Fields masks were applied to the source lists, but some artifacts and spurious sources will remain in the DEEPEST and HSC data. The DEEPEST image in SN-E2 is not deep enough to produce a meaningful test for the COADD\_TRUTH data. The data for SN-X3 and SN-C3 lie almost exactly on top of one another.}
    \label{fig:completeness}
\end{figure}

In order to estimate source completeness for our cosmology catalogue (\texttt{COADD\_TRUTH} level), we take advantage of the \texttt{DEEPEST} level of coadd images and the fact that our COSMOS field is largely overlapped by the UltraDeep level of HSC \citep{aihara2018}. For the SN-C3 and SN-X3 fields, the \texttt{COADD\_TRUTH} level of images contain just $\sim6\%$ of the $r, i$ and $z$-band data that make up the detection image of the \texttt{DEEPEST} coadds and is therefore effectively independent in noise realisation. For the SN-E2 field, the fraction is $44\%$, and thus the \texttt{DEEPEST} image in this field is not sufficiently deep nor independent enough for use in the following test. Given the similarity of the number counts in that field to SN-C3 and SN-X3, however, we expect the magnitude-dependent source completeness will also be very similar.  

The source lists from the deeper images, either HSC or our \texttt{DEEPEST} coadds, are treated as a truth table of objects that are present in the three fields. To this truth table we apply the set of masks defined for our Deep-Fields catalogue (including removal of CCDs in our ban list), described in \Sref{sec:masks}. For SN-C3 and SN-X3, the masks remove most spurious sources and areas of scattered light, though a modest number of unidentified satellite trails and other contaminants may remain. We then perform a sky position match with a tolerance of $1\arcsec$ and compute the fraction of sources that are identified in our Deep-Fields catalogue, as a function of true $i$-band model magnitude. These fractions are shown in \Fref{fig:completeness}. A strength of this method is that our estimate includes all object morphologies that exist in the Universe and as such should be a better representation of the completeness than, for instance, a simulated point-source extraction.

For the two SN fields, the completeness curve matches very well the turn-over in number counts (\Sref{sec:counts}) and shows our catalogue is highly complete ($\sim95\%$) at our reported $10\sigma$ image depth, $i=25.05$. The interpretation of the COSMOS completeness curve is less straight forward. As mentioned in \Sref{sec:counts}, the effective exposure time in the COSMOS $i$ and $z$-bands is greater than for SN-C3 or SN-X3, leading to a larger number of very faint objects. However, the exposures were not all taken with a consistent pointing which leads to depth variation due to the gaps between chips. Moreover, the source list depends on choices made during source extraction such as the minimum extent in pixels an object must have, or flux ratio threshold to deblend merged objects. Therefore, when comparing the output from two different pipelines (DES and HSC) we are in part testing the response to those choices. Finally, the seeing FWHM in the HSC image is $0.62\arcsec$, far superior to our COSMOS image at $0.94\arcsec$. Even under the same extraction configuration, we expect a greater degree of source deblending from the better seeing image. Nevertheless, our reported completeness for the COSMOS field is still $90\%$ at $i=25^{{\rm th}}$ magnitude.

\section{Star-Galaxy separation}
\label{sec:stargal}

In order to carry out star-galaxy separation at faint magnitudes, we train a machine learning classifier using external data available on the COSMOS field. We choose to make this classifier independent of morphology and use only photometric information. We make use of all 8 bands, $ugrizJHKs$, and the Hubble Space Telescope Advanced Camera for Surveys (HST-ACS) observations of \cite{2007ApJS..172..219L}. From the HST-ACS catalogue we make use of the \textsc{mu\_class} star-galaxy classification as `truth' labels. This classifier uses the high quality morphological information available from HST to identify a stellar locus in the surface brightness-magnitude plane. 

\begin{figure}
\begin{center}
\includegraphics[width=0.48\textwidth]{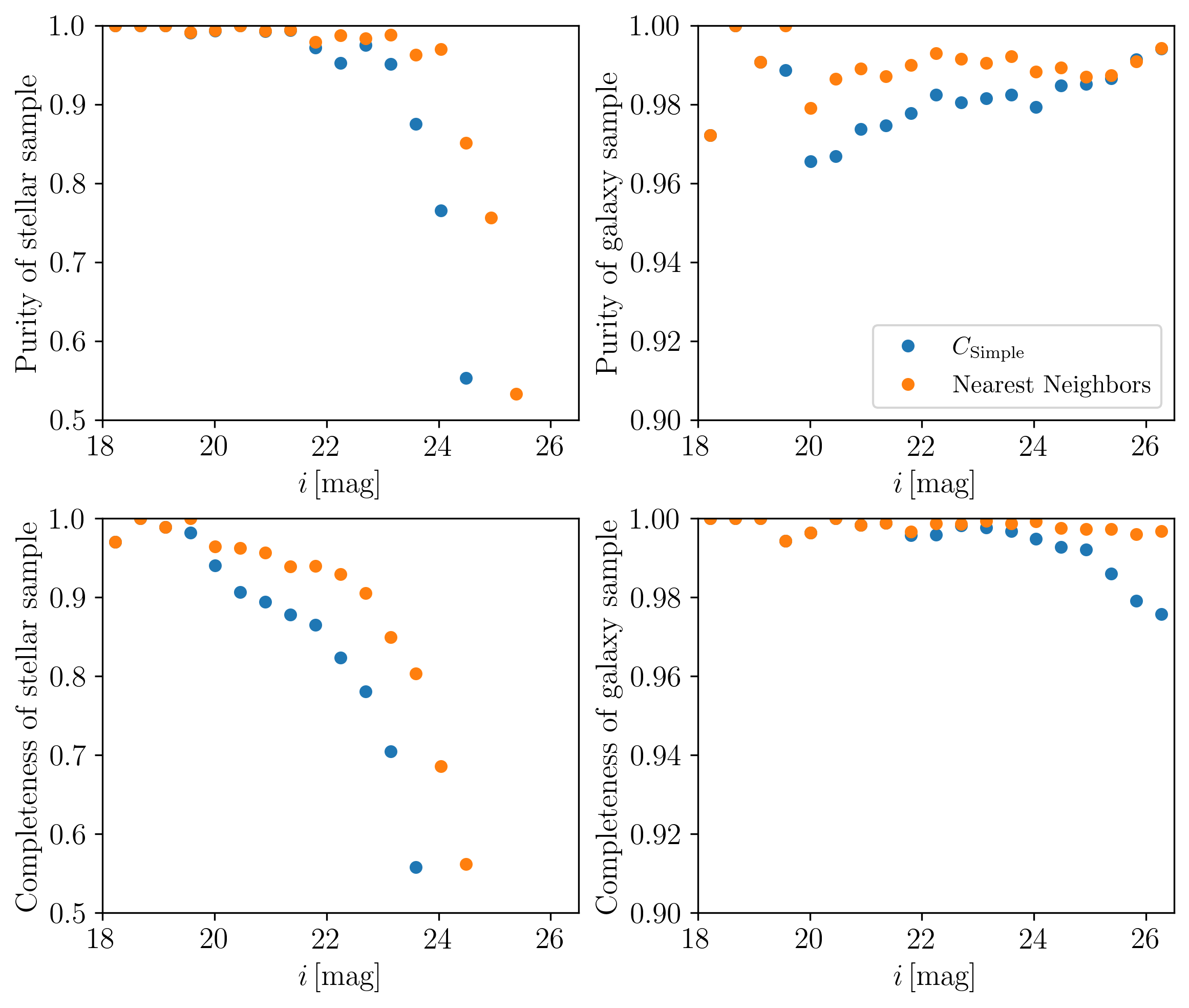}
\end{center}
\caption{Purity (upper row) and completeness (lower row) for stars (left) and galaxies (right) for the machine learning classifiers considered, along with the color classifier defined in \Eref{eqn:color_class}. Numerical scores in the legend refer to the mean accuracy score.}
\label{fig:stargal:roc}
\end{figure}

\begin{figure}
    \centering
    \includegraphics[width=\linewidth]{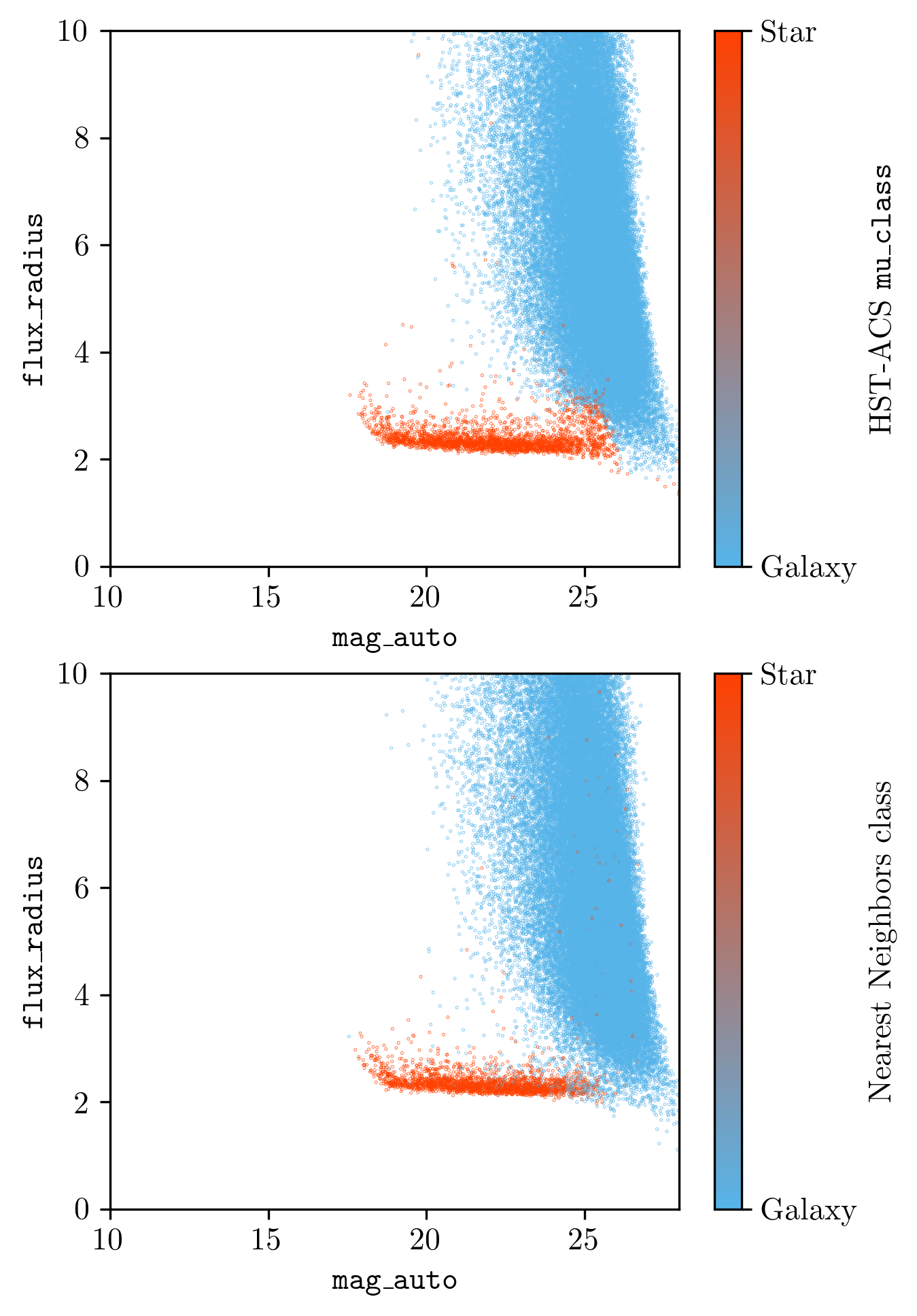}\\
    \caption{Size-magnitude diagram for the HST ACS morphological star-galaxy determination (upper panel) and kNN classifier used across all four Deep Fields.}
    \label{fig:SG_magi_morph}
\end{figure}

We match the HST-ACS catalogue to the DES Deep-Fields catalogue and randomly select a sub-sample of 20 per cent of objects to form a training set, ending up with 298338 objects in the training set. We then choose a number of supervised machine learning algorithms from \textsc{scikit-learn} and use the \textsc{mu\_class} classifications as truth labels, and available colors in $ugrizJHKs$ (colors formed from adjacent bands) as features. We find that a k-Nearest Neighbors (kNN) classifier produces the best performance, as shown in \Fref{fig:stargal:roc}. We also show the performance of a `Simple color class' classification, which takes the form:
\begin{equation}
C_{\rm Simple} = z - K - \left[1/3 (u-r) + 11/15 \right]
\label{eqn:color_class}
\end{equation}
with galaxies having $C_{\rm Simple} \geq 0$ and stars having $C_{\rm Simple} < 0$. 
\Fref{fig:SG_magi_morph} shows the resulting magnitude distributions and magnitude-size diagram for the different classes in the full catalogue. We subsequently apply the kNN classifier to the DES Deep-Fields data outside of the COSMOS field, in which HST-ACS \textsc{mu\_class} is not available, but $ugrizJHK$ are.

\section{Photometric redshifts}
\label{sec:photoz}

We employ the \texttt{EAzY} package \citep{brammer2008} to compute photometric redshift probability distribution functions (PDFs) for each object in the Deep-Fields catalogue. \texttt{EAzY} is a powerful and flexible template-fitting code, which uses linear combinations of a basis SED set derived through non-negative matrix factorisation. The combination of SED components in this way allows a greater variety of galaxies to be modelled than can be achieved with the singular or pairwise interpolation SED fitting used in, for instance, BPZ \citep{benitez2000}. The costs to this approach include a longer runtime required per galaxy and the possibility of fitting to unphysical galaxy SEDs, alhough this latter concern can be somewhat mitigated by careful selection of the allowed combinations of base SEDs. 

The greater model freedom in the linear combination approach used by \texttt{EAzY} reduces the effective number of degrees of freedom in the $\chi^2$ computation, and leads to generally flatter likelihoods in redshift space. Moreover, the base SED components do not have explicit galaxy types (Early type, late type, star-burst), and so cannot take advantage of our prior knowledge of how such types evolve in number with redshift. Instead, prior information is only included through apparent brightness. These factors make \texttt{EAzY} more suited to deriving redshifts for galaxies in richer fields, where high signal-to-noise photometric measurements are available across several bands or more \citep[e.g.,][]{hartley2013, sherman2020}. In such cases, we expect the redshift likelihoods to be fairly narrow and can use the greater model flexibility to improve precision and reduce outlier rate, relative to the single-SED case.

Our set-up with \texttt{EAzY} broadly follows the default configuration for the twelve Flexible  Stellar  Population  Synthesis (FSPS) SED components. This template set was constructed using the method described in \cite{brammer2008}, but based on the FSPS stellar population synthesis model \citep{conroy2010} and UltraVISTA photometric catalogue of \cite{muzzin2013}. We allow all possible combinations of components, include a systematic photometric error of $0.5\%$ motivated by \Sref{sec:photo_errors} and include the default extended R-band-based prior. We soften the prior very slightly via a Gaussian convolution ($\sigma=0.1$) on the high-redshift side of the peak of the prior. Our past experience with this prior has found that it penalises high-redshift solutions a little too much \citep{desprez2020}, though this change is very minor. As the systematic uncertainty on the $u$-band zero-point is far larger than for the other bands, we include it in the object catalogue for the purposes of our photo-z runs, added in quadrature with the computed photometric errors. We further perform a recalibration of the template error function (see \Sref{sec:templ_err}) and allow redshifts up to $z=8$ in intervals of $\delta z=0.01$.

\subsection{Spectroscopy in COSMOS and SN fields}
\label{sec:specz}

The performance of photo-z are typically tested against a sub-set for which true redshift values are available, in the form of high-confidence spectroscopic determinations. Such samples are inevitably biased towards brighter objects and those with clear spectral features that lie within the wavelength range of the spectrograph. Often, the chosen performance metrics are computed with a weight for each object in order to compensate for this bias and obtain results that better reflect a flux-limited selection of the photometric data set \citep{sanchez14, bonnett16}. Increasingly, very precise photo-z from surveys of many medium and/or narrow-bands are being utilised for this purpose also (\citealt{bonnett16, hoyle18, tanaka2018,alarcon2020,desprez2020}). The advantage of using photo-z is that they are by construction complete, and so don't suffer from the sort of selection biases discussed in \citet*{hartleychang2020}. On the other hand, a fraction of the photo-z are likely to be wrong, with increasing prevalence at fainter magnitudes. Between these two extremes are very low-resolution grism or prism spectra, such as those in PRIMUS \citep{cool2013} and 3D-HST \citep{momcheva2016}. These can be thought of as suffering the weaknesses of both spectroscopy and photo-z, but to generally much lesser degrees. In this work, we use these low-resolution samples to calibrate our method rather than test final performance.

In the three SN fields, SN-C3, SN-X3 and SN-E2, we use the spectroscopic compilation put together by OzDES, a partner survey of DES with the central goal to obtain SNe host galaxy redshifts. We wish to avoid biasing our performance measurements by an over-abundance of galaxies with active nuclei (AGN), which we know are not correctly modelled by the galaxy templates we employ. We therefore remove sources of spectroscopy that specifically targeted quasars or AGN, and further remove AGN hosts where they have been identified in the spectroscopy meta-data (e.g., via observer notes or specific quality flags). The sources of spectroscopy that we retain are listed in \Tref{tab:spec} together with the quality cuts we make. We also list the spectroscopic data sets that we use in the COSMOS field. As the best-studied extragalactic field, the spectroscopy in the COSMOS field is invaluable and is easily the most abundant of our Deep-Fields. We further use photo-z from the COSMOS+UltraVISTA catalogue of \citet{laigle2016} based on $30$ photometric bands, including 12 medium bands from Subaru, and the recent addition of the 40 narrow-band survey, Physics of the Accelerating Universe Survey (PAUS), in the same field, with redshifts determined by \citet{alarcon2020}.

\begin{table*}
\begin{center}
\caption{\label{tab:spec}  Sources of spectroscopic data used in photo-z performance metric assessment. $^{a}$ values of spectroscopic redshift quality used, as defined by the survey's quality flagging system.}
\begin{tabular}{|c|c|c|c|c|}  \hline \hline
Data set & Number & Fields & Flags$^{a}$ & Reference \\ \hline
2dFGRS & 297 & SN-C3,SN-E2 & $\ge3$ & \citet{colless2001} \\
2dF archive & 5,494 & SN-C3,SN-X3 & 4 & - \\
3D-HST & 6,937 & SN-C3,COSMOS & - & \citet{momcheva2016} \\
6dF & 48 & SN-C3,SN-X3,SN-E2 & 4 & \citet{jones2009} \\
ACES & 5,479 & SN-C3 & 3, 4 & \citet{cooper2012} \\
C3R2 & 2,248 & COSMOS & $\ge3.2$& \citet{masters2017} \\
DEIMOS 10K & 5931 & COSMOS & 3.x,4.x,23.x,24.x & \citet{hasinger2018} \\
FMOS COSMOS & 239 & COSMOS & 4 & \citet{silverman2015} \\
GAMA & 2,363 & SN-X3 & 4 & \citet{baldry2018} \\
GCLASS & 134 & SN-E2 & 1 & \citet{muzzin2012} \\
KMOS-3D & 326 & SN-C3,COSMOS & 0 & \citet{wisnioski2019} \\
LEGA-C & 1,405 & COSMOS & 4 & \citet{straatman2018} \\
MOSDEF & 397 & SN-C3,COSMOS & 7 & \citet{kriek2015} \\
MUSE & 494 & SN-C3 & 3 & \citet{herenz2017} \\
OzDES & 3,924 & SN-C3,SN-X3,SN-E2 & 4 & \citet{lidman2020} \\
PanSTARRS & 412 & SN-X3 & 4 & \citet{rest2014} \\
PRIMUS & 13,979 & SN-C3,SN-X3,SN-E2,COSMOS & 4 & \citet{cool2013} \\
SDSS & 2,325 & SN-X3 & 0 & \citet{abolfathi2018} \\
SNLS & 401 & SN-X3 & 1 & \citet{bazin2011} \\
VIPERS & 3,493 & SN-X3 & 3.x,4.x,23.x,24.x & \citet{garilli2014} \\
VUDS & 145 & SN-C3 & 3.x,4.x,23.x,24.x & \citet{tasca2017} \\
VVDS & 3,808 & SN-X3 & 3.x,4.x,23.x,24.x & \citet{lefevre2013} \\
zCOSMOS & 12,733 & COSMOS & 3.x,4.x,23.x,24.x & \citet{lilly2009} \\
\hline
\end{tabular}
\end{center}
\end{table*}

\subsection{Flux recovery and model error budget}

\begin{figure}
    \centering
    \includegraphics[width=0.9\linewidth]{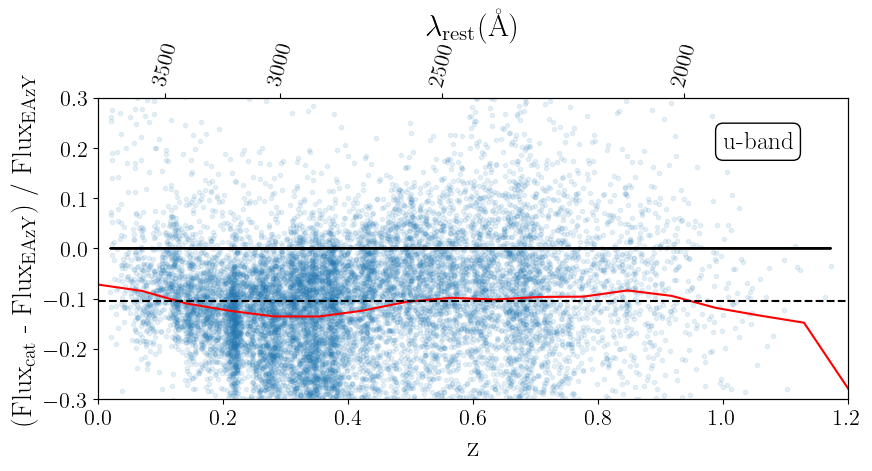}
    \includegraphics[width=0.9\linewidth]{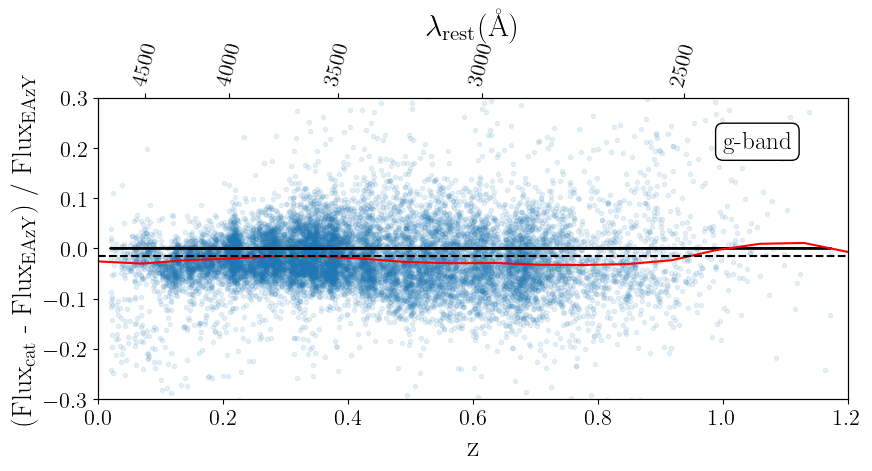}
    \includegraphics[width=0.9\linewidth]{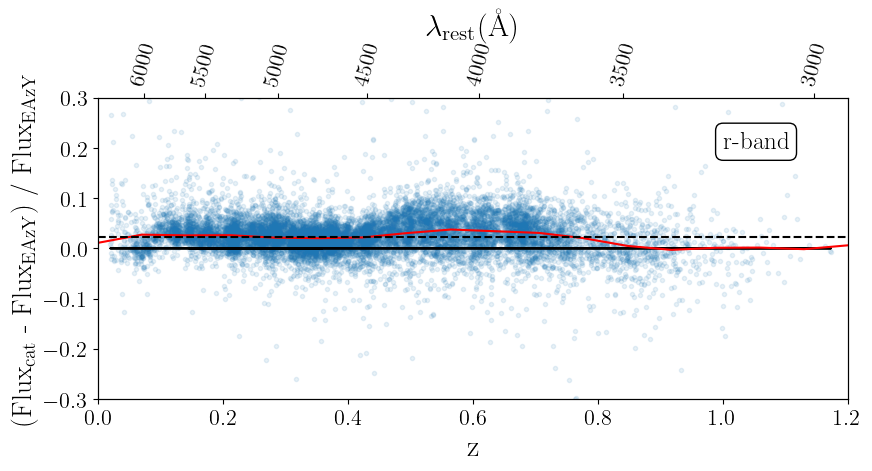}
    \includegraphics[width=0.9\linewidth]{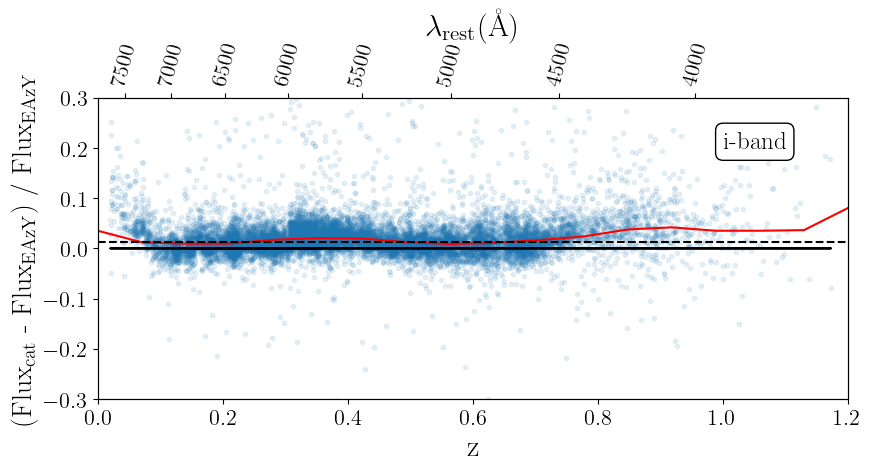}
    \includegraphics[width=0.9\linewidth]{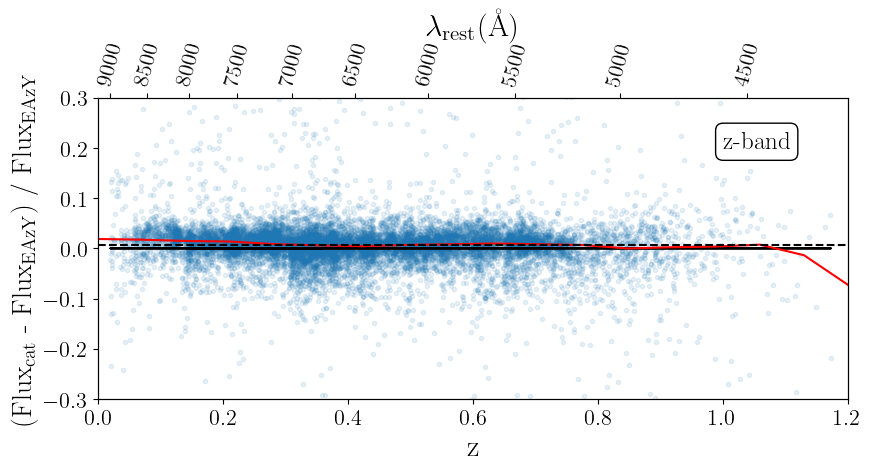}
    \caption{Fractional flux offsets between measured galaxy photometry and those predicted through \texttt{EAzY} template fits at their spectroscopic redshift. The red solid line shows the median value within a sliding window of width $\Delta z = 0.2$ and the black dashed line shows our computed zero-point correction.}
    \label{fig:eazy_zp}
\end{figure}

\begin{figure}
    \ContinuedFloat
    \centering
    \includegraphics[width=0.9\linewidth]{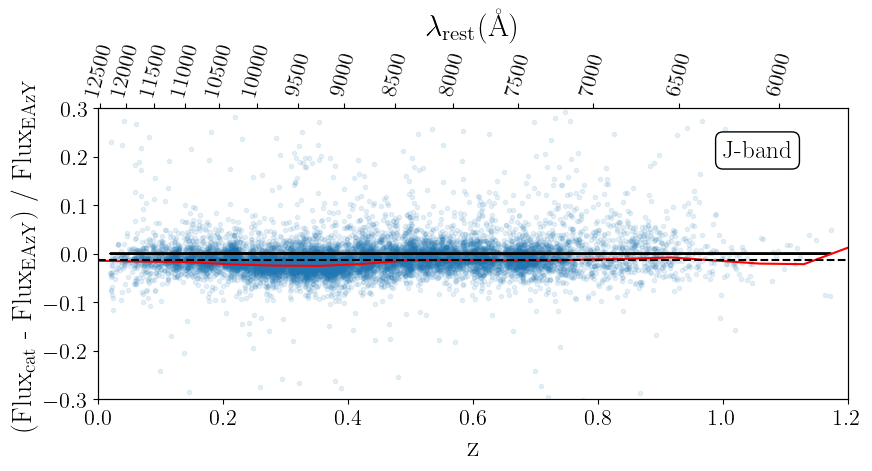}
    \includegraphics[width=0.9\linewidth]{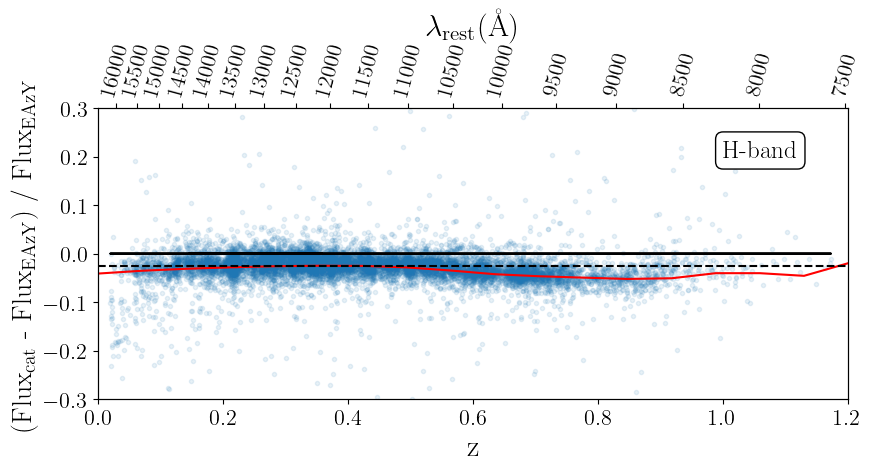}
    \includegraphics[width=0.9\linewidth]{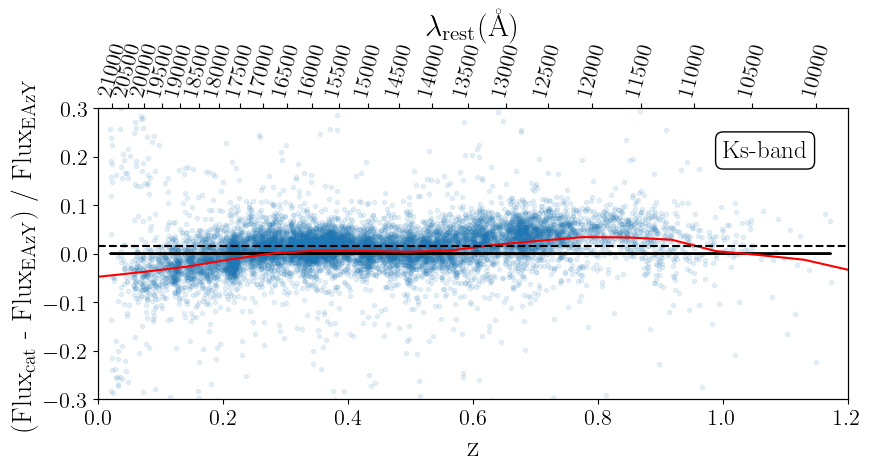}
    \caption{{\em cont.}}
    \label{fig:eazy_zp2}
\end{figure}

\begin{figure}
    \centering
    \includegraphics[width=\linewidth]{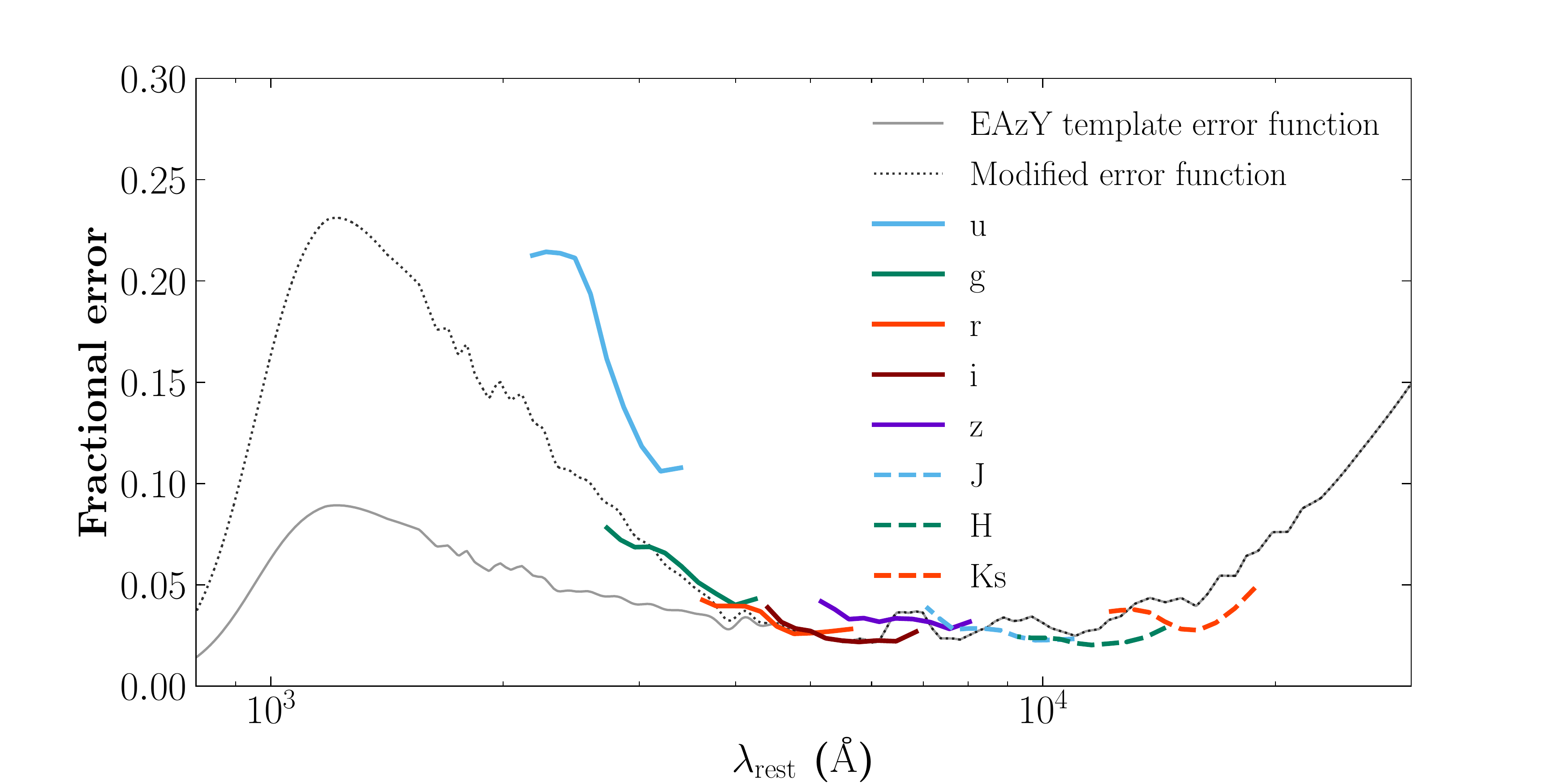}
    \caption{Dispersion in the measurements shown in \Fref{fig:eazy_zp} as a function of rest-frame wavelength, over the redshift range $0.1<z<0.8$. Coloured solid and dashed lines show the results for different photometric bands. The solid black line shows the default \texttt{EAzY} template error function for our SED component set, while the dotted line shows our adjusted version.}
    \label{fig:temp_err}
\end{figure}    
   
Not all of the spectroscopic data sets are suitable for assessing the performance of photo-z measurements. For instance, the PRIMUS data include a non-negligible fraction of incorrect redshifts, even at highest confidence, due to the nature and difficulty of reducing and analysing prism spectra. Furthermore, redshift solutions for galaxy templates above $z=1.2$ were not included in the analysis pipeline \citep{cool2013}. Nevertheless, such a large sample of galaxies (13,979, after applying cuts) where the vast majority of redshifts are indeed correct can still be of great use in calibrating zero-point corrections and a template error function, especially as the survey data include parts of all four of our Deep Fields.

\subsubsection{Zero-point calibration}
\label{sec:photoz_zp_calib}

We test and calibrate the photometric zero-points for photo-z determination through running \texttt{EAzY} with the redshift of each PRIMUS object fixed to its spectroscopic redshift. We use only objects with the most secure redshifts, that are classified as galaxies, are unflagged in our Deep-Fields catalogue and pass an $i<21.5$ cut. These cuts reduce the impact of photometric errors and colour-dependent incompleteness. The PRIMUS catalogue is $80\%$ complete under these cuts, however it remains possible that the level of incompleteness we allow could result in subtle biases (see \citealt*{hartleychang2020}). A separate run of \texttt{EAzY} is used for each of the eight photometric bands. In each run the photometric errors of the band in question are multiplied by a factor of $10^{15}$ so that the band does not contribute to the determination of the model SED coefficients. The fractional flux difference between the catalogue flux and the predicted flux from the best-fitting template combination, $f_{{\rm cat}} - f_{{\rm model}} / f_{{\rm model}}$, is computed for each object. These fractional errors are shown against spectroscopic redshift in \Fref{fig:eazy_zp}.

As noted earlier, a fraction of the redshifts will be incorrect, despite the fact that we use only the highest confidence objects. However, in using the median value within a running boxcar filter over redshift (width, $\Delta_z = 0.2$), the influence of these wrongly-assigned redshifts will be minimal. The results from the smoothing filter are shown by the red lines in \Fref{fig:eazy_zp}. Deviations from unity in these lines can be caused by a number of issues, e.g., incorrect templates, errors in filter curves or incorrect zero-point calibration. Mis-calibrated photometric zero-points will appear as a redshift-independent offset, while discrepancies between the true and model SEDs will typically show up by deviations that are quite isolated in redshift, and will be seen with a similar form but at different redshifts for different bands.

While we see evidence for a benefit in applying some small zero-point offsets, we see little evidence for the sort of clear fluctuations in flux difference between catalogue and model that would suggest an obvious problem with the template SED set. Though there are some notable features and redshift-dependent fluctuations, they are not coherent in rest wavelength between bands. The flux differences found for the $u$-band, and to a lesser extent the $g$-band, are broadly expected, given the great variance in UV spectra caused by relatively small changes in star-formation rate and internal dust extinction. This region of the spectral range has a large template uncertainty (see next subsection) in order to account for such issues. The $Ks$-band panel also stands out, having an apparent redshift-dependent slope to the model discrepancies. We could interpret this behaviour as residual imperfect background subtraction, SED miscalibration at long wavelengths, or perhaps a problem with the filter response curve. Another possibility is that it could be due in part to the fact that we do not have a longer-wavelength band to bracket the $Ks$-band with, and the $H$-band is one where the zeropoint seems to require additional calibration.

Using the results shown in \Fref{fig:eazy_zp} we derive photometric zero-point corrections for the purpose of computing photo-z with \texttt{EAzY}. For each band, an offset is computed as the median value of the boxcar-filtered data (i.e., median value of the fractional flux difference displayed by the red lines). In this way, our calibrations are not biased towards the redshifts most populated by our spectroscopic sample. These zeropoint adjustments are then applied to a new run of \texttt{EAzY}, again with redshift fixed to the spectroscopic redshifts, and the above procedure repeated. We iterate this process until stability, where the new zeropoint adjustments across $r$ to $Ks$-band are at the level of $0.5\%$ of flux or smaller\footnote{Because the $u$ and $g$-band probe the rest-frame UV the iterative process can be unstable, fitting less dusty SEDs with each iteration.}. The final zeropoint calibrations derived in this way are given in \Tref{tab:photom_tune}. Most adjustments correspond to $1-3\%$ of the object flux, with the $u$-band being the clear outlier. It is worth noting that the difference in zeropoint calibration between the H and Ks bands is similar to that found by \cite{laigle2016}, which is perhaps unsurprising given the data in common and our use of the UltraVISTA SED set. The zeropoint calibration difference between $J$ and $Ks$-bands in the COSMOS field is also reasonably close to that in \cite{laigle2016}, once our earlier field-to-field calibrations are taken into account (\Sref{sec:photo_errors}).

\subsubsection{Template error function calibration}
\label{sec:templ_err}

The variance of real galaxy SEDs is difficult, if not impossible, to fully capture with a limited template set --- even through a linear combination of 12 base SEDs. The equivalent widths of emission lines vary with redshift and also at fixed redshift and constant rest-frame colour, subtly altering even broad-band fluxes. Furthermore, inter-galactic absorption affecting the rest-frame UV spectral range is a stochastic process that would take a great many SED components to model. For these reasons (amongst others), \cite{brammer2008} introduced a template error function to the \texttt{EAzY} package to capture any wavelength-dependent uncertainty or mis-calibration in the templates. The template error function is clearly dependent on the set of SED components under consideration, and included with \texttt{EAzY} is an error function appropriate for the UltraVISTA-derived set that we use. Nevertheless, its derivation is also weakly dependent on the photometric data used, and so we assess its performance with our Deep-Fields catalogue.

We follow the procedure described in \cite{brammer2008}, but use the same fractional flux differences measured from the PRIMUS cross-matched data as earlier in this section, fixing the redshifts at their spectroscopic values. Due to the fact that we expect some outliers arising from wrong spectroscopic redshift assignments, we apply a $5\sigma$-clipping to the fractional flux differences, centred on the median and iterated five times, before computing their variance. This is again performed in a sliding window of $\Delta z=0.2$. The result of this process is shown by the short coloured lines in \Fref{fig:temp_err}, representing the interval $0.1<z<0.8$ where we have a sufficient number of objects, and with the contribution to the variance coming from the typical photometric uncertainty subtracted out. Clearly, from around $5000${\AA} onwards the template error function is already well matched to our data, with perhaps a small over estimate at the longest wavelengths. However, throughout the wavelength range probed by the $g$ and $u$-bands it is too small.

It is worth noting that, while the $g$ and $r$-bands agree well in their prediction for the required template error at $\sim4000${\AA}, the $g$ and $u$-bands are highly discrepant. In \Sref{sec:photo_errors} we estimated a systematic zeropoint error for the $u$-band of $5.5\%$. Interpreted as a coherent calibration offset uncertainty, this error would not impact our estimate of the template error function. However, its estimate owes largely to the difficulty in identifying clean sequences with which to align the different fields, and it enters our cosmology pipeline as an uncertainty on the object flux. In this sense, we feel it is appropriate to account for this source of error in our assessment of the template error function. Doing so brings the $g$ and $u$-bands into much better agreement, though they remain $\sim3\%$ apart. As a large fraction of the variance in the $u$-band measurements in \Fref{fig:eazy_zp} appears to be from the photometric calibration, and not the template error, we drop the $u$-band while calibrating the template error function.

Our modified template error function is simply a wavelength-dependent scaled version of the one shipped with \texttt{EAzY}, fit to the $g$ and $r$-band data and using the following functional form to allow a smooth extrapolation through the rest-frame UV wavelength range,

\begin{equation}
A_{{\rm new}} = A * (1 - {\rm erf}((\lambda - \lambda_0)/1000)) + 1,
\end{equation}
Where $A$ and $\lambda_0$ are free parameters to be fit and erf() is the Gauss error function. We use $A=0.8$ and $\lambda=3000${\AA} to produce our modified template error function (dotted line in \Fref{fig:temp_err}). Following the photometric redshift performance tests (\Sref{sec:photoz_performance}), we compute the predicted template error function for a sample of high-redshift galaxies with 3D-HST grism redshifts, probed by the $g, r, i$ and $z$-bands over the wavelength range $2000<\lambda<7000${\AA}. We find that, although these data are noisier than those shown for PRIMUS, the average of the four bands agrees well with our modified template error function.

The final step is to find the correct scaling of the template error function. \texttt{EAzY} returns the best $\chi^2$ value at each redshift, meaning that we cannot use a formal marginalisation over the SED component amplitudes to derive the redshift posterior. This choice improves \texttt{EAzY}'s speed and scalability, without significant detriment to the single-value best redshift estimates (point estimates). However, the approximation of the multi-dimensional likelihood into one dimension in this way causes a flattening of the final redshift PDFs, resulting in over-broadening. A compromise can be found by reducing the overall amplitude of the template error function, while retaining its influence on the relative likelihood at different redshifts. In our photo-z run we use a $50\%$ amplitude.

\subsection{Photometric redshift performance}
\label{sec:photoz_performance}

\begin{figure}
    \centering
    \includegraphics[width=0.8\linewidth]{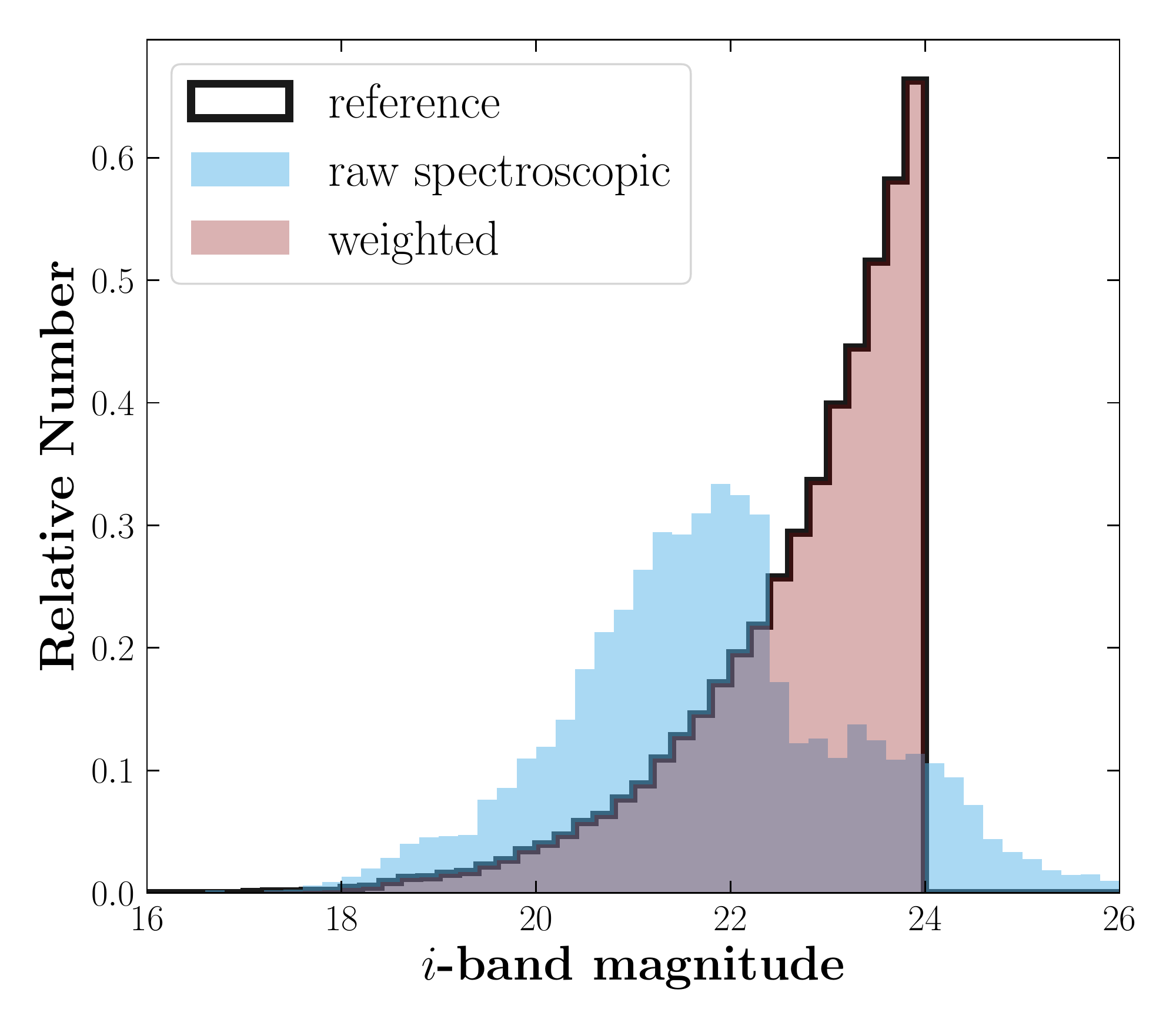}
    \caption{$i$-band magnitude distribution for objects in our collated spectroscopic redshift sample, together with the flux-limited Deep-Fields sample at $i<24$ and weighted spectroscopic distribution which matches the flux-limited sample, by construction.}
    \label{fig:spec_mag_hist}
\end{figure}

\begin{figure}
    \centering
    \includegraphics[width=0.8\linewidth]{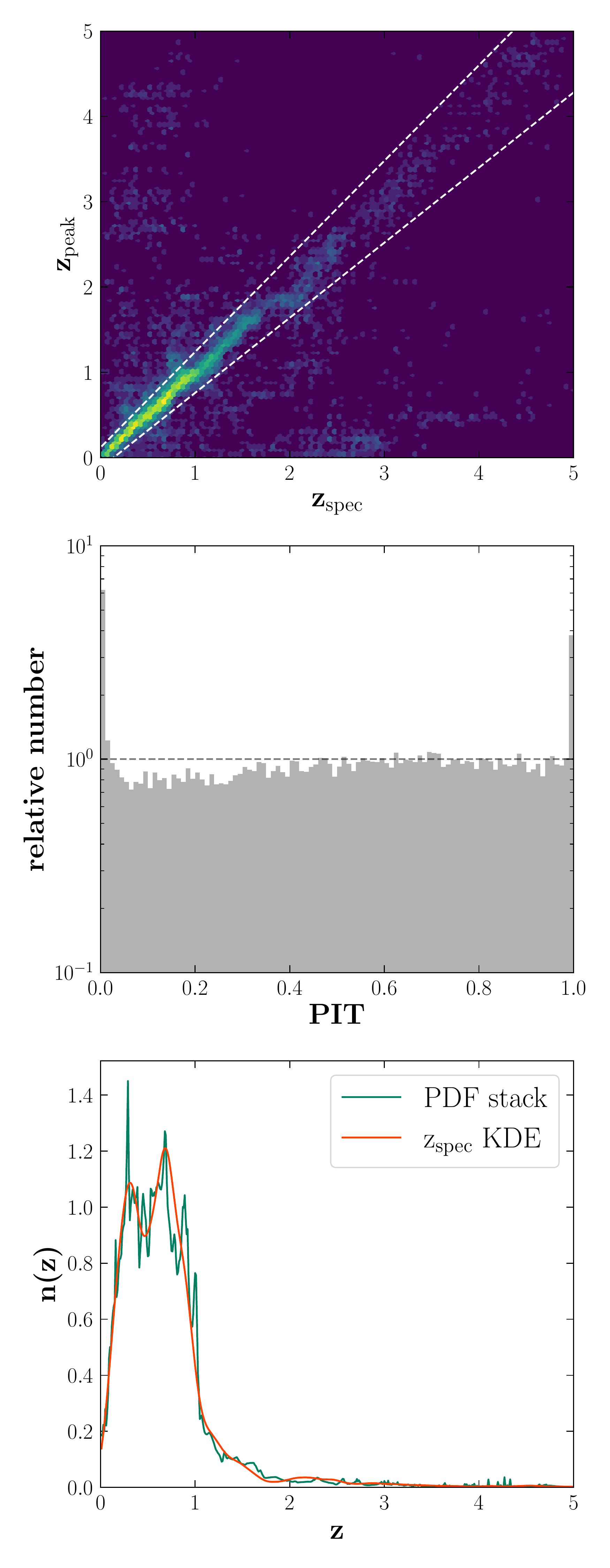}
    \caption{Photometric redshift performance for the (unweighted) spectroscopic redshift sample. {\em Top:} comparison of spectroscopic and best photometric redshift estimate. The colour scale corresponds to the log density of points, i.e., the vast majority of objects lie at $z<1$ and have very accurate photo-z estimates. {\em Middle:} Probability Integral Transform histogram. A flat histogram represents a probabilistically-calibrated data set. Deviations from flatness indicate over or under-dispersed PDFs, while a slope may indicate an overall bias. Peaks at either end of the histogram quantify the fraction of outliers, where the spectroscopic redshift lies entirely outside of the photo-z PDF. Our results compare very favourably to similar plots in the literature (\citealt{schmidt2020,desprez2020}) {\em Bottom:} n($z$) recovery via coaddition of photo-z PDFs, compared with a Kernel Density Estimate of their spectroscopic redshifts.}
    \label{fig:photoz_results}
\end{figure}

\begin{table*}
\begin{center}
\caption{\label{tab:photz}  Photo-z metrics: Mean bias $((z_{ph}-z_{sp})/(1+z_{sp}))$, Normalised Median Absolute Deviation, outlier rate and Kullback-Leibler divergence of the probability integral transform histogram with respect to an ideal flat distribution (see text for details).}
\begin{tabular}{|p{0.15\textwidth}|p{0.12\textwidth}|p{0.16\textwidth}|p{0.12\textwidth}|p{0.12\textwidth}|p{0.12\textwidth}|}  \hline \hline
 & Spectroscopic & Weighted spectroscopic, \newline $17<i<24$ & Alarcon20, \newline $17<i^{\prime}<23$ & Laigle16, \newline $17<i<24$ & Laigle16, \newline $17<i<26$ \\ \hline
Overall NMAD & 0.0223 & 0.0301 & 0.0228 & 0.0339 & 0.0736\\
NMAD, no outliers & 0.0205 & 0.0264 & 0.0219 & 0.0302 & 0.0467\\
Outlier rate - point & $6.3\%$ & $9.5\%$ & $2.9\%$& $8.4\%$ & $26.1\%$\\
Mean bias, $\mathbb{E}~{\Delta z}$ & 0.0147 & 0.0097 & 0.0136 & 0.0131 & 0.0551\\
Mean bias, no outliers & 0.0027 & 0.0057 & 0.0019 & 0.0094 & 0.0096\\
\hline
PIT KL-divergence & $1.2\times10^{-4}~{\rm Hart}$& $3.2\times10^{-4}~{\rm Hart}$ & $1.2\times10^{-4}~{\rm Hart}$ & $5.0\times10^{-4}~{\rm Hart}$ & $9.3\times10^{-4}~{\rm Hart}$\\
Outlier rate - PDF & $5.0\%$ & $6.2\%$ & $3.3\%$ & $5.8\%$ & $6.2\%$\\ \hline
N objects & 34,948 & 32,956 & 30,737 & 72,391 & 222,231 \\
Mean redshift & 0.693 & 0.898 & 0.642 & 0.846 & 1.294 \\
Mean i-band mag. & 21.49 & 22.72 & 21.87 & 22.75 & 24.33 \\
\hline
\end{tabular}    
\end{center}
\end{table*}

Ideal photo-z will be both highly predictive, i.e., allow a point redshift estimate that is close to the true value, and have meaningful and accurate PDFs. The former has traditionally been the focus in assessing photo-z performance for deep extragalactic science \citep[e.g.,][]{dahlen2013}, though the use of redshift PDFs in extragalactic science has become increasingly common \citep{wake2011,hartley2013,etherington2017} and with it a necessity to assess the accuracy of them. A number of metrics have been developed to measure and compare different photo-z predictions (see, e.g., \citealt{zhan2006,bordoloi2010,dahlen2013,sanchez14,bonnett16,schmidt2020,desprez2020}). Most require a sample of objects with high-confidence spectroscopic redshifts, though methods have also been developed to infer performance purely statistically \citep{quadri2008}. The most suitable performance metrics to measure will clearly depend upon the science goals of interest. It is beyond the scope of this work to be exhaustive in our tests, but to aid comparison with other data sets we choose the most commonly used subset that nevertheless cover the broad classes of use cases. The remainder of this subsection is split into those that concern how predictive our point redshifts are, and those that estimate the accuracy of our PDFs.

\subsubsection{Point prediction performance}

We assess the quality of our best single-value redshift estimates through three metrics: mean bias ($\mathbb{E}~\Delta_z$, where $\Delta_z = (z_{{\rm ph}} - z_{{\rm sp}}) / (1+z_{{\rm sp}})$), Normalised Absolute Median Deviation (NMAD),
\begin{equation}
{\rm NMAD} = 1.48 \times {\rm median}(|\Delta_z|)
\end{equation}
and outlier fraction. Outliers are defined as $|\Delta_z| \ge 0.12$. We also compute bias and NMAD with outliers excluded. Results are shown in \Tref{tab:photz} for five different samples:
\begin{itemize}
\item The high-confidence spectroscopic sample described in \Sref{sec:specz}, excluding PRIMUS, 3D-HST and objects flagged as AGN.
\item This same spectroscopic sample, but with a set of weights applied based on i-band magnitude, such that their weighted magnitude histogram matches the Deep-Fields catalogue cut to $17<i<24$ (see \Fref{fig:spec_mag_hist}).
\item A flux limited sample, $17<i^{\prime}<23$, with selection magnitude and reference redshift taken from \cite{alarcon2020}.
\item A flux limited sample, $17<i<24$, with reference redshift taken from \cite{laigle2016}.
\item A flux limited sample, $17<i<26$, with reference redshift taken from \cite{laigle2016}.
\end{itemize}
The total number used, mean redshift and mean i-band magnitude are also reported and we show a comparison of the point redshift estimate against spectroscopic redshift in the upper panel of \Fref{fig:photoz_results}.

Performance is very good up to redshift, $z=1.5$, and down to $24^{\rm th}$ magnitude, with a very tight core of objects quantified by an overall NMAD of less than $3.5\%$, and well-controlled mean bias. However, the fraction of outliers rises quite quickly between $23^{{\rm rd}}$ and $24^{{\rm th}}$ magnitude, and even more so at progressively fainter magnitudes. Beyond $z=1.5$, precise redshifts require an intermediate band between the $z$-band and $J$-band to constrain the $4000\AA$ break wavelength (e.g., VISTA $Y$-band), which is not included in our catalogue. Performance is improved slightly beyond $z\sim2$, when the break comfortably enters the $J$-band, but by these redshifts there are few objects with strong $4000${\AA} break features. As a result the dispersion in $z_{{\rm ph}} - z_{{\rm sp}}$ becomes larger at $z>1.5$, until galaxies bright in the $i$-band become $u$-band drop-outs and an effective Lyman break selection becomes possible ($z>3$). At these high redshifts we see the typical degeneracy caused by confusion between the Lyman break and $4000${\AA} break in what are otherwise largely featureless SEDs. Although outliers in the sense of point redshift estimates, many of these objects will have redshift PDFs that capture the degeneracy.

It is interesting to compare the relative metric performances between the weighted spectroscopic sample and the L16 sample under the same magnitude cuts. The spectroscopic sample is highly heterogeneous, with selections targetting emission line galaxies, luminous red galaxies, high-redshift galaxies and other particular galaxy subsamples. Moreover, despite the fact that we have removed AGN where labelled, it is by no means certain that we have reduced their number to a fair sampling rate. In addition, a small fraction of the spectroscopic objects are likely to be blended sources, perhaps with redshift determined from emission lines that do not represent the bulk of the galaxy light \citep{masters2019}, and may contain other subtle selection biases \citep*{hartleychang2020}. Conversely, a small fraction of the L16 redshifts are incorrect, even for these relatively bright sources \citep[see ][ Sec. 4.3]{laigle2016}. Though the differences in the metric results are not large, they suggest that perhaps the spectroscopic sample is under-represented by a galaxy sub-population that have relatively weak predictions, but over-represented by the type of objects that cause catastrophic failures in redshift determination. Meanwhile, the results of the sample matched to \citet{alarcon2020} show that our point estimates are exceptional for an 8-band catalogue at $i<23$.

\subsubsection{PDF performance}

To assess the accuracy of the \texttt{EAzY} PDFs we turn to the Probability Integral Transform (PIT) distribution \citep{dawid1984}. PIT is defined as the value of the cumulative distribution function evaluated at the true (spectroscopic / many-band photo-z) redshift,
\begin{equation}
PIT = \int_{-\infty}^{z_{{\rm sp}}} p(z) dz
\end{equation}
and its distribution over a sample is being increasingly used as a key test of the accuracy of redshift PDF calibration. It can also be used as part of an additional post-processed calibration if the PDFs are found to be inaccurate \citep{bordoloi2010,hoyle18}, but we do not use it for this purpose. If a set of one-dimensional PDFs are well calibrated then their distribution of PIT values should be indistinguishable from a set of random draws from a uniform distribution, $U(0,1)$. We show the PIT distribution for the (unweighted) spectroscopic sample in the middle panel of \Fref{fig:photoz_results}. For some objects the spectroscopic redshift lies entirely outside of the photo-z PDF, either at higher or lower redshift. These objects we define as outliers and contribute to the extreme ends of the histogram, but we exclude them when quantifying the shape of the PIT distribution. Our metrics in testing our photo-z PDFs are the outlier fraction, and the Kullback-Leibler (KL) divergence of the PIT histogram, relative to perfect performance --- i.e., relative to a flat distribution:
\begin{equation}
KL = \sum p~log_{10}(p/q)
\end{equation}
where p is the histogram of PIT values, and q a flat distribution. The KL divergence is a measure of information which, in the case of log base-10, carries the unit, Hart. As a final, visual, test of our PDFs, we show the redshift distribution estimated via summing the individual redshift PDFs and comparing it with a kernel density estimate (KDE) of their spectroscopic redshifts. The bandwidth of the KDE was chosen such that the smoothing kernel reflects the value of NMAD. While a sum of PDFs is by no means the best estimator of the redshift distribution \citep[for a discussion, see ][]{malz2020}, it is nevertheless commonly used in scientific analyses and thus warrants inclusion here. 

The results of the two PDF metrics are shown in \Tref{tab:photz}. Two trends stand out: the PDF outlier fraction is fairly insensitive to the sample used and doesn't climb at fainter magnitudes; the PIT KL-divergence increases as the sample magnitude limit is made fainter. The first observation tells us that the large outlier fraction found in the point redshift metrics at faint magnitudes ($26.1\%$ for the deepest sample) is not due to catastrophic outliers of galaxies without appropriate models in the photo-z set up, but rather represents a simple decrease in predictive power caused by a lack of information. In other words, the high point-redshift outlier fraction in these deeper samples does not suggest a fault in redshift determination, but a limit to the precision that can be achieved at faint magnitudes with our 8-band data.

The rise in KL-divergence for faint samples represents a possible weakness in the accounting of the error budget of these objects. Indeed, the PIT histogram for our faintest sample (not shown) has a clear but shallow concave shape, which is typically due to slightly under-dispersed PDFs (i.e., under-estimated uncertainties), and evidence of a small bias in the sense that the PDFs are shifted towards higher redshift. The PIT histogram for our combined spectroscopic sample shows the opposite behaviour: the slight deficit of objects with PIT$\sim0.1-0.3$ indicates a very small bias in the sense that the PDFs are systematically too low in redshift. Note that this possible PDF bias for the spectroscopic sample is in the opposite sense of the point redshift bias. The solution to this minor calibration defect at faint magnitudes is unclear, and it is likely that we would not be able to incorporate it within the existing \texttt{EAzY} software. Possible solutions may include fully incorporating the band-to-band flux covariance, computing a magnitude dependent template error function or applying ad-hoc contributions to the flux errors as a function of magnitude.

\section{Conclusions}
\label{sec:conclusions}

In this paper, we have described the construction of the DES Deep-Fields images and the catalogue drawn from a subset of those images for the survey's main three-year cosmology analysis (covering four fields of $\sim 1.5$ sq. deg. apiece). Through combining observations from our SN survey, community data and additional dedicated observations of the COSMOS field and SN fields, we have constructed images that comprise ten times the exposure time of the main survey, and that have seeing FWHM better than $50\%$ or more of the WS data set. We have combined these images with near-IR data from two VISTA programmes: VIDEO and UltraVISTA, to produce a data set spanning the $u$ to $Ks$ wavelength range. The final catalogue is based on detection from the average of the $r$, $i$ and $z$-bands, numbering $2.8$ million sources, which is reduced to $1.7$ million after extensive and careful masking is applied. Deblended source photometry and forced photometry measurements of individual bands were performed with the \texttt{fitvd} model-fitting code, which ensures consistent colours for our extracted sources.

We have presented tests of the source extraction, finding $>90\%$ completeness to $i=25^{{\rm th}}$ magnitude in each field and agreement in number counts at the level of $90\%$ or better across the magnitude range of relevance for DES Y3 cosmology. Our PSF modelling accuracy has been demonstrated to be within $1\%$ error across the required range for PSF construction in all bands, which is crucial for measuring accurate source colours. We have further performed a fine tuning of the relative colours and photometric zero-points between our four fields, and anchored the calibration to our main survey data. Our final estimated zero-point uncertainties are found to be $0.5\%$ in the key DES bands ($g$, $r$, $i$ and $z$), $0.8\%$ in the VISTA bands ($J$, $H$ and $Ks$), and $5.5\%$ in the $u$-band.

With our calibrated catalogue we have shown the performance of a colour-based star-galaxy separation method, using morphology-determined stars in the HST COSMOS data as ground truth. We have found that we are able to efficiently separate the two object classes to magnitudes as faint as $i \sim 22$, with stellar completeness degrading to 90\% by $i=22.5$. Finally, we have detailed the production of photometric redshifts, using the \texttt{EAzY} code, including additional zero-point corrections that are specific to the template set employed, and a small calibration adjustment to the template error function used. The resulting photo-z show excellent performance with respect to spectroscopic redshifts and highly-accurate photo-z from COSMOS \citep{laigle2016} and PAUS \citep{alarcon2020}, to magnitudes as faint as $i=24$. At fainter magnitudes, the performance understandably declines, but remains useful.

The DES Deep-Fields catalogue is suitable for use in a plethora of stand-alone scientific analyses, from exploration of the evolution of the stellar mass function (Gschwend et al., in prep.) to studies of the host galaxies of transient events (Palmese et al., in prep.) and multi-dimensional derivation of galaxy property posteriors via machine learning \citep{Mucesh2020}. However, its primary motivation is for use in the DES cosmology analysis, combining weak lensing shear, galaxy clustering and galaxy-galaxy lensing \citep{y3-3x2ptkp}. The principal roles of the Deep-Fields data for the cosmology analysis are: 1) measuring the transfer function of the survey --- i.e., the characteristics of the sources we extract from our main DES survey data relative to the input truth \citep[for the Y3 implementation, see][]{y3-balrog}; 2) building a high-dimensional self-organising map (SOM), in which the redshift distributions of each cell are intrinsically narrow with respect to the SOM built from the 4-band data of the main survey \citep*[for the Y3 implementation, see][]{y3-sompz}. The Y3 image simulations used to calibrate the weak lensing shear also utilises a variant of the Deep-Fields catalogue based on detections and fits to morphological parameters on HST imaging in COSMOS and fluxes estimated on the DECam $griz$ imaging \citep{y3-imagesims}.  Other advantages in building these data include being able to use a data-driven prior on the moments of galaxy light distributions for galaxy shear measurement \citep{bernsteinarmstrong2014,bernstein2016}, possible because of our selection of input images with good seeing ($<1.0\arcsec$ in $riz$ bands).

The coming few years will see the second full public release of DES data (DR2), including data from all six years of DES observations. Shortly thereafter will follow our cosmology analyses with these data, and as part of that we will need to build a Deep-Fields data set that is able to support the increased statistical precision that the final survey data will provide. With respect to the catalogue presented in this work, that data set will need to be incrementally deeper, but crucially also cover a wider area. Taking advantage of our other SN fields and the recent VISTA Extragalactic Infrared Legacy Survey\footnote{https://people.ast.cam.ac.uk/~mbanerji/VEILS} (VEILS; Banerji et al., in prep.), we will be able to double our area, thereby reducing sample variance uncertainties and adding additional sources of spectroscopy. Finally, in the spirit of establishing a legacy value for the DES Deep Fields, we will produce a catalogue based on the DEEPEST level of SN coadd images, across all ten SN pointings.

\section*{Acknowledgements}
We thank our referee, Adriano Fontana, for their careful reading of our manuscript and suggestions that improved the clarity of our text.

AC acknowledges support from NASA grant 15-WFIRST15-0008. IH, acknowledges support from the European Research Council in the form of a Consolidator Grant with number 681431 and from the Beecroft Trust.

Funding for the DES Projects has been provided by the U.S. Department of Energy, the U.S. National Science Foundation, the Ministry of Science and Education of Spain, 
the Science and Technology Facilities Council of the United Kingdom, the Higher Education Funding Council for England, the National Center for Supercomputing 
Applications at the University of Illinois at Urbana-Champaign, the Kavli Institute of Cosmological Physics at the University of Chicago, 
the Center for Cosmology and Astro-Particle Physics at the Ohio State University,
the Mitchell Institute for Fundamental Physics and Astronomy at Texas A\&M University, Financiadora de Estudos e Projetos, 
Funda{\c c}{\~a}o Carlos Chagas Filho de Amparo {\`a} Pesquisa do Estado do Rio de Janeiro, Conselho Nacional de Desenvolvimento Cient{\'i}fico e Tecnol{\'o}gico and 
the Minist{\'e}rio da Ci{\^e}ncia, Tecnologia e Inova{\c c}{\~a}o, the Deutsche Forschungsgemeinschaft and the Collaborating Institutions in the Dark Energy Survey. 

The Collaborating Institutions are Argonne National Laboratory, the University of California at Santa Cruz, the University of Cambridge, Centro de Investigaciones Energ{\'e}ticas, 
Medioambientales y Tecnol{\'o}gicas-Madrid, the University of Chicago, University College London, the DES-Brazil Consortium, the University of Edinburgh, 
the Eidgen{\"o}ssische Technische Hochschule (ETH) Z{\"u}rich, 
Fermi National Accelerator Laboratory, the University of Illinois at Urbana-Champaign, the Institut de Ci{\`e}ncies de l'Espai (IEEC/CSIC), 
the Institut de F{\'i}sica d'Altes Energies, Lawrence Berkeley National Laboratory, the Ludwig-Maximilians Universit{\"a}t M{\"u}nchen and the associated Excellence Cluster Universe, 
the University of Michigan, NFS's NOIRLab, the University of Nottingham, The Ohio State University, the University of Pennsylvania, the University of Portsmouth, 
SLAC National Accelerator Laboratory, Stanford University, the University of Sussex, Texas A\&M University, and the OzDES Membership Consortium.

Based in part on observations at Cerro Tololo Inter-American Observatory at NSF’s NOIRLab (NOIRLab Prop. ID 2012B-0001; PI: J. Frieman), which is managed by the Association of Universities for Research in Astronomy (AURA) under a cooperative agreement with the National Science Foundation.

The DES data management system is supported by the National Science Foundation under Grant Numbers AST-1138766 and AST-1536171.
The DES participants from Spanish institutions are partially supported by MICINN under grants ESP2017-89838, PGC2018-094773, PGC2018-102021, SEV-2016-0588, SEV-2016-0597, and MDM-2015-0509, some of which include ERDF funds from the European Union. IFAE is partially funded by the CERCA program of the Generalitat de Catalunya.
Research leading to these results has received funding from the European Research
Council under the European Union's Seventh Framework Program (FP7/2007-2013) including ERC grant agreements 240672, 291329, and 306478.
We  acknowledge support from the Brazilian Instituto Nacional de Ci\^encia
e Tecnologia (INCT) do e-Universo (CNPq grant 465376/2014-2).

This manuscript has been authored by Fermi Research Alliance, LLC under Contract No. DE-AC02-07CH11359 with the U.S. Department of Energy, Office of Science, Office of High Energy Physics.

\textit{Software}: \SExtractor \citep{Bertin:1996}, \PSFEx \citep{Bertin:2011}, \scamp \citep{Bertin:2006}, \swarp \citep{Bertin:2002,Bertin:2010}

\section*{Data Availability}

The data underlying this article and that were used to build the processed data products were accessed from NOAO (\url{https://www.noao.edu/}). The derived data will be released publicly through the DES collaboration at \url{https://des.ncsa.illinois.edu/releases/} upon the completion of the cosmology analysis for which they are used.



\bibliographystyle{mnras_2author}
\bibliography{deepfields.bib,des_y3kp.bib} 



\afterpage{%
\clearpage
\begin{sidewaystable*}
\footnotesize
\begin{centering}
\caption{\label{tab:full_table}Systematics for Deep Fields}
\begin{tabular}{|l|l|r|r|r|r|r|r| r|r|r|r|r|r| r|r|r|r|r|r|} \hline\hline
        &       & \multicolumn{6}{c}{$\sum{T_{\rm exptime}}$ [s]} & \multicolumn{6}{c}{Avg PSF [arcsec]} & \multicolumn{6}{c}{Magnitude Limit} \\ 
 Flavor & Field & $u$ & $g$ & $r$ & $i$ & $z$ & $Y$ &  $u$ & $g$ & $r$ & $i$ & $z$ & $Y$ & $u$ & $g$ & $r$ & $i$ & $z$ & $Y$ \\ 
\hline 
 COADD\_TRUTH & COSMOS    &    45380  &    23364  &     9742  &    18916  &    23950  &    12150   &  1.17  &  1.22  &  0.93  &  0.94  &  0.97  &  0.90   &  25.94  &  26.46  &  25.73  &  25.54  &  24.97  &  23.70  \\ 
\hline 
 SE\_TRUTH    & SN-C1     &  ---  &     2800  &     2700  &     1800  &     1200  & ---   & ---  &  0.91  &  0.84  &  0.76  &  0.67  & ---   & ---  &  25.29  &  24.90  &  24.26  &  23.47  & ---  \\ 
 COADD\_TRUTH & SN-C1     &     3300  &    11725  &     8550  &     8000  &     5800  &      900   &  1.01  &  1.09  &  0.95  &  0.84  &  0.73  &  1.26   &  24.47  &  26.04  &  25.63  &  25.06  &  24.31  &  22.48  \\ 
 DEEPEST      & SN-C1     &     3300  &    14875  &    10500  &    17600  &    30800  &      900   &  1.01  &  1.17  &  1.00  &  0.98  &  0.90  &  1.26   &  24.47  &  26.20  &  25.72  &  25.53  &  25.20  &  22.48  \\ 
\hline 
 SE\_TRUTH    & SN-C2     & ---  &     2975  &     2700  &     2200  &     1600  & ---   & ---  &  0.88  &  0.80  &  0.74  &  0.67  & ---   & ---  &  25.28  &  24.95  &  24.29  &  23.59  & ---  \\ 
 COADD\_TRUTH & SN-C2     &     3300  &    11025  &     9600  &     7600  &     6200  &      900   &  0.99  &  1.06  &  0.95  &  0.82  &  0.71  &  1.33   &  24.45  &  26.03  &  25.70  &  25.06  &  24.32  &  22.46  \\ 
 DEEPEST      & SN-C2     &     3300  &    15750  &    11250  &    17200  &    30600  &      900   &  0.99  &  1.18  &  0.99  &  0.98  &  0.89  &  1.33   &  24.45  &  26.22  &  25.78  &  25.51  &  25.20  &  22.46  \\ 
\hline 
 SE\_TRUTH    & SN-C3     & ---  &     2200  &     4000  &     3600  &     1320  & ---   & ---  &  0.78  &  0.74  &  0.70  &  0.66  & ---   & ---  &  25.22  &  24.85  &  24.27  &  23.46  & ---  \\ 
 COADD\_TRUTH & SN-C3     &    23200  &     9200  &    11200  &    10080  &     5610  &     7000   &  1.10  &  0.86  &  0.78  &  0.72  &  0.68  &  0.78   &  25.29  &  26.05  &  25.67  &  25.06  &  24.28  &  23.40  \\ 
 DEEPEST      & SN-C3     &    23200  &    51800  &    82800  &   159840  &   301290  &     7000   &  1.10  &  1.19  &  0.99  &  1.00  &  0.90  &  0.78   &  25.29  &  26.95  &  26.90  &  26.75  &  26.42  &  23.40  \\ 
\hline 
 SE\_TRUTH    & SN-E1     & ---  &     2625  &     1950  &     1800  &     1200  & ---   & ---  &  0.94  &  0.83  &  0.74  &  0.66  & ---   & ---  &  25.05  &  24.78  &  24.33  &  23.52  & ---  \\ 
 COADD\_TRUTH & SN-E1     &     3300  &    10150  &     7200  &     8400  &     6200  & ---   &  1.46  &  1.10  &  0.96  &  0.86  &  0.72  & ---   &  24.35  &  25.84  &  25.47  &  25.05  &  24.30  & ---  \\ 
 DEEPEST      & SN-E1     &     3300  &    13825  &     9450  &    16800  &    30400  & ---   &  1.46  &  1.17  &  1.00  &  0.99  &  0.90  & ---   &  24.35  &  26.03  &  25.60  &  25.44  &  25.16  & ---  \\ 
\hline 
 SE\_TRUTH    & SN-E2     & ---  &     2800  &     1950  &     1400  &     1200  & ---   & ---  &  0.92  &  0.82  &  0.74  &  0.67  & ---   & ---  &  25.30  &  24.92  &  24.25  &  23.56  & ---  \\ 
 COADD\_TRUTH & SN-E2     &    18000  &    11025  &     8250  &     7000  &     6200  & ---   &  1.31  &  1.09  &  0.96  &  0.83  &  0.73  & ---   &  25.52  &  26.03  &  25.64  &  25.06  &  24.32  & ---  \\ 
 DEEPEST      & SN-E2     &    18000  &    15750  &    10500  &    18200  &    30200  & ---   &  1.31  &  1.20  &  1.01  &  1.01  &  0.90  & ---   &  25.52  &  26.20  &  25.74  &  25.53  &  25.16  & ---  \\ 
\hline 
 SE\_TRUTH    & SN-S1     & ---  &     3325  &     2100  &     1800  &     1400  & ---   & ---  &  0.94  &  0.83  &  0.75  &  0.68  & ---   & ---  &  25.23  &  24.82  &  24.29  &  23.51  & ---  \\ 
 COADD\_TRUTH & SN-S1     &     3300  &    10325  &     7350  &     8200  &     5800  & ---   &  1.27  &  1.10  &  0.96  &  0.86  &  0.74  & ---   &  24.30  &  25.87  &  25.47  &  25.05  &  24.29  & ---  \\ 
 DEEPEST      & SN-S1     &     3300  &    12950  &     8700  &    14600  &    25600  & ---   &  1.27  &  1.18  &  0.99  &  0.99  &  0.90  & ---   &  24.30  &  26.00  &  25.57  &  25.36  &  25.04  & ---  \\ 
\hline 
 SE\_TRUTH    & SN-S2     & ---  &     2975  &     2550  &     1800  &     1400  & ---   & ---  &  0.95  &  0.84  &  0.75  &  0.68  & ---   & ---  &  25.17  &  24.86  &  24.32  &  23.55  & ---  \\ 
 COADD\_TRUTH & SN-S2     &     3300  &     8400  &     7050  &     8600  &     6000  & ---   &  1.22  &  1.10  &  0.96  &  0.88  &  0.73  & ---   &  24.34  &  25.74  &  25.41  &  25.06  &  24.32  & ---  \\ 
 DEEPEST      & SN-S2     &     3300  &    11900  &     8400  &    13600  &    24000  & ---   &  1.22  &  1.18  &  1.00  &  0.99  &  0.90  & ---   &  24.34  &  25.93  &  25.51  &  25.32  &  25.00  & ---  \\ 
\hline 
 SE\_TRUTH    & SN-X1     & ---  &     2275  &     1800  &     1800  &     1200  & ---   & ---  &  0.93  &  0.83  &  0.77  &  0.67  & ---   & ---  &  25.02  &  24.72  &  24.29  &  23.50  & ---  \\ 
 COADD\_TRUTH & SN-X1     &     3300  &     8750  &     6150  &     7600  &     6200  &     5700   &  1.26  &  1.11  &  0.97  &  0.89  &  0.76  &  1.22   &  24.33  &  25.73  &  25.34  &  24.97  &  24.30  &  23.30  \\ 
 DEEPEST      & SN-X1     &     3300  &    12075  &     7350  &    13200  &    21600  &     5700   &  1.26  &  1.19  &  1.00  &  0.99  &  0.90  &  1.22   &  24.33  &  25.93  &  25.43  &  25.29  &  24.93  &  23.30  \\ 
\hline 
 SE\_TRUTH    & SN-X2     & ---  &     1575  &     1500  &     1200  &     1800  & ---   & ---  &  0.91  &  0.83  &  0.76  &  0.71  & ---   & ---  &  24.83  &  24.54  &  23.92  &  23.57  & ---  \\ 
 COADD\_TRUTH & SN-X2     &     3300  &     8575  &     6150  &     6800  &     7400  &      900   &  1.26  &  1.10  &  0.96  &  0.88  &  0.79  &  1.15   &  24.30  &  25.66  &  25.27  &  24.90  &  24.32  &  22.22  \\ 
 DEEPEST      & SN-X2     &     3300  &    11375  &     7500  &    14200  &    23200  &      900   &  1.26  &  1.17  &  0.99  &  0.99  &  0.90  &  1.15   &  24.30  &  25.88  &  25.40  &  25.34  &  24.96  &  22.22  \\ 
\hline 
 SE\_TRUTH    & SN-X3     & ---  &     2400  &     2000  &     2880  &     1320  & ---   & ---  &  0.87  &  0.80  &  0.74  &  0.65  & ---   & ---  &  25.28  &  24.77  &  24.26  &  23.41  & ---  \\ 
 COADD\_TRUTH & SN-X3     &    13700  &    11600  &    10000  &    10440  &     5610  &     1400   &  1.50  &  0.96  &  0.83  &  0.77  &  0.67  &  1.22   &  25.21  &  26.05  &  25.68  &  25.04  &  24.29  &  22.48  \\ 
 DEEPEST      & SN-X3     &    13700  &    46200  &    64000  &   124200  &   228360  &     1400   &  1.50  &  1.20  &  0.99  &  0.99  &  0.90  &  1.22   &  25.21  &  26.78  &  26.68  &  26.53  &  26.25  &  22.48  \\ 
\hline 
\end{tabular}
\end{centering}
\end{sidewaystable*}
\clearpage
}

\appendix

\section{Astrometry test with Gaussian Process}\label{sec:gp_astrom}
\begin{figure*}
    \centering
    \includegraphics[width=1.0\textwidth]{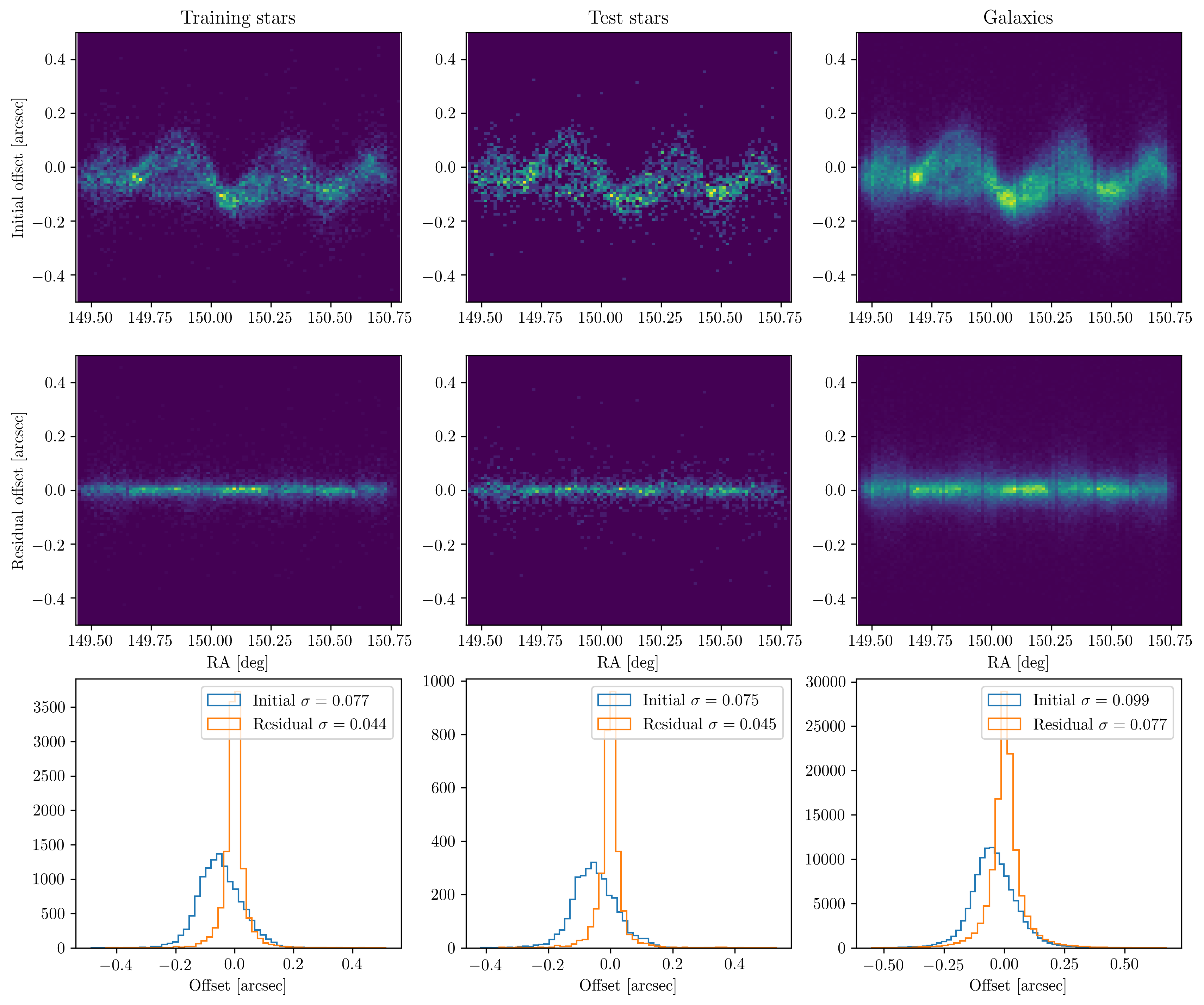}
    \caption{Residuals between UltraVISTA DR3 positions and DECam positions in the COSMOS field. \emph{Top} row shows pre-correction offsets projected in the RA direction, \emph{middle} row shows residuals following correction with the Gaussian Process model, \emph{lower} row shows one dimensional histograms of the offsets both pre- and post-correction. \emph{Left} column shows offsets for the sample of stars used to train the Gaussian Process, \emph{centre} column shows offsets for the reserved test stars, and \emph{right} column for the galaxy sources.}
    \label{fig:gp-astrometry-residuals}
\end{figure*}

At the start of this project, UltraVISTA data from the DR3 release\footnote{\url{http://www.eso.org/sci/observing/phase3/data_releases/uvista_dr3.pdf}} was used. When comparing matched source positions between the UltraVISTA DR3 data set and the DECam images of the COSMOS field, significant offsets were found, as can be seen in the upper panels of \Fref{fig:gp-astrometry-residuals}. In order to place both data sets into a common astrometric reference frame, we correct the offsets using a Gaussian Process (GP) model \citep[a supervised machine learning method, see e.g.,][]{Rasmussen2006Gaussian}. We select bright stars from the matched UltraVISTA-DECam catalogue with the condition:
\begin{align}
    \tt{spread\_model} + \frac{5}{3} \tt{spreaderr\_model} < 0.002,
\end{align}
where \texttt{spread\_model} and \texttt{spreaderr\_model} are the corresponding columns from the DECam \textsc{SExtractor} catalogue. We then take eighty per cent of these sources and train a two dimensional Gaussian process for each RA offset and Declination offset between the sources' positions in the DECam and UltraVISTA images. The result of applying the correction from this Gaussian Process model for the two dimensional surfaces in RA and Declination offset are shown in the left hand column of \Fref{fig:gp-astrometry-residuals}. The centre column of \Fref{fig:gp-astrometry-residuals} also shows the offsets for the remaining twenty per cent of the star sources which were reserved as a test set (i.e., which were not used in training the GP). As can be seen from the residual offsets, after correction with the GP, (centre row), the correction has significantly reduced the systematic offsets between the UltraVISTA and DECam positions. This can also be seen in the galaxy sample, in the right most column of \Fref{fig:gp-astrometry-residuals}. The GP used was implemented using GPy\footnote{\url{https://github.com/SheffieldML/GPy}}, with a Matern 5/2 kernel. Separate GPs were trained in the stripes in RA defined by the UltraVISTA deep and ultra-deep regions. We implemented a number of improvements on this baseline configuration, including adding a white noise kernel to allow for stellar motions; including bright, compact galaxies as extra training points; and a number of different appropriate kernels. None of these led to significant improvements in the residual offsets after correction. In particular, tuning of kernels to allow for smaller scale variations in the offset correction surface led to over-training, with smaller residuals for training sources, but larger residuals for the test sources. A summary of the improvements in the astrometry can be seen in the lower left panel of \Fref{fig:gp-astrometry-residuals}.

Upon the release of UltraVISTA DR4\footnote{\url{http://ultravista.org/release4/dr4_release.pdf}}, the initial offsets between UltraVISTA and DECam data were reduced to the same level as the post-correction residuals for DR3. We therefore proceed with the DR4 data without GP correction.

\section{Additional data}
\label{sec:append_data}
We expand on the description of the data in \Sref{sec:optobs} by summarising the quantities related to depth and image quality for the full set of 10 deep fields in \Tref{tab:full_table}.

\begin{figure}
    \centering
    \includegraphics[width=0.9\linewidth]{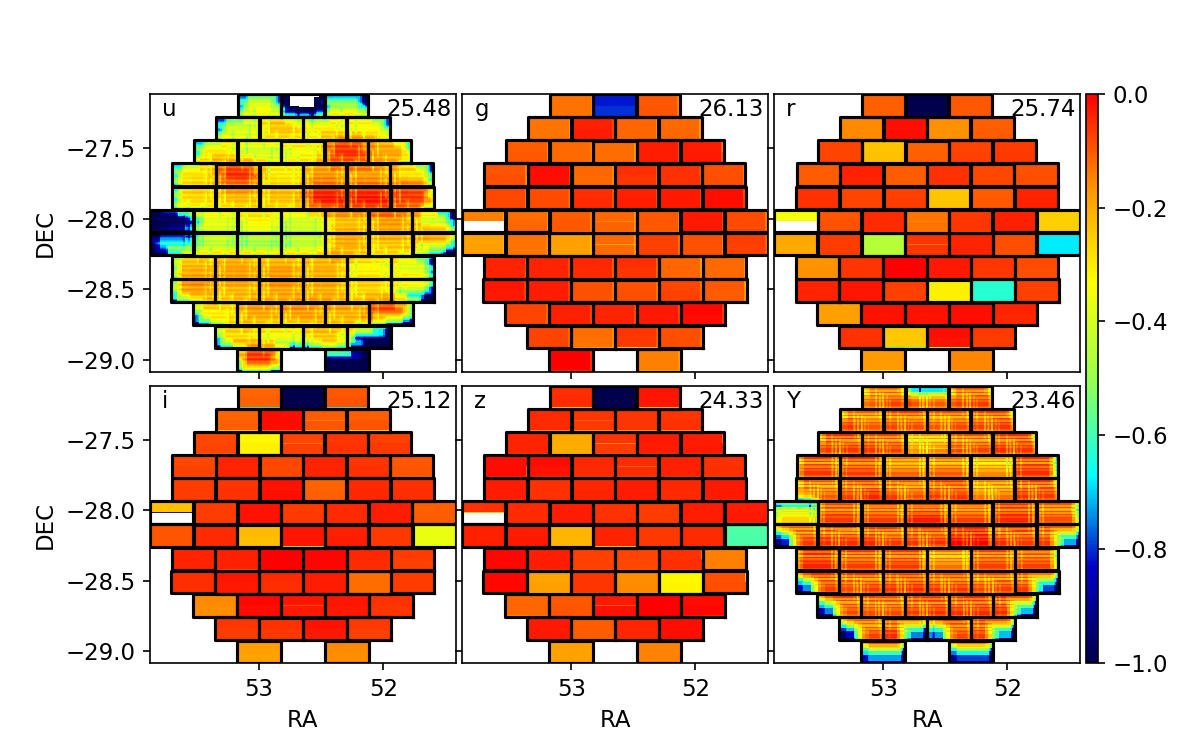}
    \includegraphics[width=0.9\linewidth]{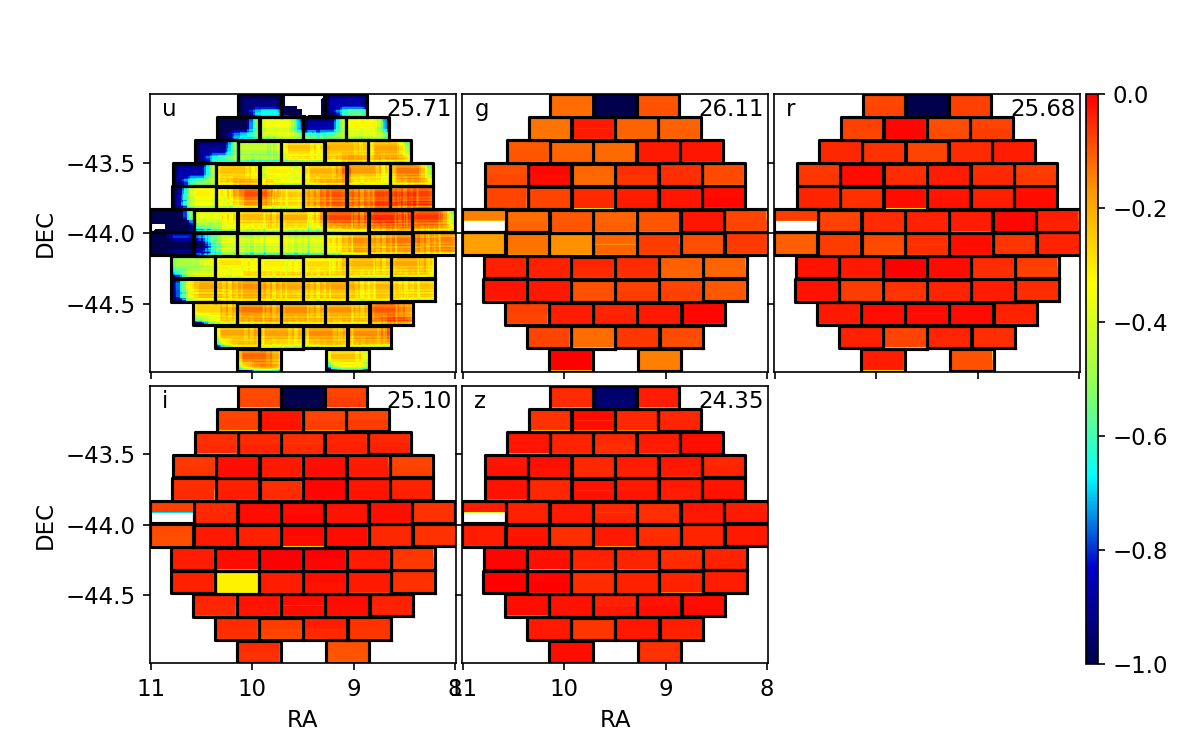}
    \includegraphics[width=0.9\linewidth]{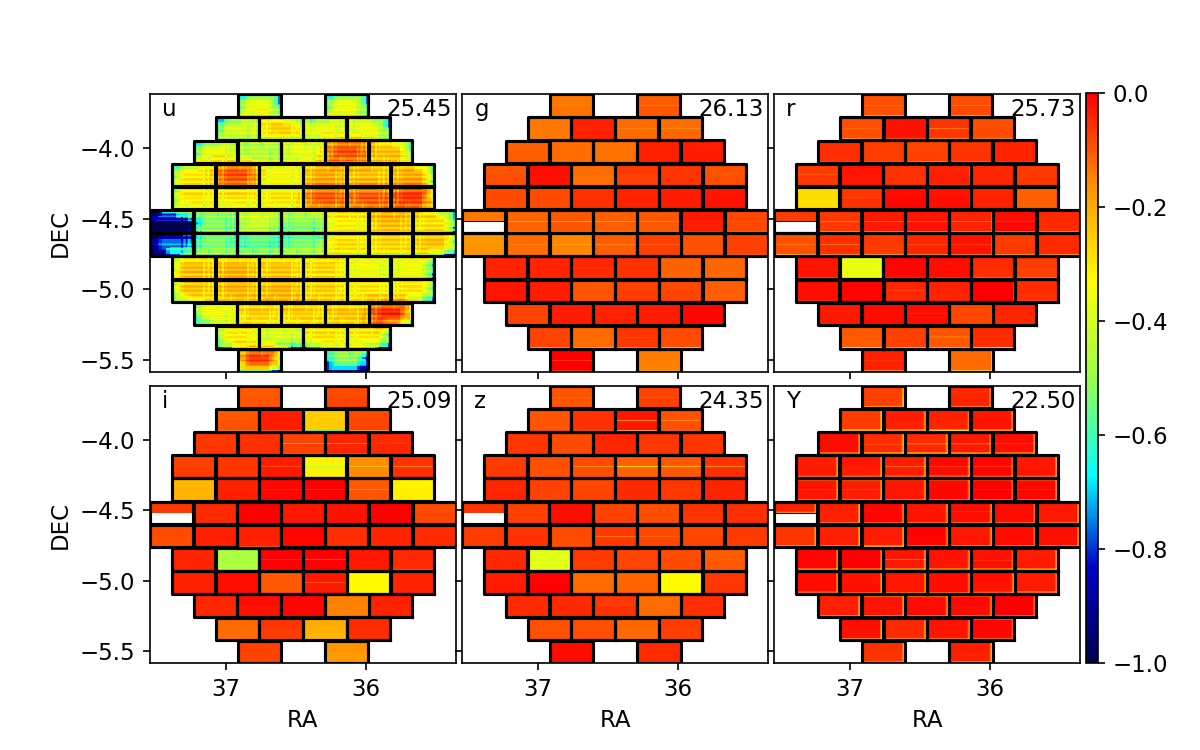}
    \includegraphics[width=0.9\linewidth]{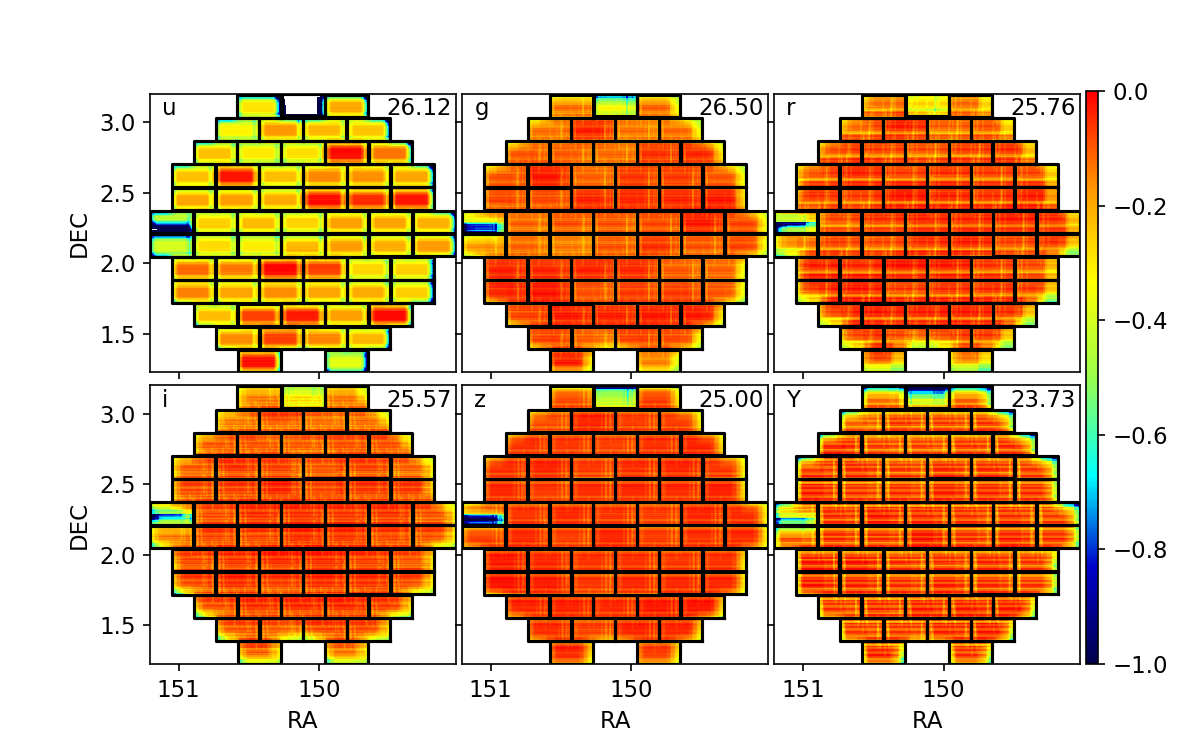}
    \caption{Magnitude limits calculated across each field. The fields (top to bottom) are SN-C3, SN-E2, SN-X3, and COSMOS, respectively.  The range shown for each panel is set to show variations (within 1 magnitude) of the maximum depth for a given field/band combination (the value for the maximum depth is given in the upper right corner of each panel).}
    \label{fig:maglim_maps}
\end{figure}

In \Fref{fig:maglim_maps} we visualise the magnitude limits for the DECam data over each of the four fields chosen for the cosmology catalogue and for each of the filters $ugrizY$.

\begin{figure}
    \centering
    \includegraphics[width=0.9\linewidth]{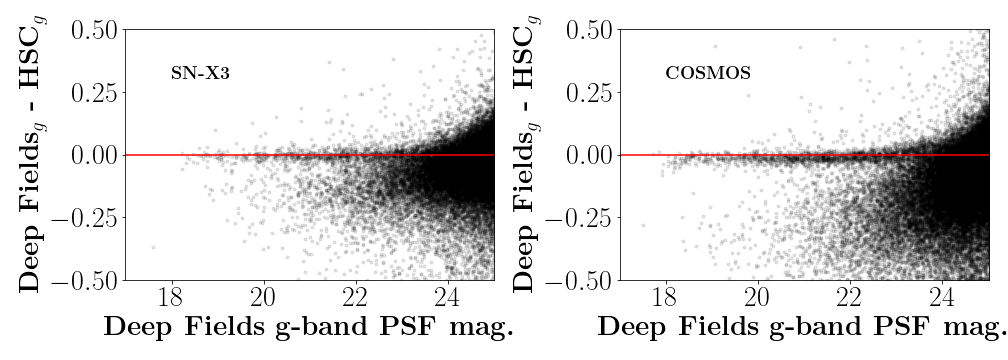}
    \includegraphics[width=0.9\linewidth]{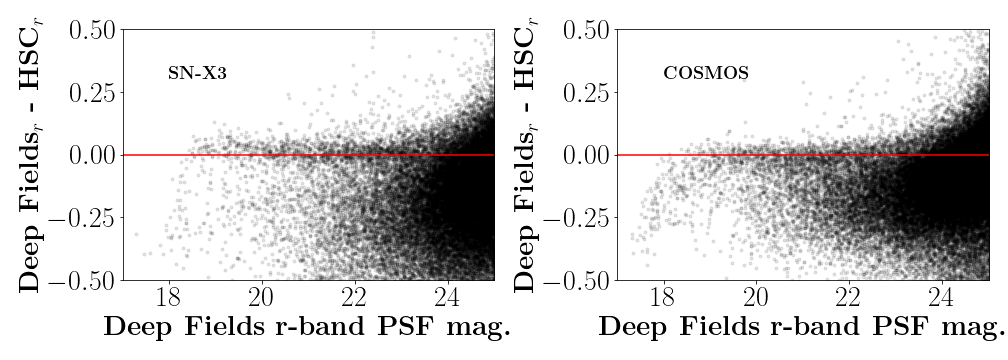}
    \includegraphics[width=0.9\linewidth]{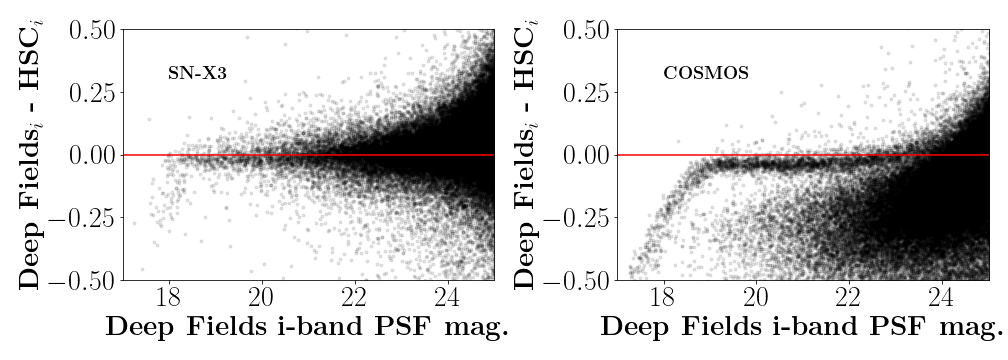}
    \includegraphics[width=0.9\linewidth]{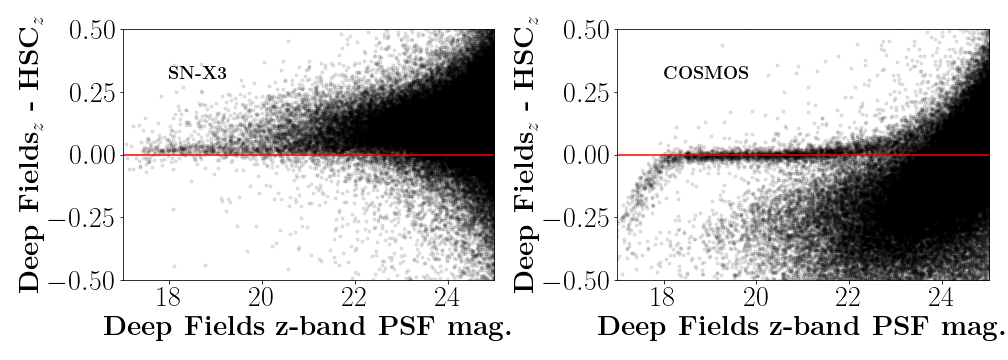}
    \caption{Comparison of PSF magnitudes between DES Deep Fields and HSC. DES magnitudes have been converted to the HSC filter system.}
    \label{fig:HSC_comparison}
\end{figure}

In \Fref{fig:HSC_comparison} we plot the difference between the Deep-Fields catalogue PSF magnitudes in $griz$ and the same objects found in HSC for SN-X3 and COSMOS.  In order to make the conversion between the DECam filters and the HSC filters, we used the \citet{laigle2016} catalogue and \citet{2014ApJS..212...18B} templates to perform a guided interpolation from Subaru broad bands to each of DES and HSC (see the appendix of \citealt{rudnick2003} for a description of the method used). With the magnitude differences and DECam colours per object, we fit a first-order polynomial to convert DECam bands to their HSC counterparts. The r-band conversion was poorly represented by a simple polynomial, with considerable dispersion at fixed colour.

For stellar sources, the $g,r$ and $z$ bands match within the expected uncertainties ($0.5\%$ for the Deep-Fields catalogue and $1\%$ for HSC). The For the $i$-band, the COSMOS field show a $4\%$ difference between the two catalogues, the origin of which is unclear. There is a known issue in the HSC Cosmos photometry \citep{aihara2018}, though this is not expected to result in an offset larger than $1.5\%$. We use PSF magnitudes for this comparison and so do not expect the magnitudes of galaxies to match.

\newpage

\section{Catalog Release Columns}

\onecolumn
\begin{longtable}[c]{l c l}
\caption{Selected Deep Field Catalogue Columns. {\em Note:} names ending ``*'' show that the quantity is available for UGRIZJHKs filter bands; names ending ``**'' show that the quantity is available for redshift values from $z = 0.01$ to $z = 8$. Full details will be available upon release at \url{https://des.ncsa.illinois.edu/releases}. \label{tab:DF_catalog_columns}} \\ \hline \hline \\ 
Catalogue Column  & Units & Description \\ \hline 
ID           & - & Object identifier \\
TILENAME           & - & DECam field and chip number string \\
RA          & Degrees & Equatorial Coordinates (J2000) \\
DEC        & Degrees & Equatorial Coordinates (J2000) \\
 &  & \\
BDF\_FLUX\_*       & 3.63~nJy & Bulge+disk model flux with AB zero-point$=30$ \\
BDF\_FLUX\_ERR\_*           & 3.63~nJy &  Bulge+disk model flux uncertainty \\
BDF\_FLUX\_DERED\_CALIB\_*      & 3.63~nJy & Corrected bulge+disk model flux with AB zero-point$=30$ \\
BDF\_FLUX\_ERR\_DERED\_CALIB\_*          & 3.63~nJy & Corrected bulge+disk model flux uncertainty \\
BDF\_MAG\_DERED\_CALIB\_*      & AB mag & Corrected bulge+disk model magnitude \\
BDF\_MAG\_ERR\_DERED\_CALIB\_*          & AB mag & Corrected bulge+disk model magnitude uncertainty \\
 &  & \\
BDF\_T  & arcsec & Pre-seeing bulge+disk model size \\
BDF\_T\_RATIO & - & Bulge+disk model axis ratio \\
BDF\_FRACDEV  & - & Fraction of light contained in the bulge model component \\
 &  & \\
PSF\_FLUX\_*        & 3.63~nJy & PSF model flux with AB zero-point$=30$ \\
PSF\_FLUX\_ERR\_*            & 3.63~nJy & PSF model flux uncertainty \\
PSF\_MAG\_*         & AB mag & PSF model magnitude \\
PSF\_MAG\_ERR\_* & AB mag & PSF model magnitude uncertainty \\
PSF\_MAG\_DERED\_CALIB\_*       & AB mag & Corrected PSF model magnitude \\
PSF\_MAG\_ERR\_DERED\_CALIB\_*           & AB mag & Corrected PSF model magnitude uncertainty \\
PSF\_T   & arcsec & Model PSF size \\
 &  & \\
GAP\_FLUX\_*       & 3.63~nJy & Gaussian aperture flux with AB zero-point$=30$ \\
GAP\_FLUX\_ERR\_*           & 3.63~nJy & Gaussian aperture flux uncertainty \\
 &  & \\
EBV\_SFD98        & - & $E(B-V)$ value from \citet{1998ApJ...500..525S} \\
MASK\_FLAGS    & - & Flag containing mask information for DECam data: Unmasked sources=0 \\
MASK\_FLAGS\_NIR         & - & Flag containing mask information for VIRCam data: Unmasked sources=0 \\
FLAGS    & - & Flag identifier for DECam data \\
FLAGSTR              & - & Flag information string for DECam data \\
FLAGS\_NIR        & - & Flag identifier for VIRCam data \\
FLAGSTR\_NIR   & - & Flag information string for VIRCam data \\
FOF\_ID               & - & Friends-of-friends group identifier \\
FOF\_SIZE            & - & Number of objects in friends-of-friends group \\
 &  & \\
Z\_P & - & Maximum posterior redshift value \\
Z\_PEAK & - & Mean redshift of the most probable redshift peak \\
Z\_MC & - & Random draw from probability distribution function \\
PDZ\_** & - & Redshift probability density at $z=**$ \\
 &  & \\
KNN\_CLASS & - & Star-galaxy classification: Galaxy=1, Star=2, Ambiguous=3, Unclassified=0 \\

\hline
\end{longtable}
\twocolumn

\Tref{tab:DF_catalog_columns} provides descriptions of selected columns from our catalogue which will appear alongside other data products released as part of the DES Y3 cosmology analysis at \url{https://des.ncsa.illinois.edu/releases}.

\section*{Affiliations}

$^{1}$ Department of Astronomy, University of Geneva, ch. d’Ecogia 16, CH-1290 Versoix, Switzerland\\
$^{2}$ Center for Cosmology and Astro-Particle Physics, The Ohio State University, Columbus, OH 43210, USA\\
$^{3}$ Kavli Institute for Particle Astrophysics \& Cosmology, P. O. Box 2450, Stanford University, Stanford, CA 94305, USA\\
$^{4}$ Department of Astronomy, University of Illinois at Urbana-Champaign, 1002 W. Green Street, Urbana, IL 61801, USA\\
$^{5}$ National Center for Supercomputing Applications, 1205 West Clark St., Urbana, IL 61801, USA\\
$^{6}$ Brookhaven National Laboratory, Bldg 510, Upton, NY 11973, USA\\
$^{7}$ Department of Physics, University of Oxford, Denys Wilkinson Building, Keble Road, Oxford OX1 3RH, UK\\
$^{8}$ Jodrell Bank Center for Astrophysics, School of Physics and Astronomy, University of Manchester, Oxford Road, Manchester, M13 9PL, UK\\
$^{9}$ Department of Physics and Astronomy, University of Pennsylvania, Philadelphia, PA 19104, USA\\
$^{10}$ Centro de Investigaciones Energ\'eticas, Medioambientales y Tecnol\'ogicas (CIEMAT), Madrid, Spain\\
$^{11}$ Fermi National Accelerator Laboratory, P. O. Box 500, Batavia, IL 60510, USA\\
$^{12}$ Argonne National Laboratory, 9700 South Cass Avenue, Lemont, IL 60439, USA\\
$^{13}$ Institute of Astronomy, University of Cambridge, Madingley Road, Cambridge CB3 0HA, UK\\
$^{14}$ Kavli Institute for Cosmology, University of Cambridge, Madingley Road, Cambridge CB3 0HA, UK\\
$^{15}$ Physics Department, 2320 Chamberlin Hall, University of Wisconsin-Madison, 1150 University Avenue Madison, WI  53706-1390\\
$^{16}$ SLAC National Accelerator Laboratory, Menlo Park, CA 94025, USA\\
$^{17}$ George P. and Cynthia Woods Mitchell Institute for Fundamental Physics and Astronomy, and Department of Physics and Astronomy, Texas A\&M University, College Station, TX 77843,  USA\\
$^{18}$ University of Nottingham, School of Physics and Astronomy, Nottingham NG7 2RD, UK\\
$^{19}$ School of Mathematics and Physics, University of Queensland,  Brisbane, QLD 4072, Australia\\
$^{20}$ Department of Physics, Carnegie Mellon University, Pittsburgh, Pennsylvania 15312, USA\\
$^{21}$ Department of Astronomy and Astrophysics, University of Chicago, Chicago, IL 60637, USA\\
$^{22}$ Kavli Institute for Cosmological Physics, University of Chicago, Chicago, IL 60637, USA\\
$^{23}$ Santa Cruz Institute for Particle Physics, Santa Cruz, CA 95064, USA\\
$^{24}$ Jet Propulsion Laboratory, California Institute of Technology, 4800 Oak Grove Dr., Pasadena, CA 91109, USA\\
$^{25}$ Department of Physics, Stanford University, 382 Via Pueblo Mall, Stanford, CA 94305, USA\\
$^{26}$ Department of Physics, The Ohio State University, Columbus, OH 43210, USA\\
$^{27}$ Department of Applied Mathematics and Theoretical Physics, University of Cambridge, Cambridge CB3 0WA, UK\\
$^{28}$ Department of Physics, University of Michigan, Ann Arbor, MI 48109, USA\\
$^{29}$ Department of Astrophysical Sciences, Princeton University, Peyton Hall, Princeton, NJ 08544, USA\\
$^{30}$ Department of Physics, ETH Zurich, Wolfgang-Pauli-Strasse 16, CH-8093 Zurich, Switzerland\\
$^{31}$ Department of Physics, Duke University Durham, NC 27708, USA\\
$^{32}$ Cerro Tololo Inter-American Observatory, NSF's National Optical-Infrared Astronomy Research Laboratory, Casilla 603, La Serena, Chile\\
$^{33}$ Departamento de F\'isica Matem\'atica, Instituto de F\'isica, Universidade de S\~ao Paulo, CP 66318, S\~ao Paulo, SP, 05314-970, Brazil\\
$^{34}$ Laborat\'orio Interinstitucional de e-Astronomia - LIneA, Rua Gal. Jos\'e Cristino 77, Rio de Janeiro, RJ - 20921-400, Brazil\\
$^{35}$ Institute of Cosmology and Gravitation, University of Portsmouth, Portsmouth, PO1 3FX, UK\\
$^{36}$ CNRS, UMR 7095, Institut d'Astrophysique de Paris, F-75014, Paris, France\\
$^{37}$ Sorbonne Universit\'es, UPMC Univ Paris 06, UMR 7095, Institut d'Astrophysique de Paris, F-75014, Paris, France\\
$^{38}$ Department of Physics and Astronomy, Pevensey Building, University of Sussex, Brighton, BN1 9QH, UK\\
$^{39}$ Department of Physics \& Astronomy, University College London, Gower Street, London, WC1E 6BT, UK\\
$^{40}$ Instituto de Astrofisica de Canarias, E-38205 La Laguna, Tenerife, Spain\\
$^{41}$ Universidad de La Laguna, Dpto. Astrofisica, E-38206 La Laguna, Tenerife, Spain\\
$^{42}$ Institut de F\'{\i}sica d'Altes Energies (IFAE), The Barcelona Institute of Science and Technology, Campus UAB, 08193 Bellaterra (Barcelona) Spain\\
$^{43}$ Institut d'Estudis Espacials de Catalunya (IEEC), 08034 Barcelona, Spain\\
$^{44}$ Institute of Space Sciences (ICE, CSIC),  Campus UAB, Carrer de Can Magrans, s/n,  08193 Barcelona, Spain\\
$^{45}$ INAF-Osservatorio Astronomico di Trieste, via G. B. Tiepolo 11, I-34143 Trieste, Italy\\
$^{46}$ Institute for Fundamental Physics of the Universe, Via Beirut 2, 34014 Trieste, Italy\\
$^{47}$ Observat\'orio Nacional, Rua Gal. Jos\'e Cristino 77, Rio de Janeiro, RJ - 20921-400, Brazil\\
$^{48}$ Department of Astronomy, University of California, Berkeley,  501 Campbell Hall, Berkeley, CA 94720, USA\\
$^{49}$ Department of Physics, IIT Hyderabad, Kandi, Telangana 502285, India\\
$^{50}$ Faculty of Physics, Ludwig-Maximilians-Universit\"at, Scheinerstr. 1, 81679 Munich, Germany\\
$^{51}$ Department of Astronomy/Steward Observatory, University of Arizona, 933 North Cherry Avenue, Tucson, AZ 85721-0065, USA\\
$^{52}$ Institute of Theoretical Astrophysics, University of Oslo. P.O. Box 1029 Blindern, NO-0315 Oslo, Norway\\
$^{53}$ Instituto de Fisica Teorica UAM/CSIC, Universidad Autonoma de Madrid, 28049 Madrid, Spain\\
$^{54}$ Department of Astronomy, University of Michigan, Ann Arbor, MI 48109, USA\\
$^{55}$ Center for Astrophysics $\vert$ Harvard \& Smithsonian, 60 Garden Street, Cambridge, MA 02138, USA\\
$^{56}$ Australian Astronomical Optics, Macquarie University, North Ryde, NSW 2113, Australia\\
$^{57}$ Lowell Observatory, 1400 Mars Hill Rd, Flagstaff, AZ 86001, USA\\
$^{58}$ Department of Astronomy, The Ohio State University, Columbus, OH 43210, USA\\
$^{59}$ Radcliffe Institute for Advanced Study, Harvard University, Cambridge, MA 02138\\
$^{60}$ Instituci\'o Catalana de Recerca i Estudis Avan\c{c}ats, E-08010 Barcelona, Spain\\
$^{61}$ Max Planck Institute for Extraterrestrial Physics, Giessenbachstrasse, 85748 Garching, Germany\\
$^{62}$ School of Physics and Astronomy, University of Southampton,  Southampton, SO17 1BJ, UK\\
$^{63}$ Computer Science and Mathematics Division, Oak Ridge National Laboratory, Oak Ridge, TN 37831\\
$^{64}$ Universit\"ats-Sternwarte, Fakult\"at f\"ur Physik, Ludwig-Maximilians Universit\"at M\"unchen, Scheinerstr. 1, 81679 M\"unchen, Germany\\
$^{65}$ Institute for Astronomy, University of Edinburgh, Edinburgh EH9 3HJ, UK\\

\bsp	
\label{lastpage}
\end{document}